\documentclass[aps,prx,superscriptaddress,twocolumn,amsmath]{revtex4-2}

\usepackage{graphicx, color}
\usepackage{xcolor}
\usepackage{amsfonts,amssymb,amsmath}
\usepackage{siunitx}
\usepackage{bm}
\usepackage{bbold}
\usepackage{braket}
\usepackage{mathtools}
\usepackage{dsfont}
\usepackage{lineno}
\usepackage[normalem]{ulem}

\UseRawInputEncoding
\begin{document}

\title{Gigantic magnetochiral anisotropy in the topological semimetal ZrTe$_5$}

\author{Yongjian Wang}
\affiliation{Physics Institute II, University of Cologne, Z\"ulpicher Str. 77, 50937 K\"oln, Germany}

\author{Henry F. Legg}
\affiliation{Institute for Theoretical Physics, University of Cologne, Z\"ulpicher Str. 77, 50937 K\"oln, Germany}
\affiliation{Department of Physics, University of Basel, Klingelbergstrasse 82, CH-4056 Basel, Switzerland}

\author{Thomas B\"omerich}
\affiliation{Institute for Theoretical Physics, University of Cologne, Z\"ulpicher Str. 77, 50937 K\"oln, Germany}

\author{Jinhong Park}
\affiliation{Institute for Theoretical Physics, University of Cologne, Z\"ulpicher Str. 77, 50937 K\"oln, Germany}

\author{Sebastian~Biesenkamp}
\affiliation{Physics Institute II, University of Cologne, Z\"ulpicher Str. 77, 50937 K\"oln, Germany}

\author{A. A. Taskin}
\affiliation{Physics Institute II, University of Cologne, Z\"ulpicher Str. 77, 50937 K\"oln, Germany}

\author{Markus Braden}
\affiliation{Physics Institute II, University of Cologne, Z\"ulpicher Str. 77, 50937 K\"oln, Germany}

\author{Achim Rosch}
\affiliation{Institute for Theoretical Physics, University of Cologne, Z\"ulpicher Str. 77, 50937 K\"oln, Germany}

\author{Yoichi Ando}
\email[]{ando@ph2.uni-koeln.de}
\affiliation{Physics Institute II, University of Cologne, Z\"ulpicher Str. 77, 50937 K\"oln, Germany}

\begin{abstract}
Topological materials with broken inversion symmetry can give rise to nonreciprocal responses, such as the current rectification controlled by magnetic fields via magnetochiral anisotropy. Bulk nonreciprocal responses usually stem from relativistic corrections and are always very small. Here we report our discovery that ZrTe$_5$ crystals in proximity to a topological quantum phase transition present gigantic magnetochiral anisotropy, which is the largest ever observed to date. We argue that a very low carrier density, inhomogeneities, and a torus-shaped Fermi surface induced by breaking of inversion symmetry in a Dirac material are central to explain this extraordinary property.
\end{abstract}
\maketitle


The magnetochiral anisotropy (MCA) is a nonreciprocal transport effect induced by an external magnetic field in a chiral or polar system without inversion symmetry. Nonreciprocal response means that the resistance $R$ of a material is different for electrical current $\mathbf{I}$ flowing to the right ($+I$) and to the left ($-I$), which immediately implies broken inversion symmetry. 
Remarkably, nonreciprocal transport can be triggered and controlled by external magnetic fields. Depending on the mechanism, there are two possible types of the nonreciprocal resistance   \cite{Tokura2018}: one is the inner-product type \cite{Rikken2001} expressed by $R = R_0[1+\gamma(\mathbf{B} \cdot \mathbf{I})]$ (where $R_0$ is the reciprocal resistance and $\gamma$ is a numerical coefficient), and the other is the vector-product type \cite{Rikken2005} expressed by $R = R_0[1+\gamma (\hat{\mathbf{P}} \times \mathbf{B}) \cdot \mathbf{I}]$, where $\hat{\mathbf{P}}$ is a unit vector which characterizes the axis of the nonreciprocal effect. The spin-texture of Fermi surfaces in topological materials can give rise to such MCAs, with known examples of both types \cite{Tokura2018}. 

The coefficient $\gamma \equiv [(R/R_0)-1]/(|B|\cdot|I|)$, obtained for $\mathbf{B} \parallel \mathbf{I}$ for the inner-product type and for $\mathbf{B} \perp \mathbf{I}$ with $(\mathbf{B} \times \mathbf{I}) \perp \hat{\mathbf{P}}$ for the vector-product type, is usually used as a measure of the MCA \cite{Tokura2018}. However, this $\gamma$ depends on the shape/size of the specimen used for the measurement, and a better measure for a bulk material is the normalized coefficient $\gamma' \equiv A_\perp \gamma$, where $A_\perp$ is the cross-section of the specimen \cite{Ideue2017}. As a materials property, the MCA is usually of relativistic origin and has been ubiquitously found to be very small. Recently, tellurium was shown to have an inner-product type MCA with $|\gamma'|$ of 10$^{-8}$ m$^2$T$^{-1}$A$^{-1}$  \cite{Rikken2019}, which is the largest reported as a bulk property. It was theoretically predicted that the chiral anomaly in Weyl semimetals may lead to a large MCA of the inner-product type \cite{Morimoto2016}, but there has been no confirmation. In this Letter, we report that  topological semimetal ZrTe$_5$ presents a vector-product type MCA with $|\gamma'|$ of up to $4 \times 10^{-7}$ m$^2$T$^{-1}$A$^{-1}$ as its bulk property.

ZrTe$_5$ has an orthorhombic layered structure which nominally belongs to the $Cmcm$ ($D^{17}_{2h})$ space group \cite{Weng2014} (the actual symmetry is, however, lower, see below). The crystal structure consists of two-dimensional (2D) layers stacked along the $b$ axis via van-der-Waals interactions (Fig. 1a). In each layer (i.e. $ac$ plane), ZrTe$_3$ chains running along the $a$ axis, proving the highest conductivity along the $a$ axis. In transport studies, the principal crystal axes $a$, $c$ and $b$ correspond to the directions $x$, $y$ and $z$, respectively \cite{Weng2014}. Bulk single crystals of ZrTe$_5$ have been a focus of significant interest in recent years \cite{Xu2018, RYChen2015, Li2016, YZhang2017, Liang2018, HWang2018, Shahi2018, Tang2019, Sun2020}, with major discoveries such as chiral magnetic effects \cite{Li2016}, unconventional anomalous Hall effect \cite{Liang2018}, and three-dimensional (3D) quantum Hall effect \cite{Tang2019}. While initially there was a debate about the electronic structure realized in ZrTe$_5$, it is now generally believed that in most samples there is a temperature-driven transition from a strong 3D topological insulator (TI) phase to a weak 3D TI phase with increasing temperature and that a pronounced resistivity peak marks a gapless semimetal realised between the two gapped TI phases \cite{Xu2018}, although there are still other interpretations \cite{Fu2020, Wang2021}. In this work, we focus on ZrTe$_5$ crystals (grown by a Te-flux method \cite{SM}) whose resistivity is maximum at base temperature (Fig. 1b), suggesting that the system has been tuned to a semimetalic state. Our detailed data discussed below indeed support the realization of the semimetallic state.

\begin{figure}[t]
	\centering
	\includegraphics[width=8.6cm]{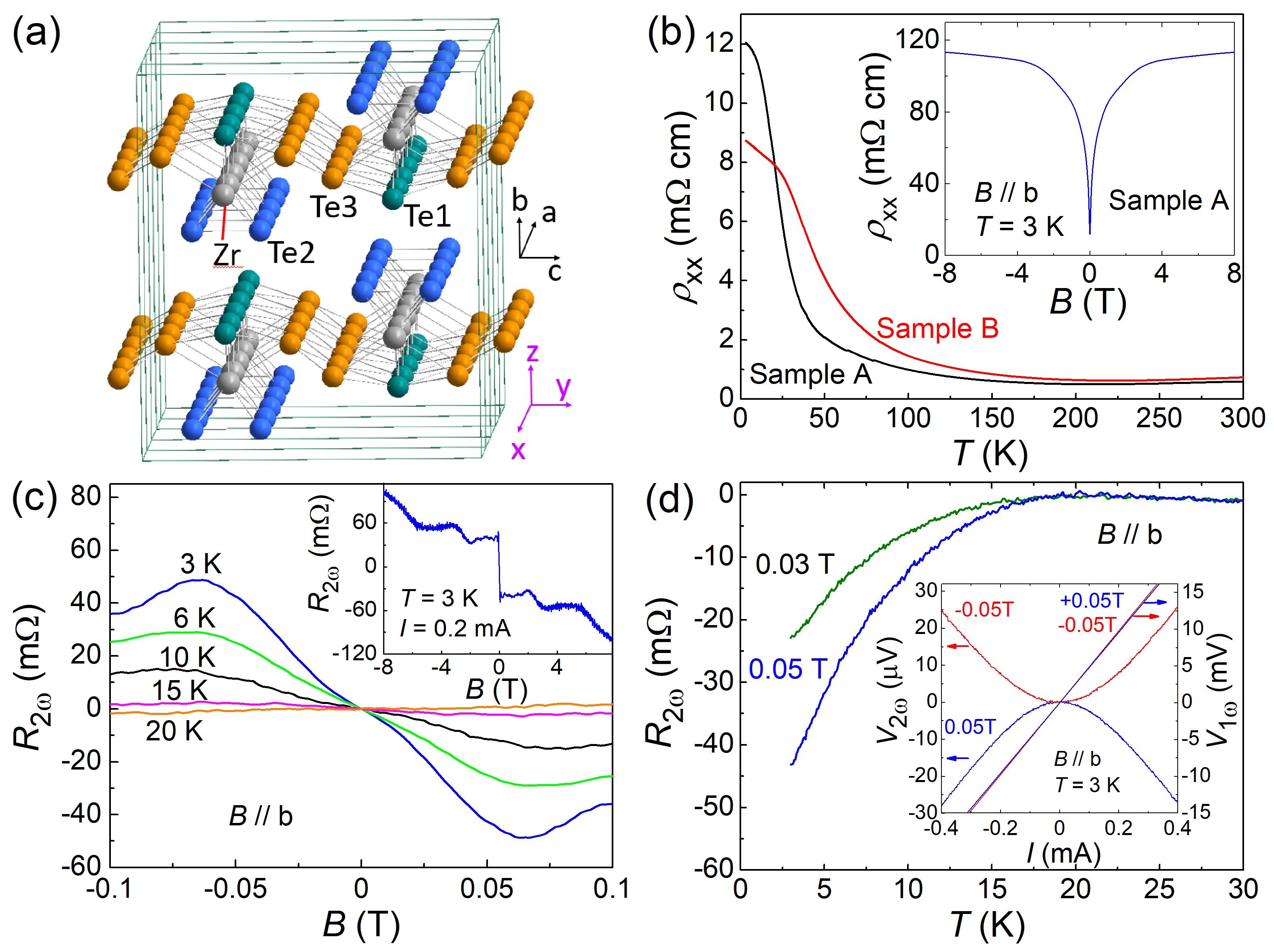}
	\caption{Structure and transport properties of ZrTe$_5$. (a) Layered crystal structure of ZrTe$_5$ (b) Temperature dependence of the resistivity $\rho_{xx}$ of samples A and B measured with $I \parallel a$. Inset: magnetoresistance of sample A for $B \parallel b$ at 3 K. (c) Magnetic-field dependence of the second-harmonic component of the resistance, $R_{2\omega}$, of sample A in $B \parallel b$ for low fields at various temperatures; inset shows the data at 3 K for a wider range of $B$ up to $\pm8$ T. (d) Temperature dependence of $R_{2\omega}$ of sample A measured in $B \parallel b$. Inset: Current dependencies of the first- and second-harmonic voltages, $V_{1\omega}$ and $V_{2\omega}$, in $+0.05$ T and $-0.05$ T at 3 K; the slight difference in $V_{1\omega}$ for opposite $B$ is due to a small admixture of Hall voltage. Throughout this paper, whenever $R_{2\omega}$ is shown, it was measured with $I_{\rm ac}$ = $I_0/\sqrt{2}$ = 0.2 mA.
	}
	\label{fig:1}
\end{figure}

As was already reported \cite{HWang2018, Shahi2018}, these semimetal samples in perpendicular magnetic fields present unconventional magnetoresistance, which is singular at low fields and saturates in high fields, as shown in the inset of Fig. 1(b) for sample A. We measured the resistivity $\rho_{xx}$ with a low-frequency AC excitation $I = I_0 \sin \omega t$ along the $a$-axis and, when the second-harmonic component $R_{2\omega}$ was probed, we discovered an unusually large signal [Fig. 1(c) inset] whose magnetic-field ($B$) dependence is totally different from that of the first harmonic.  As discussed in \cite{SM}, this $R_{2\omega}$ directly reflects $\gamma$. Note that the physics behind $R_{2\omega}$ is totally different from the second-harmonic generation in the optical range \cite{Hafez2018, Sun2019, Kovalev2020, Cheng2020}, which is a photonic process at much higher energy \cite{Tokura2018}.
The main panel of Fig. 1(c) shows that $|R_{2\omega}|$ grows rapidly and almost linearly with $B$ in a narrow range of $|B| \lesssim 0.06$ T. This component shows up only below 20 K (Fig. 1(d)). The second-harmonic voltage $V_{2\omega}$ depends quadratically on the current $I$, is observed for $\mathbf{B} \perp \mathbf{I}$, and is antisymmetric with respect to $B$ [Fig. 1(d) inset], which is the behaviour expected for the vector-product type, $V_{2\omega} = \gamma R_0 I (\hat{\mathbf{P}} \times \mathbf{B}) \cdot \mathbf{I}$. In contrast, the first-harmonic voltage $V_{1\omega}$ is linear in $I$ and symmetric with respect to $B$ [Fig. 1(d) inset].

\begin{figure*}[t]
	\centering
	\includegraphics[width=12.6cm]{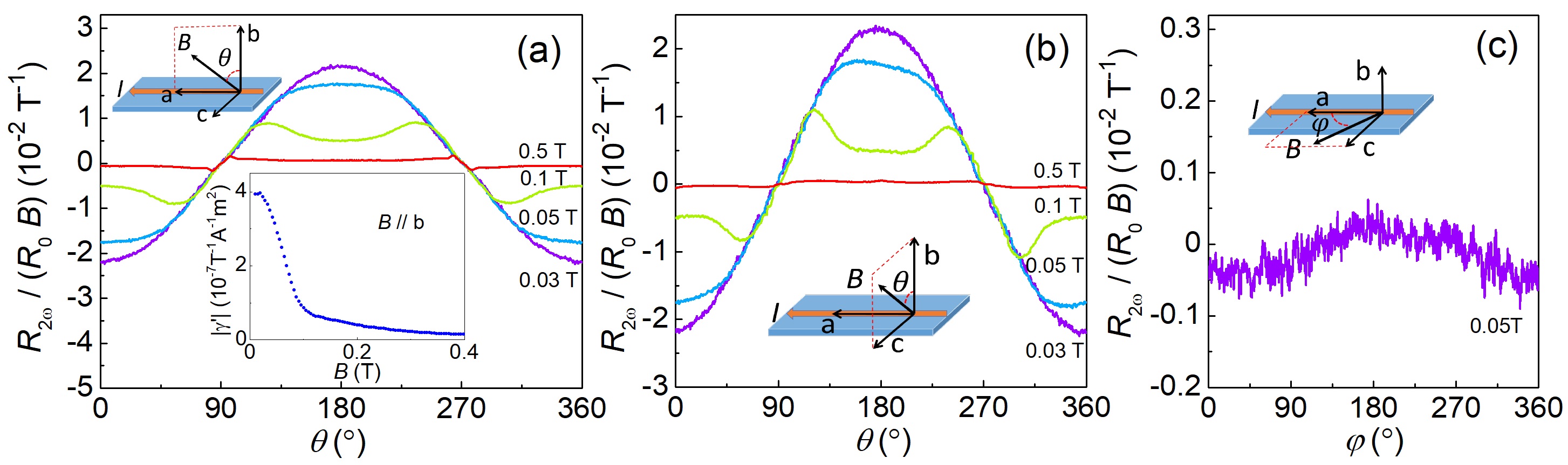}
	\caption{Symmetry of the second-harmonic signal. (a-c) Magnetic-field-orientation dependencies of $R_{2\omega}/(R_0 B)$ in sample A measured at 3 K in 0.03, 0.05, 0.1, and 0.5 T (except for (c) which is only for 0.05 T) as the magnetic field was rotated in the $ab$, $bc$, and $ac$ planes. The rotation plane and the definition of the angle ($\theta$ or $\varphi$) are shown in each panel. The lower inset of (a) shows the $B$-dependence of $|\gamma'|$ [$\equiv 2A_\perp | R_{2\omega}/(R_0 B I_0)| $].
	}
	\label{fig:2}
\end{figure*}

To identify the axis of the characteristic unit vector $\hat{\mathbf{P}}$, we have performed the measurements of $R_{2\omega}$ in varying orientations of the magnetic field rotated in the $ab$, $bc$, and $ac$ planes. The results are shown in Fig. 2, where $R_{2\omega}$ is normalized by $R_0 B$ to factor out the change in the reciprocal response \cite{SM}. In both the $ab$- and $bc$-plane rotations, $R_{2\omega}/(R_0 B)$ at very low field, 0.03 T, shows a $\cos \theta$ dependence ($\theta$ is measured from the $b$ axis), while $R_{2\omega}/ (R_0 B)$ remains essentially zero in the $ac$-plane rotation. Since $\mathbf{I}$ is along the $a$ axis, this result indicates that $\hat{\mathbf{P}}$ is along the $c$ axis. 
Detailed magnetic-field-orientation dependencies of $\rho_{xx}$ \cite{SM} suggest that inversion symmetry is broken and, in particular, $ab$ and $ac$ are not mirror planes while $bc$ is likely still a mirror plane. This suggests the lowering of the crystal symmetry from the space group $Cmcm$ to $Cm$. 
To corroborate this conclusion, we performed comprehensive single-crystal X-ray diffraction (XRD) studies, which actually detected broken inversion symmetry at room temperature \cite{SM}. The main distortion to break the inversion symmetry was found to be staggered displacements of Te3 atoms along the $c$ axis \cite{SM}. No further symmetry breaking was detected between 30 to 300~K.

As shown in the inset of Fig. 2(a), the value of $|\gamma'|$ = $2A_\perp | R_{2\omega}/(R_0 B I_0) |$ for $B \parallel b$-axis \cite{SM} is strongly enhanced at low fields and reaches $4 \times 10^{-7}$ m$^2$T$^{-1}$A$^{-1}$, which is gigantic \cite{Tokura2018}. In the following, we focus on the behavior at low fields. The behavior of $R_{2\omega}$ in high magnetic fields (in the ultra-quantum limit) is much more complicated and requires a separate study \cite {SM}.

The MCA is triggered by the combined effect of crystalline symmetry breaking and an external magnetic field. To explore whether the gigantic effect can be explained within existing theories \cite{Ideue2017} which focus on effects arising from the field-induced deformation of the Fermi surface, it is essential to identify both experimentally and theoretically the relevant band structure. The topological semimetal state of ZrTe$_5$ is usually considered to be a 3D Dirac semimetal in zero magnetic field \cite{RYChen2015} and it was claimed, based on the observation of negative longitudinal magnetoresistance \cite{Li2016} and anomalous Hall effect \cite{Liang2018}, that a Weyl semimetal state is realised in magnetic field. 
To derive an effective low-energy Hamiltonian, we start from the Dirac semimetal obtained in Ref. \citenum{RYChen2015} based on symmetry arguments and comparison with band-structure calculations \cite{Weng2014} assuming a high-symmetry $Cmcm$ space group. It is formulated using the basis states $( |\Psi^{\uparrow}_+\rangle, |\Psi^{\uparrow}_-\rangle, |\Psi^{\downarrow}_+\rangle, |\Psi^{\downarrow}_-\rangle)$, 
where the $\pm$ index describes linear combinations of Te $p_y$ orbitals of even and odd parity \cite{RYChen2015}. Taking all experimentally observed symmetry breaking into account, we arrive at the following minimal model to describe ZrTe$_5$ \cite{SM}
\begin{equation}
\begin{split}
H=m \mathbb{1} \otimes \tau_z+\hbar (v_a k_a \sigma_z \otimes \tau_x +v_b k_b \sigma_x \otimes \tau_x \\
 + v_c k_c \mathbb{1} \otimes \tau_y)+\Delta \mathbb{1} \otimes \tau_x + \xi \sigma_x \otimes \tau_y -\mu \mathbb{1}. \label{Hamiltonian2}
\end{split}
\end{equation}

Here the space of the four lowest bands is spanned by $4\times 4$ matrices of the form $\sigma_{\alpha} \otimes \tau_{\beta}$ where the Pauli matrices $\sigma_{\alpha}$ and $\tau_\beta$ act on the spin and parity space, respectively. 
The mass of the Dirac bands, $m$, is approximately tuned to zero in our samples of ZrTe$_5$ such that we consider $m=0$ throughout. Importantly, the constant terms $\Delta$ and $\xi$ describe the effect of $ab$- and $ac$-mirror symmetry breaking (respectively) as indicated by our experimental probes. A finite $\Delta$ or $\xi$ splits the Dirac point into two massive bands and a nodal line \cite{SM}. The nodal line lies in a plane rotated about the $a$-axis from the $ab$-plane by the angle $\theta_{\rm tilt}$, defined via $\cos\theta_{\rm tilt}=\frac{\Delta}{\sqrt{\Delta^2+v_b^2 \xi^2/v_c^2}}\approx1-\frac{v_b^2 \xi^2}{2\Delta^2 v_c^2}$. However, in ZrTe$_5$ the Fermi-velocities satisfy $v_c \gg v_b$ and so the angle $\theta_{\rm tilt}$ is likely very small (indeed we determine experimentally that $\theta_{\rm tilt}\lesssim1^{\rm o}$, see below). Upon doping the system slightly, one obtains a Fermi surface with a torus shape wrapping around the nodal line, see Fig. S15 in \cite{SM}.

\begin{figure}[t]
	\centering
	\includegraphics[width=8.6cm]{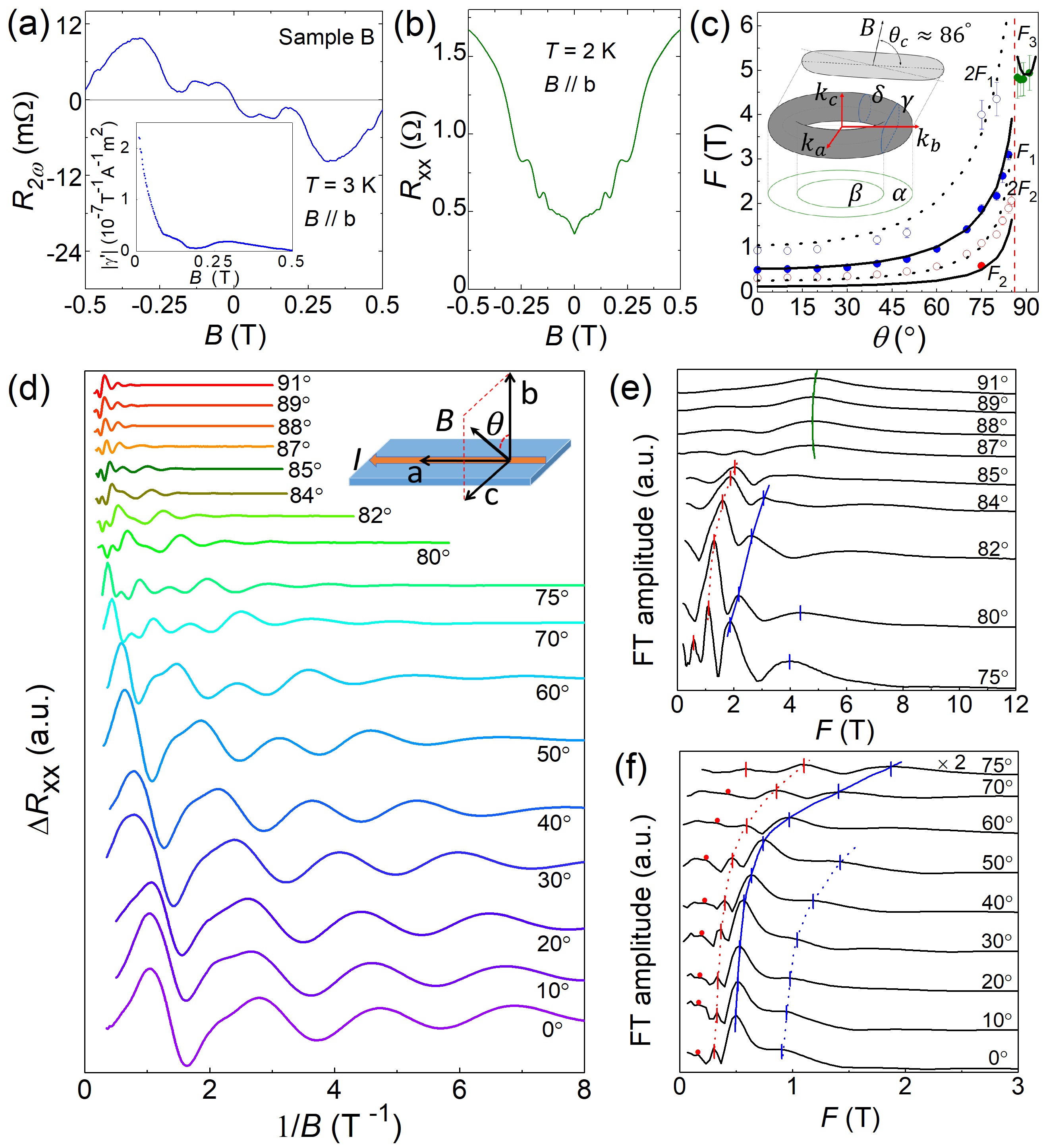}
	\caption{\linespread{1.05}\selectfont{}
		Shubnikov-de Haas (SdH) oscillations. (a) Magnetic-field dependence of $R_{2\omega}$ for sample B measured in $B \parallel b$ at 3 K; inset shows the $B$-dependence of the calculated $|\gamma'|$. (b) Resistance component $R_{xx}$ measured across a transverse electrode pair \cite{SM}.
		(c) Experimentally-observed SdH frequencies (symbols) and the theoretical fits based on the torus Fermi surface (lines); inset shows the schematic diagram of a torus Fermi surface and its extremal orbits $\alpha$, $\beta$, $\delta$, and $\gamma$. As discussed in detail in \cite{SM}, the frequencies $F_1$, $F_2$, and $F_3$ correspond to $\gamma$, $\delta$, and $\beta$ orbits, respectively, and $2F_1$ ($2F_2$) is the second harmonic of $F_1$ ($F_2$); error bars are shown only when they are larger than the symbol size.
		(d) SdH oscillations in magnetic fields rotated in the $bc$ plane, measured in $R_{xx}$ after subtracting a smooth background.	(e,f) Results of Fourier transforms of the SdH oscillations; ticks mark obvious peaks, and red dots mark the expected position of $F_2$ based on its 2nd harmonic, $2F_2$.
	}
	\label{fig:3}
\end{figure}

We now consider how the torus Fermi-surface can explain the presence of MCA. In the experimentally relevant case $\theta_{\rm tilt}\approx0$, the spin is locked to the momentum such that its orientation is $-(\hat{\mathbf{a}}\sin\varphi+ \hat{\mathbf{b}}\cos\varphi)$, i.e. always parallel to the nodal line plane, and  the chirality of this texture is controlled by the sign of $\Delta$. A magnetic field in the $b$-direction ($z$-axis) provides additional Zeeman energy that leads to a distortion of the Fermi surface necessary for obtaining nonreciprocal transport where the vector $\hat{\mathbf{P}}$ is set by the spin-texture such that it is parallel to the normal of the plane containing the nodal line (i.e. parallel to the $c$-direction).
We have adapted the theory of Ref. \citenum{Ideue2017} to this situation and also explored a novel mechanism for nonreciprocal transport due to the anisotropic scattering resulting from the matrix-element effects \cite{SM}.
In both cases, we obtain  a nonreciprocal response of the form
\begin{equation}
|\gamma'|
\approx \eta \frac{3 g_b \mu_B}{8 \pi e v_a n \Delta} \label{gammaPAna}\,\,,
\end{equation}
where $g_b \approx$ 20 is the $g$-factor for a field in $b$ direction \cite{RYChen2015, Liu2016, Sun2020}, $e$ the electron charge, and $\Delta\gg \mu$ such that there is only a single Fermi surface. We find $\eta=1$ for the mechanism of Ref. \citenum{Ideue2017} and $\eta=3$ for the anisotropic scattering. For both mechanisms, $\gamma'$ will be strongly enhanced in the limit of small symmetry breaking  $\Delta$ and small carrier doping with density $n$. In other words, a substantial MCA is expected only when both $\Delta$ and $\mu$ are very small.

Because the torus shape of the Fermi surface predicted from the broken mirror symmetries is crucial for the presence of MCA, we have performed quantum oscillation experiments, which would allow us to estimate the parameters for Eq.~\eqref{gammaPAna}.
For this purpose, we have grown a new batch of single crystals that are cleaner than sample A to observe quantum oscillations. One of such crystals (sample B) not only reproduced the gigantic $R_{2\omega}$ [Fig. 3(a)] but also presented clear Shubnikov-de Haas (SdH) oscillations [Figs. 3(a) and 3(b)]; the oscillations were observed only at low fields, because the Fermi surface is extremely small and the system enters the ultra-quantum limit already at $\sim$1 T for $B \parallel b$-axis. 
The evolution of the SdH-oscillation data when the direction of the $B$ field was rotated within the $bc$ plane is shown in Fig. 3(d), with their Fourier transforms presented in Figs. 3(e) and 3(f) (see \cite{SM} for details). 
Since the putative torus Fermi surface is expected to lie approximately in the $ab$ plane, one would expect a switching of the extremal orbits (from $\delta$ and $\gamma$ to $\alpha$ and $\beta$ illustrated in Fig. 3(c) inset) above a critical angle when the $B$-field direction approaches the $c$ axis \cite{Kwan2020}.
In fact, multiple frequencies were observed for most of the field orientations and their angle-dependencies show a break between 85$^{\circ}$ and 87$^{\circ}$ [Fig. 3(c)]; both observations are at odds with an elliptical Fermi surface but consistent with a torus Fermi surface \cite{Kwan2020}. From our fits we obtain a tiny electron density $n \approx 2.3\times 10^{16}$\,cm$^{-3}$ corresponding to the chemical potential $\mu$ = 4.9 meV, a small value for $\Delta\approx 19.1$\,meV, and determine $\theta_{\rm tilt}\lesssim1^{\rm o}$ \cite{SM}. The extremely small $\mu$ implies that the Fermi surface is thermally smeared already at $\sim$50 K, explaining why MCA diminishes with increasing $T$ in Fig. 1(d). 

Using the parameters that explain the dispersion in the SdH-oscillation data \cite{SM}, we find Eq.~\eqref{gammaPAna} predicts $|\gamma'| \sim 1 \times 10^{-11}\;{\rm m^2 A^{-1}T^{-1}}$. This is a relatively large value compared to other materials but four orders of magnitude smaller than our measured value. We conclude that the deformation of the Fermi surface by the Zeeman effect is not sufficient to explain the gigantic MCA. We have also checked \cite{SM} that orbital effects of the magnetic field and further perturbations of the minimal model Eq.~\eqref{Hamiltonian2} cannot naturally explain such a large effect.

A likely mechanism giving rise to the giant enhancement of nonlinear transport in ZrTe$_5$ are large-scale fluctuations in the electronic density as they may arise due to the unavoidable presence of charged impurities \cite{Skinner2012, Borgwardt2016}. In regions of low density, local electric fields and therefore nonlinear effects can be strongly enhanced \cite{SM}. Two experimental  observations strongly support such a scenario in ZrTe$_5$. First, inhomogeneities triggered by charged impurities may naturally form in ZrTe$_5$ due to the extremely small  carrier density of only $5\times 10^{-6}$ electrons per formula unit (2.3$\times$10$^{16}$ cm$^{-3}$), which also suppresses screening. Second, more directly, a comparison of our quantum oscillation data to resistivity reveals that the measured resistivity is much higher than that expected for a homogeneous material; namely, we found that the transport scattering rate extracted from the resistivity is almost an order of magnitude larger than the scattering rates obtained from the decay of SdH oscillations \cite{SM}. This is naturally understood by assuming that transport is forced to occur through regions with high resistivity, while quantum oscillations arise from areas with fewer scattering events and lower resistivity.
The anisotropic Fermi velocities characteristic for ZrTe$_5$ and the resulting quasi-one-dimensional transport are also of relevance for this effect as it suppresses electron flow around obstacles.

The reproducibility of this striking phenomenon is confirmed in 10 more samples showing the resistivity maximum close to 0 K \cite{SM}, which all presented $|\gamma'|$ of similar order. Nevertheless, its exact value varied among samples and we found no clear correlation between the residual resistivity $\rho_0$ and $|\gamma'|$; such a strong sample dependence is consistent with the puddle scenario. Note that we did not intentionally introduce impurities and their distribution is random. The sign of $\gamma'$ was also sample dependent, suggesting that the sign of the $\hat{\mathbf{P}}$ vector is randomly fixed, possibly by an anisotropic strain created upon cooling. In samples having the resistivity-peak temperature $T_{\rm p}$ of 15 -- 50 K, a finite $|\gamma'|$ which decreases with $T$ was observed, but $|\gamma'|$ was no longer discernible in samples with $T_{\rm p} \simeq$ 130 K \cite{SM}. 
The suppression of $|\gamma'|$ in higher $T_{\rm p}$ samples most likely originates from an increased carrier density of those samples which also suppresses large density fluctuations.
  
In conclusion, close to the topological phase transition, ZrTe$_5$ is a topological semi-metal with a torus-shaped Fermi-surface. As a result of the proximity to the topological phase transition, this Fermi-surface possesses a spin texture that specifies the $\hat{\mathbf{P}}$ vector responsible for a large MCA, which is further enhanced by large-scale electron density fluctuations in ZrTe$_5$. This intriguing finding points to rich physics in nonreciprocal transport taking place in topological materials with extremely low carrier density.

\acknowledgements{This project has received funding from the European Research Council (ERC) under the European Union's Horizon 2020 research and innovation programme (grant agreement No 741121) and was also funded by the Deutsche Forschungsgemeinschaft (DFG, German Research Foundation) under CRC 1238 - 277146847 (Subprojects A02, A04 and C02) as well as under Germany's Excellence Strategy - Cluster of Excellence Matter and Light for Quantum Computing (ML4Q) EXC 2004/1 - 390534769. }

\clearpage
\onecolumngrid

\renewcommand{\thefigure}{S\arabic{figure}} 
\renewcommand{\thetable}{\Roman{table}} 

\setcounter{figure}{0}

{\bf{\large Supplemental Material}
}

\section{Supplemental Data and Discussions}

\subsection{Methods}

{\bf Crystal growth.} 
Single crystals of ZrTe$_5$ with the resistivity-peak temperature $T_{\rm p}$ = 0 K were grown by a Te-flux method. High-purity zirconium (99.8\% for sample A and 99.9\% for sample B) and tellurium (99.9999\%) were loaded in an quartz tube with a molar ratio of Zr:Te = 1:70. The sealed quartz tube was heated to 860 $^{\circ}$C and kept for 24 h with intermittent shaking to ensure a homogeneity of the melt, followed by cooling rapidly to 660 $^{\circ}$C. The tube was then cooled to 460 $^{\circ}$C in 200 h. The ZrTe$_5$ crystals were isolated from the Te flux by centrifuging at 460 $^{\circ}$C. The samples with $T_{\rm p}$ = 15 -- 50 K were grown with the same method except for the molar ratio of Zr:Te = 1:40.
The samples with $T_{\rm p}$ = 138 K were grown by a chemical vapor transport method by using I$_2$ as transport agent; high-purity raw materials were loaded in an quartz tube with a molar ratio of Zr:Te = 1:5.5, and the tube was placed in a two-temperature-zone furnace with $T_{\rm high}$ = 530 $^{\circ}$C and $T_{\rm low}$ = 480 $^{\circ}$C for 1 week.

{\bf Second-harmonic resistance $R_{2\omega}$.}
The voltage is given by $V = R_0 I (1+\gamma BI)$ for $I \parallel a$ and $B \parallel b$, with which the nonreciprocal response is maximal. For an AC current $I = I_0 \sin \omega t$, this becomes $V = R_0 I_0 \sin \omega t + \frac{1}{2} \gamma R_0 B I_0^2 [1 + \sin (2\omega t - \frac{\pi}{2})]$. Therefore, we identify $R_{2\omega} = \frac{1}{2} \gamma R_0 B I_0$ from the out-of-phase component of the AC voltage at the frequency of $2\omega$.

{\bf Transport measurements.}
To make good electrical contacts, the surface of a bulk single crystal was cleaned by Argon plasma to remove the oxidized layer and gold contact electrodes were sputter-deposited. The relevant dimensions of sample A (B) were the thickness 14 (23) $\mu$m, the width 172 (100) $\mu$m, and the voltage-contact distance 421 (238) $\mu$m. Transport measurements were performed in a Quantum Design Physical Properties Measurement System (PPMS) with a rotating sample holder. Both the first- and second-harmonic signals of the resistance and the Hall resistance were measured in the four-terminal configuration using a low-frequency (13.777 Hz) AC lock-in technique. During the AC resistance measurements, the phase of the first- and second-harmonic signals were confirmed to be approximately 0$^{\circ}$ and 90$^{\circ}$, respectively.

{\bf X-ray diffraction.}
Complete sets of Bragg reflection intensities were taken at room temperature and at 100 and 30\,K using a dual flow nitrogen and helium gas cooler n-Helix on an x-ray single-crystal diffractometer Bruker X8 Apex equipped with a CCD detector (Mo K$_\alpha$ radiation).

{\bf Theory.}
The low-energy Hamiltonian Eq.~(1) and mirror-symmetry breaking term were obtained from a small momentum expansion, utilising the symmetries of ${\rm ZrTe_5}$. First and second harmonic transport coefficients were calculated using the Boltzmann equation to second order in the electric field, including orbital effects, and using various approximations to the collision integral.

\begin{figure}[b]
	\centering
	\includegraphics[width=\textwidth]{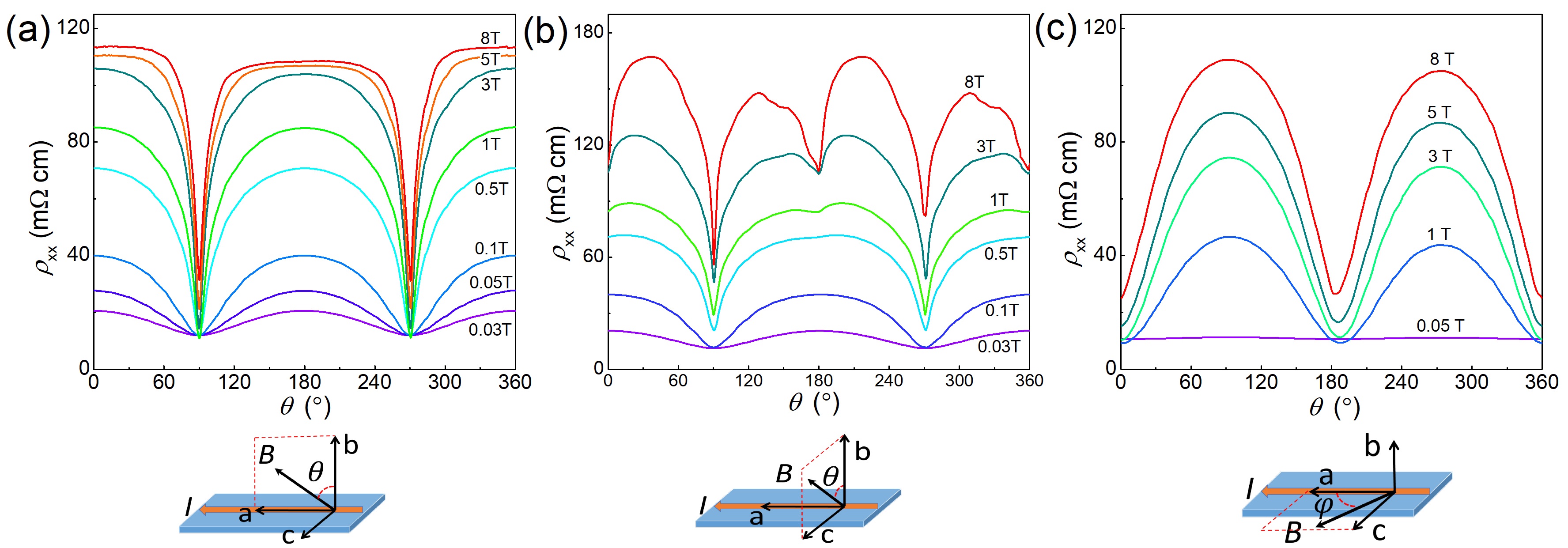}
	\caption{\textbf{Symmetry of the magnetoresistance in sample A.} (a-c) Magnetic-field-orientation dependencies of the magnetoresistance measured at 3 K when the magnetic field is rotated in the $ab$, $bc$, and $ac$ planes. The data are shown for some selected strengths of the magnetic field indicated in each panel. The schematic diagrams below the plots depict the measurement configurations.
	}
	\label{fig:S1}
\end{figure}

\subsection{Magnetic-field-orientation dependencies of $\rho_{xx}$}

The dependence of the resistivity $\rho_{xx}$ of sample A on the orientation of the applied magnetic field was measured with the magnetic field $B$ rotated in the $ab$, $bc$ and $ac$ planes (Figs. S1(a)--S1(c)). These angular dependencies indicate that the mirror symmetry is broken with respect to $ab$ and $ac$ planes (see theoretical section below for details).
We note that the alignment of the experimental rotation planes of the magnetic field had a small but finite misalignment (less than 1$^{\circ}$) with respect to the exact crystallographic planes. This misalignment caused a small asymmetry in the data for the $ac$-plane rotation.

\subsection{Negative longitudinal magnetoresistance}

\begin{figure}[h]
	\centering
	\includegraphics[width=0.7\textwidth]{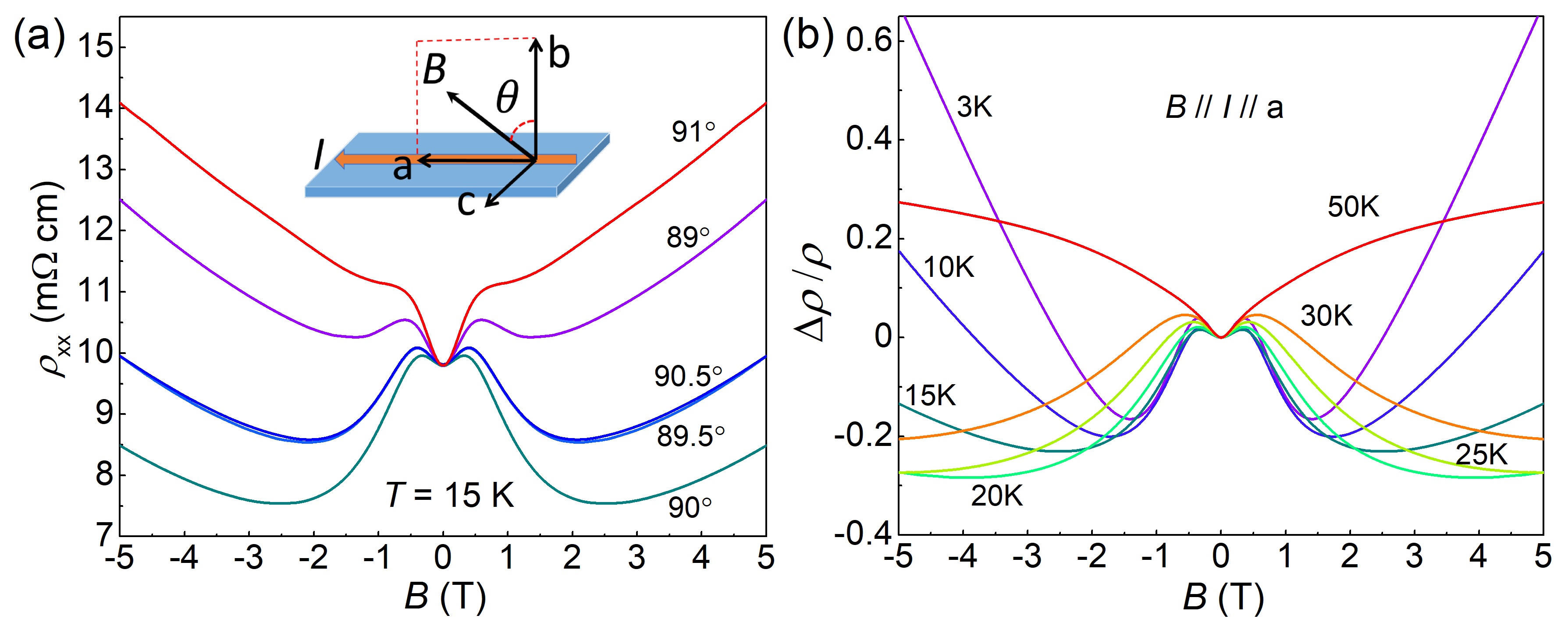}
	\caption{\textbf{Negative longitudinal magnetoresistance in sample A.} (a) $\rho_{xx}(B)$ behabior of sample A at 15 K measured at selected magnetic-field angles close to the longitudinal configuration, $I \parallel B$, which corresponds to $\theta$ = 90$^{\circ}$; inset depicts the measurement configuration. (b) $\rho_{xx}(B)$ behavior for $I \parallel B$ at various temperatures from 3 to 50 K.
	}
	\label{fig:S2}
\end{figure}

The chiral anomaly originates from the non-conservation of particle numbers of the Weyl fermions of opposite chirality under parallel electric and magnetic fields, which leads to a negative magnetoresistance in the longitudinal configuration, $I \parallel B$ \cite{Li2016, Armitage2018}.
The observation of this effect has been reported for ZrTe$_5$ samples in the gapped strong 3D tpological insulator phase \cite{Li2016} and in the semimetallic state similar to ours \cite{Liang2018}. Our samples reproduced these previous reports of the negative longitudinal magnetoresistance (LMR), which shows up only in a very narrow range of the magnetic-field orientation, within $\pm$1$^{\circ}$ of the exact $I \parallel B$ situation (Fig. S2(a)). This effect is observed up to $\sim$30 K (Fig. S2(b)). Since our result shows that the low-energy physics of ZrTe$_5$ is governed {\it not} by Weyl nodes but by a nodal-line loop (which gives rise to the torus Fermi surface for a finite doping), the interpretation of the negative LMR in ZrTe$_5$ as evidence for the chiral magnetic current needs to be revisited.

\subsection{Definition of the amplitudes $\gamma$ and $\gamma'$}

Since there are conflicting conventions within the literature on how to parametrize the magnetochiral anisotropy (MCA), we explicitly discuss our definitions of $\gamma$ and $\gamma'$ in this section.

In the limit $B\rightarrow 0$, the size of the MCA is unambiguously defined as $R=R_0(1+\gamma B I)$, with $R_0$ the reciprocal linear resistivity which is measured with the original excitation frequency (i.e. the first harmonic) in the AC measurement. As $R$ depends on the total current $I$, it also depends on the cross-section of the sample. To measure the intrinsic nonlinearity of bulk transport, one should therefore instead consider $\gamma'=A_\perp \gamma$, with $A_\perp$ the cross-sectional area of the sample. One therefore obtains in the limit of small magnetic field
$R=R_0(1+\gamma' B j)$, where $j=I/A_\perp$ is the current density.
Any experiment measuring $\gamma'$ is done at a finite magnetic field. In a finite magnetic field, we define $\gamma'(B)$ using the relations $R=R_0(B)[1+\gamma'(B) B j]$ with $R_0(B)$ the first-harmonic signal at finite field.

Within transport theory,  one usually calculates currents as functions of electric fields, $j=\sigma^{(1)} E+\sigma^{(2)} E^2$. In this case, the value of $\gamma'$ for $B\to 0$ is obtained \cite{Ideue2017} with the formula
\begin{equation}
|\gamma'|=\frac{|\sigma^{(2)}|}{|B| (\sigma^{(1)})^2}.
\end{equation}

\subsection{Reproducibility of the gigantic magnetochiral anisotropy in many samples of ZrTe$_5$}

\begin{figure}[b]
	\centering
	\includegraphics[width=1\textwidth]{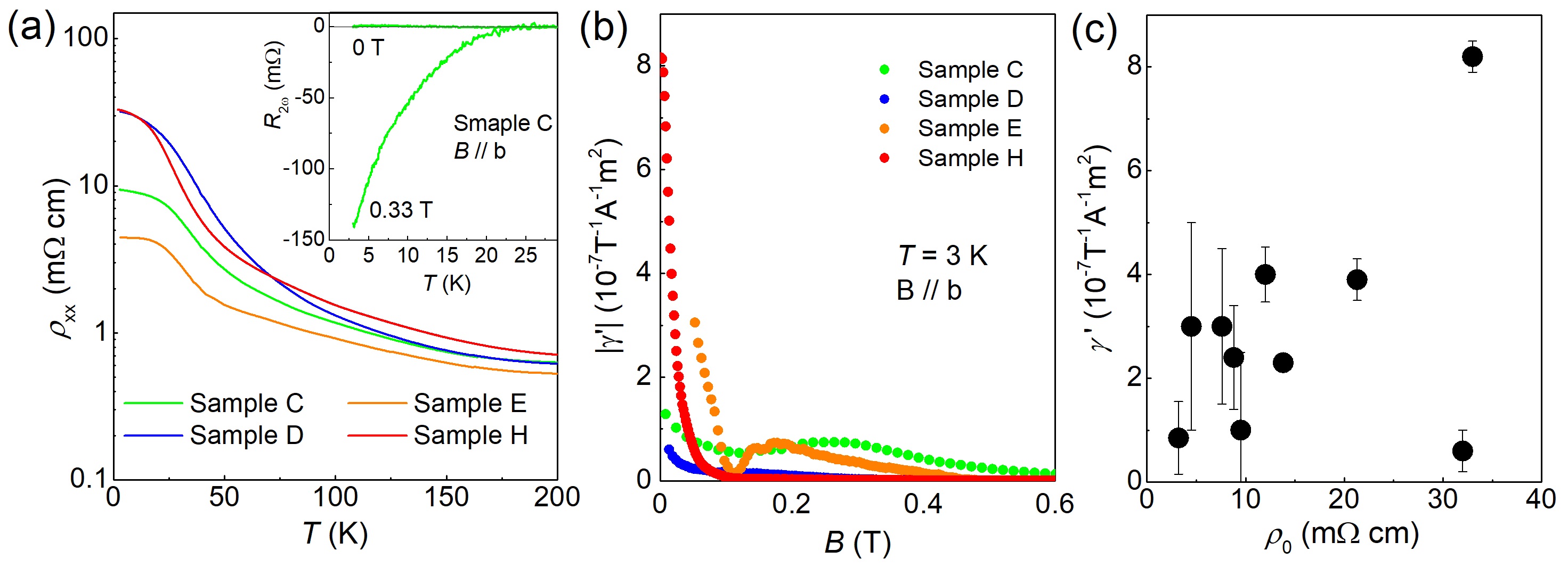}
	\caption{\textbf{Reproducibility of the gigantic magnetochiral anisotropy.} (a) Temperature dependencies of $\rho_{xx}$ for four representative samples, which are all in the semimetallic state at low temperature. Inset shows the temperature dependence of $R_{2\omega}$ in $B \parallel b$ for sample C, which reproduces that of sample A reported in the main text.
(b) Magnetic-field dependencies of the normalized coefficient $|\gamma'|$ of the magnetochiral anisotropy of the four samples at 3 K.
(c) Plot of $|\gamma'|$ vs the residual resistivity $\rho_0$ for all 10 samples having $T_{\rm p}$ = 0 K.
	}
	\label{fig:S3}
\end{figure}

We have confirmed the gigantic MCA in many samples of ZrTe$_5$ that were grown with the same condition as samples A or B. The $\rho_{xx}(T)$ and $|\gamma'(B)|$ curves for some of them are shown in Figs. S3(a) and S3(b). One can see that the value of $\rho_{xx}$ at low temperature is very much sample dependent even when all the samples are in the semimetallic state (i.e. the resistivity maximum is located at 0 K). Nevertheless, all these samples presented the large MCA at low magnetic fields (Fig. S3(b)). The maximum value of $|\gamma'|$ for each sample, which lies in the range  $0.6 \times 10^{-7}$ -- $8.2 \times 10^{-7}$ m$^2$T$^{-1}$A$^{-1}$ is plotted against the residual resistivity $\rho_0$ in Fig. S3(c), where there is no clear correlation. Table I summarizes the results of all the samples used in the present study: samples A -- J had the resistivity-peak temperature $T_{\rm p}$ = 0 K, and we additionally measured samples having a finite $T_{\rm p}$ that will be discussed in Sec. \ref{sec:finiteTp}

\begin{table}[!t]
		\caption{Summary of the resistivity-peak temperature $T_{\rm p}$, residual resistivity $\rho_0$, $|\gamma'|$ [$\equiv 2A_\perp |R_{2\omega}/(R_0 B I_0)| $], and the sign of $\gamma'$ for all the samples used in the present study. The results of sample A-E, K and N were obtained at $T$ = 3 K, while others were at $T$ = 2 K.}
		\label{table:1}
{\scriptsize
	\begin{center}
		\begin{tabular}{cccccc}
			\hline			
			& sample\,\,\, & \,\,\,$T_{\rm p}$ (K)\,\,\, & \,\,\,$\rho_0$ (m$\Omega$cm)\,\,\, & \,\,\,$|\gamma'|$ (10 $^{-7}$m$^2$ T$^{-1}$A$^{-1}$)\,\,\, & \,\,\,sign of $\gamma'$  \\
			\hline
			 & A & 0 & 12 & 4 & $-$  \\  
			 & B & 0 & 8.8 & 2.4 & $-$  \\  
			 & C & 0 & 9.5 & 1 & $-$  \\  
			 & D & 0 & 32 & 0.6 & $-$  \\  
			 & E & 0 & 4.5 & 3 & $-$  \\  
			 & F & 0 & 7.6 & 3 & $-$  \\  
			 & G & 0 & 13.8 & 2.3 & $+$  \\  
		     & H & 0 & 33 & 8.2 & $-$  \\  
			 & I & 0 & 3.2 & 0.85 & $+$  \\  
			 & J & 0 & 21.3 & 3.9 & $+$  \\  
			 & K & 15 & 3.2 & 2.8 & $-$  \\  
			 & L & 29 & 2.1 & 0.77 & $+$  \\  
			 & M & 35 & 2.66 & 1.27 & $+$  \\  
			 & N & 38 & 2.3 & 0.96 & $+$  \\  
			 & O & 44 & 2.8 & 0.77 & $+$  \\  
			 & P & 49 & 2.3 & 0.34 & $+$  \\  
			 & Q & 138 & 0.16 & 0 & N.A.  \\  
			 & R & 138 & 0.11 & 0 & N.A.  \\  
			\hline
		\end{tabular}
	\end{center}
}
\end{table}

\subsection{Behaviour of $R_{2\omega}$ in high magnetic fields}

\begin{figure}[h]
	\centering
\includegraphics[width=0.75\textwidth]{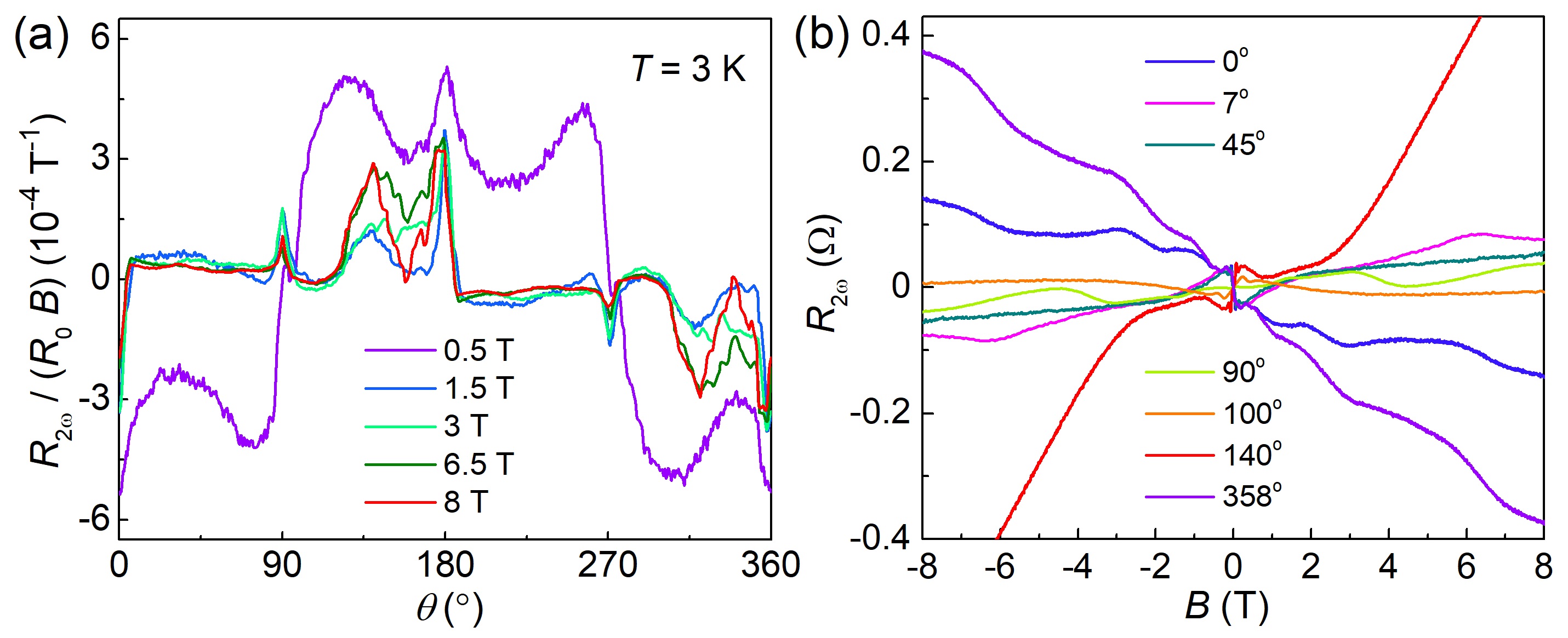}
	\caption{\textbf{Magnetic-field-orientation dependence of the second-harmonic signal at high magnetic fields.} (a) $R_{2\omega}/(R_0 B)$ vs $\theta$ behaviour in $B$ = 0.5, 1.5, 3, 6.5, and 8 T when $B$ is rotated in the $bc$ plane.
 (b) Magnetic-field dependencies of $R_{2\omega}$ up to 8 T for selected magnetic-field orientations. Reflecting the complex $R_{2\omega}(\theta)$ behaviour, the $R_{2\omega}(B)$ behaviour changes significantly with $\theta$.
	}
	\label{fig:S4}
\end{figure}

While the 2nd harmonic signal $R_{2\omega}/(R_0 B)$ presents a simple $\cos \theta$ dependence at low magnetic fields when $B$ is rotated in the $ab$- or $bc$-plane (see Fig. 2 of the main text), the angular dependence at high magnetic fields becomes very complicated as shown in Fig. S4. Understanding its origin is an interesting topic of future studies.

\subsection{Shubnikov-de Haas (SdH) oscillations}\label{S:sdh}

\subsubsection{Analysis of the SdH oscillations in $B \parallel b$}

\begin{figure}[h]
	\centering
	\includegraphics[width=0.75\textwidth]{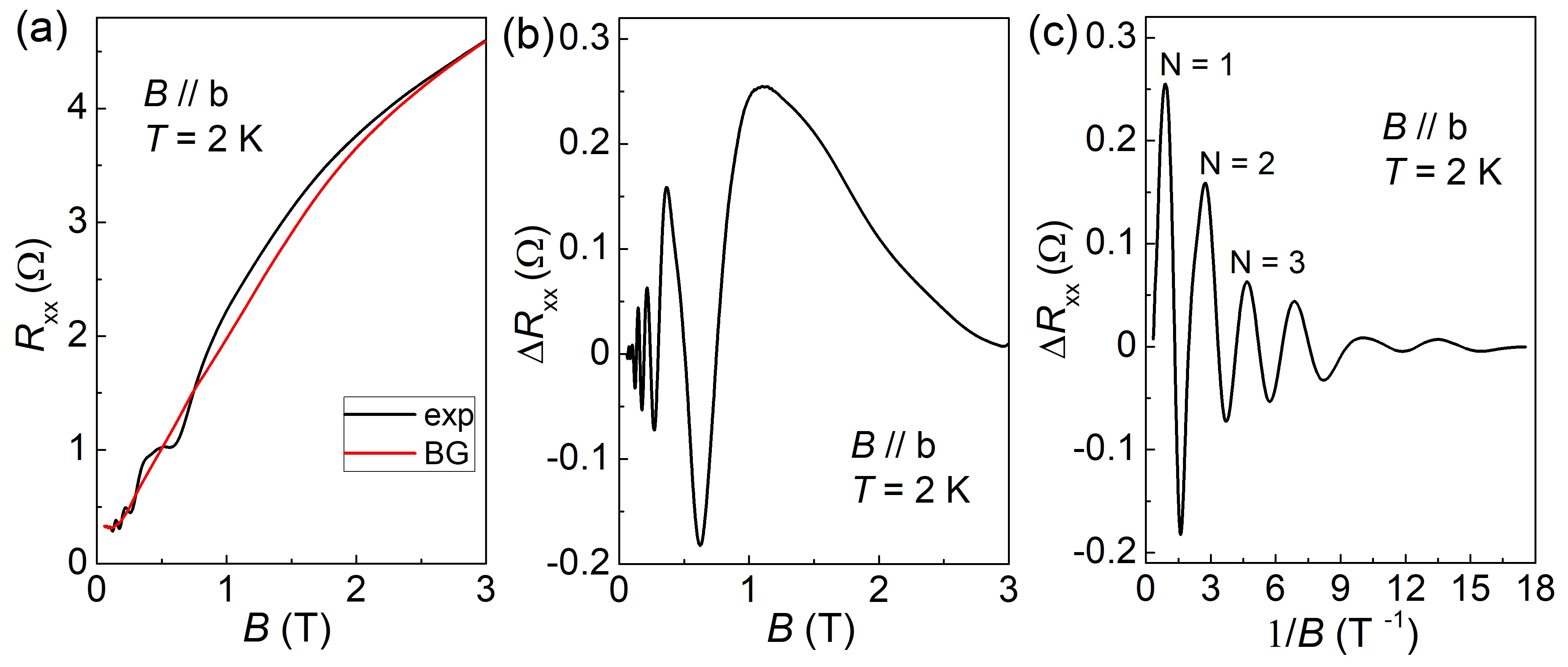}
	\caption{\textbf{SdH oscillations in sample B for $B \parallel b$.} (a) $R_{xx}(B)$ data (black curve) measured in $B \parallel b$ at 2 K and its approximate background (red curve). (b) The oscillating component $\Delta R_{xx}(B)$ obtained after subracting the background from $R_{xx}(B)$. (c) Plot of $\Delta R_{xx}$ vs $1/B$.
}
	\label{fig:S5}
\end{figure}

For the analysis of the Shubnikov-de Haas (SdH) oscillations, we used the resistance component $R_{xx}$ measured on a transverse voltage-contact pair. This is essentially the longitudinal voltage mixed into the Hall-voltage measurement due to a finite misalignment of the transverse voltage-contact pair. Since the Hall-resistance component $R_{yx}$ is antisymmetric with $B$, one can easily separate $R_{xx}$ as the component that is symmetric with $B$.
It turned out that this $R_{xx}(B)$ gives the best SdH-oscillation data due to the relatively benign background change at low fields. The oscillating component $\Delta R_{xx}$, shown in Fig. S5(b), is obtained by subtracting a smooth background described by a polynomial (red curve in Fig. S5(a)) from the $R_{xx}(B)$ data. The plot of $\Delta R_{xx}$ vs $1/B$ (Fig. S5(c)) clearly shows that the oscillations are periodic in $1/B$ and hence are SdH oscillations. The Fourier transform (FT) is performed for the $\Delta R_{xx}(B^{-1})$ data, and the result is shown in Fig. 3(f) of the main text.



\subsubsection{Extraction of the cyclotron mass for $B \parallel b$}

\begin{figure}[h]
	\centering
	\includegraphics[width=0.65\textwidth]{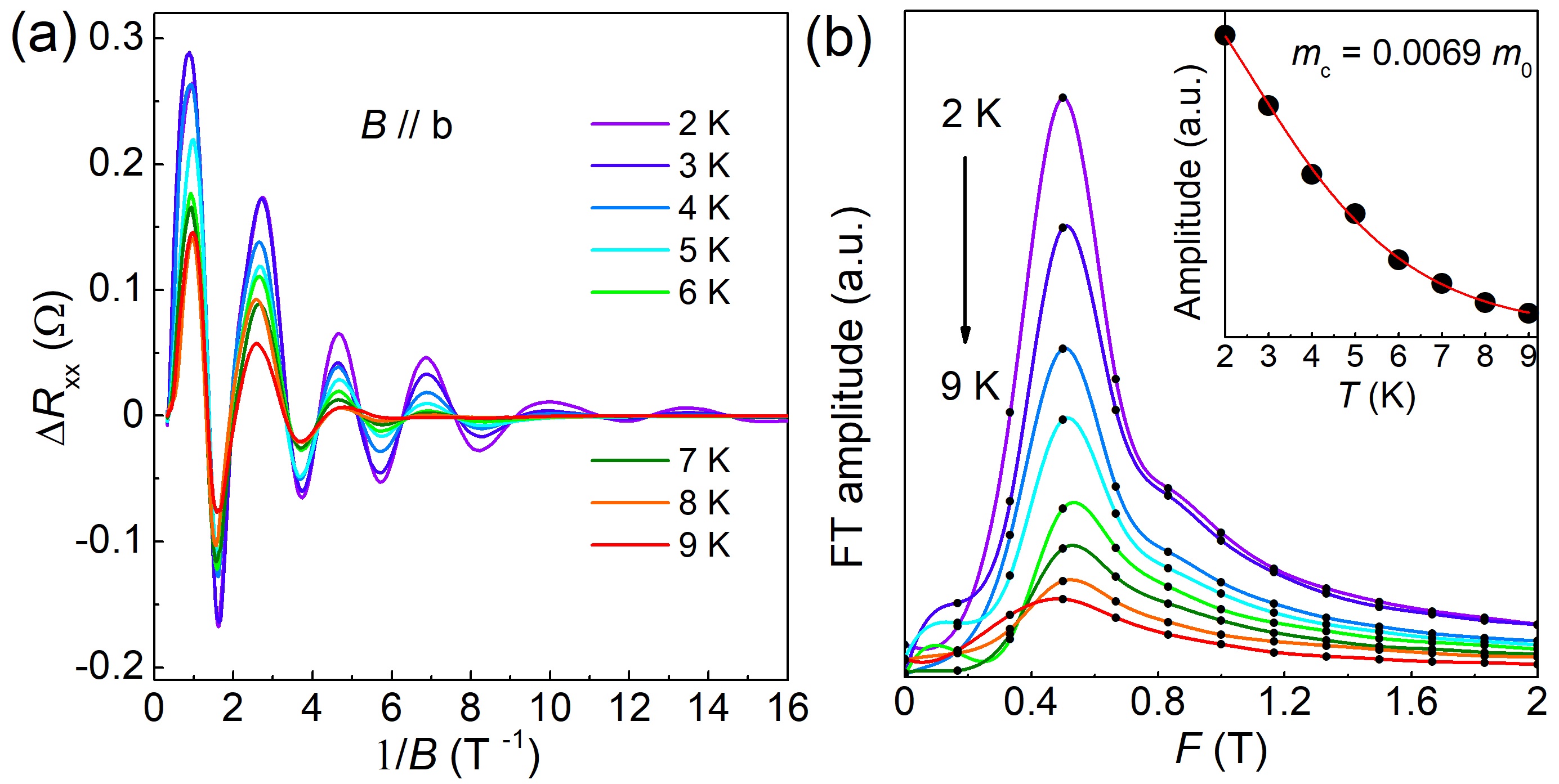}
	\caption{\textbf{Extraction of the cyclotron mass for $B \parallel b$.} (a) $\Delta R_{xx}$ vs $B^{-1}$ at various temperatures from 2 to 9 K. (b) FT spectra of the data calculated by using the $B^{-1}$ range of 3 -- 9 T$^{-1}$. Inset: Temperature dependence of the amplitude of the main FT peak and its fit to the LK formula Eq. (S2), giving $m_c$ = $0.0069 m_0$. In the main text, the frequency of this peak is named $F_1$.
	}
	\label{fig:S6}
\end{figure}

SdH oscillations are expressed in the Lifshitz-Kosevich (LK) theory \cite{Ando2013, Shoenberg2009} as
\begin{equation}
\Delta R_{x x} \propto R_{T} R_{D} R_{S} \cos \left[2\pi\left(\frac{F}{B}-\frac{1}{2}+\beta \pm \frac{1}{8}\right)\right]
\end{equation}
with $R_{T}=2 \pi^{2}\left(k_{B} T / \hbar \omega_{c}\right) / \sinh \left[2 \pi^{2}\left(k_{B} T / \hbar \omega_{c}\right)\right]$,
$R_{D}=\exp \left[-2 \pi^{2}\left(k_{B} T_{D} / \hbar \omega_{c}\right)\right]$, and
$R_{S}=\cos \left(\frac{1}{2} \pi g m_{c} / m_{0}\right)$,
which are called temperature, Dingle, and spin damping factors, respectively. Here, $\omega_c$ is the cyclotron frequency, $T_D$ is the Dingle temperature, $g$ is the electron $g$-factor, $m_c$ is the cyclotron mass, and $m_0$ is the bare electron mass. The SdH oscillations in $B \parallel b$ at different temperatures are shown in Fig. S6(a). The amplitude of the oscillations decreases with increasing temperature. The corresponding FT spectra are shown in Fig S6(b). To make the FT analyses consistent for different temperatures by avoiding the complications coming from the secondary frequency, we restricted the $B^{-1}$ range to 3 -- 9 T$^{-1}$ for these FT analyses. The temperature dependence of the oscillation amplitude, defined by the FT peak height, is shown in the inset of Fig. S6(b) with a fitting to the LK theory giving the cyclotron mass $m_c$ = $0.0069 m_0$ for $B \parallel b$.

\subsubsection{Magnetic-field rotation in the $bc$-plane}

\begin{figure}[h]
	\centering
	\includegraphics[width=0.65\textwidth]{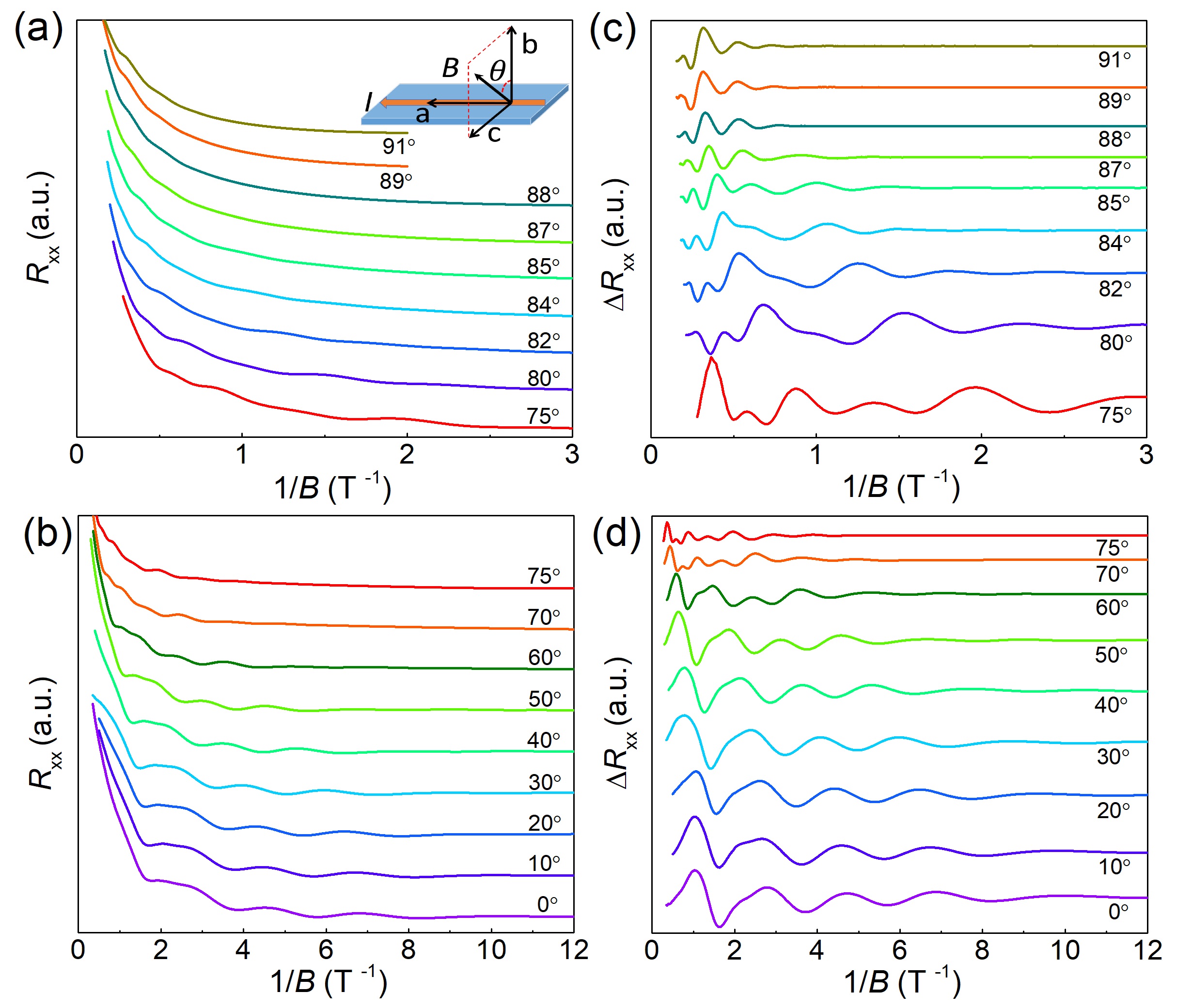}
	\caption{\textbf{SdH oscillations in sample B for $B$ rotated in the $bc$-plane.} (a),(b) $R_{xx}$ vs $B^{-1}$ at 2 K for various angles of the magnetic field rotated in the $bc$-plane; the definition of $\theta$ is shown in the inset of (a). (c),(d) Oscillating component $\Delta R_{xx}$ obtained after subtracting the background.
}
	\label{fig:S7}
\end{figure}

As explained in the main text, to elucidate the torus-shaped Fermi surface, systematic SdH-oscillation data for varying magnetic-field orientation rotated in the $bc$-plane were obtained. The $B^{-1}$ dependence of $R_{xx}$ at 2 K for different angles are shown in Figs. S7(a) and S7(b). The background subtractions give the $\Delta R_{xx}$ vs $B^{-1}$ curves shown in Figs. S7(c) and S7(d). For almost all angles, the existence of multiple frequencies is recognizable in the oscillations. To extract a clear view of this evolution, the FT analyses are performed for different angles by fixing the width of the $B^{-1}$ range to 30 T$^{-1}$; note that, to secure a sufficient number of points in the FT spectrum, the $B^{-1}$ range was extended to 30 T$^{-1}$  for the analysis. The same extension of the $B^{-1}$ range was also done for the $ab$-rotation data shown later. In addition, always the same number of data points (4096) were used for the FT calculations.
The resulting FT spectra are shown in the main text in Figs. 3(e) and 3(f). The oscillation frequencies are determined from the peaks in the FT spectra, and their errors are defined as the half-span of the spectrum at 95\% of the maximum.

\subsubsection{Extraction of the cyclotron mass for $B \parallel c$}

\begin{figure}[h]
	\centering
	\includegraphics[width=0.65\textwidth]{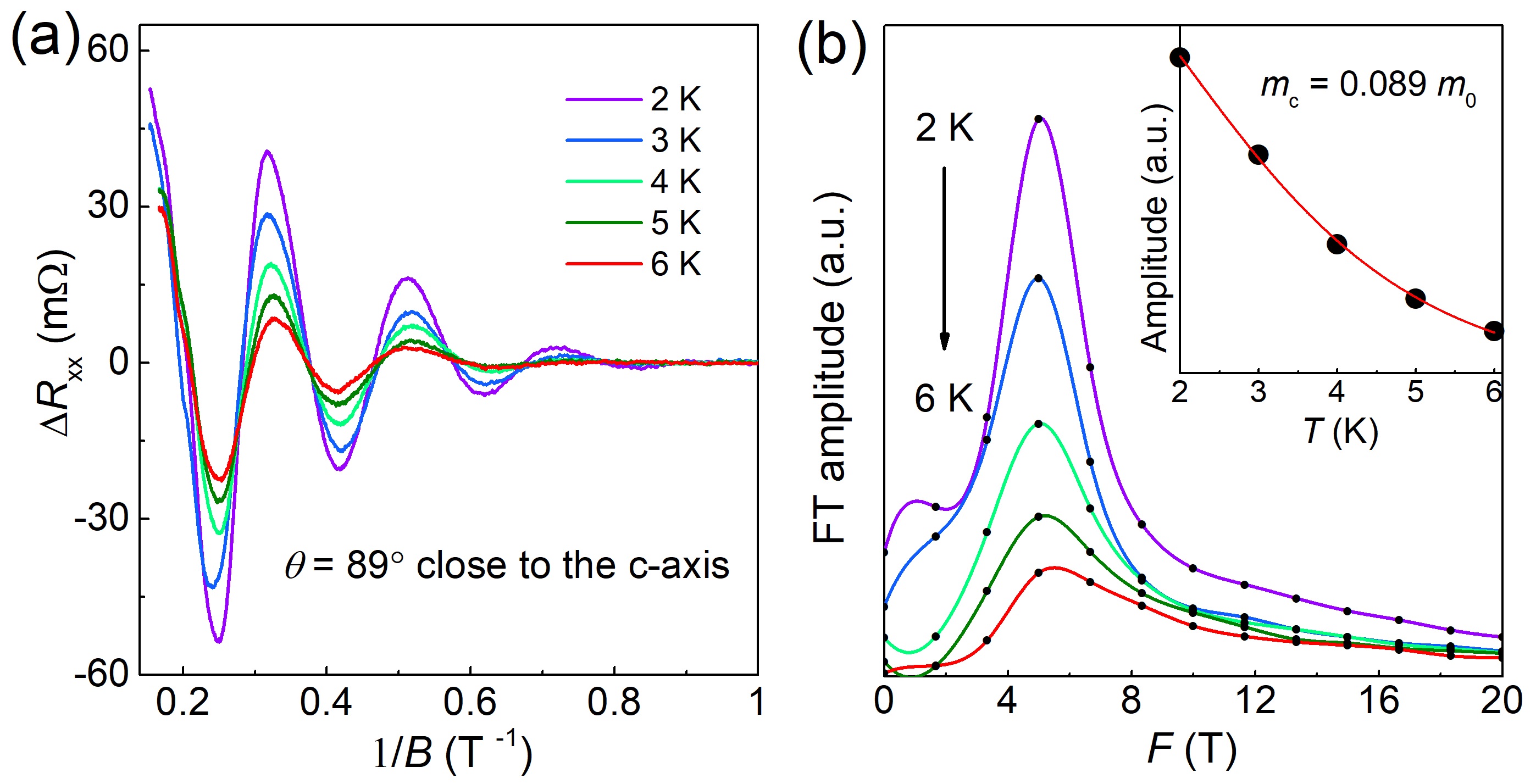}
	\caption{\textbf{SdH oscillations in sample B for $B$ close to the $c$-axis.} (a) $\Delta R_{xx}$ vs $B^{-1}$ at 2, 3, 4, 5, and 6 K for $\theta$ = 89$^{\circ}$, which is close to the $c$-axis. (b) FT spectra of the data calculated by using the $B^{-1}$ range of 0.27 -- 0.87 T$^{-1}$. Inset: Temperature dependence of the amplitude of the main FT peak and its fit to the LK formula Eq. (S2), giving $m_c$ = $0.089 m_0$. In the main text, the frequency of this peak is named $F_3$.
	}
	\label{fig:S8}
\end{figure}

The SdH oscillations in magnetic fields applied almost along the $c$-axis ($\theta$ = 89$^{\circ}$) are shown in Fig. S8(a). The FT analysis performed for the $B^{-1}$ range of 0.27 -- 0.87 T$^{-1}$ gives the main frequency $F_3$ = 5.0 T (Fig. S8(b)). The LK analysis of the temperature dependence of the peak amplitude  (inset of Fig. S8(b)) gives the cyclotron mass $m_c$ = $0.089 m_0$.

\subsubsection{Magnetic-field rotation in the $ab$-plane}

\begin{figure}[h]
	\centering
	\includegraphics[width=0.95\textwidth]{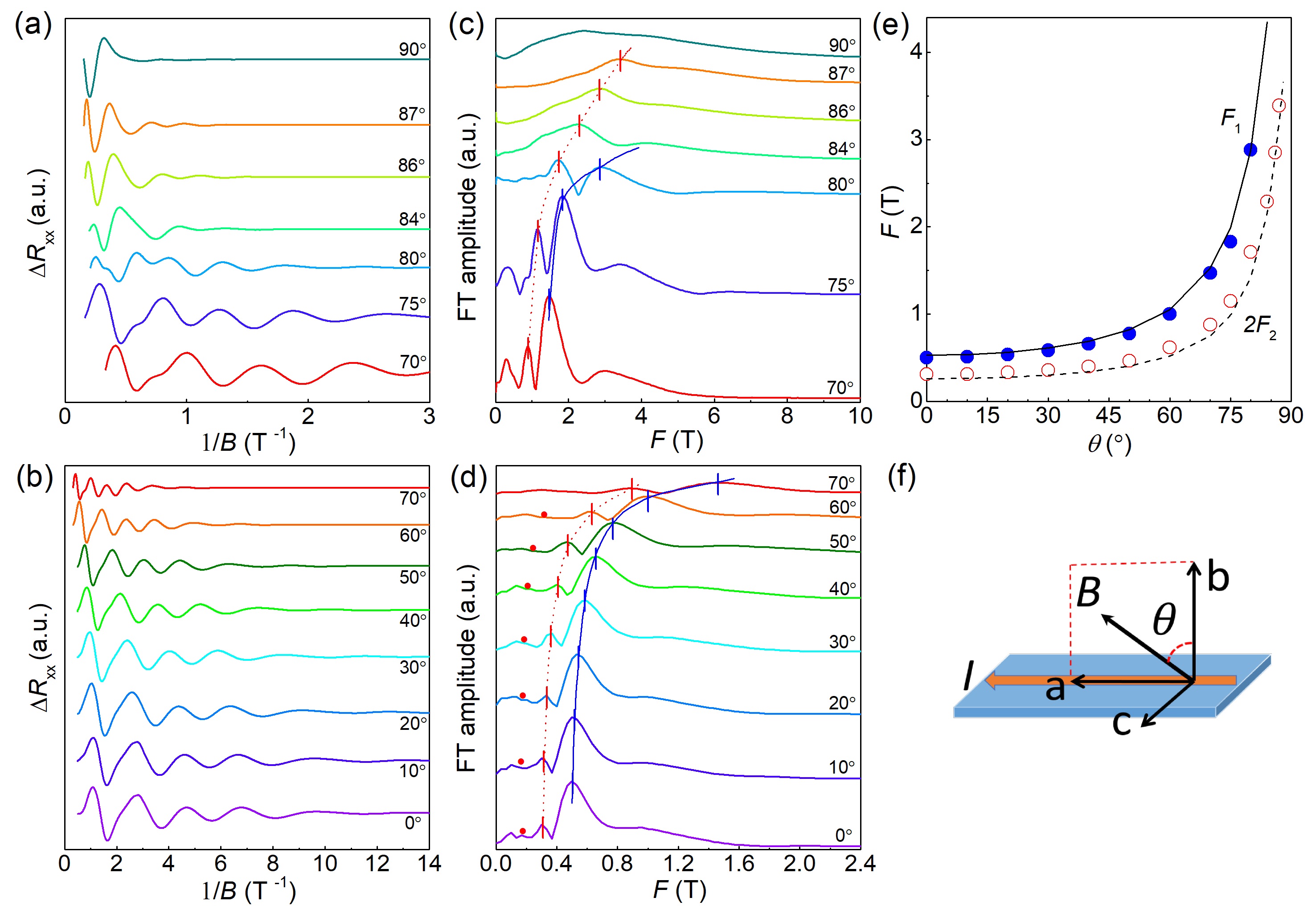}
	\caption{\textbf{SdH oscillations in sample B for $B$ rotated in the $ab$-plane.} (a),(b) $\Delta R_{xx}$ vs $B^{-1}$ at 2 K for various angles of the magnetic field rotated in the $ab$-plane; the definition of $\theta$ is shown in panel (f). (c),(d) Results of the FT analyses of the data shown in panels (a) and (b); ticks mark obvious peaks, corresponding to $F_1$ and $2F_2$, and dots mark the expected position of $F_2$ based on its 2nd harmonic, $2F_2$. (e) Angular dependencies of the oscillation frequencies $F_1$ and $2F_2$ obtained from the FT analyses. The lines are the theoretically-calculated $F_1$ and $2F_2$ frequencies for the orbits around the torus Fermi surface using the parameters discussed in Sec. \ref{sec:vels}.
	}
	\label{fig:S9}
\end{figure}

The $B^{-1}$ dependence of $\Delta R_{xx}$ for a series of $\theta$ varied in the $ab$-plane at 2 K are shown in Figs. S9(a) and S9(b). Here, $\theta$ is defined as the magnetic-field angle measured from the $b$-axis in the $ab$-plane (see Fig. S9(f)). The FT spectra obtained for these oscillations are shown in Figs. S9(c) and S9(d), where the width of the $B^{-1}$ range used for the FT analysis was fixed at 30 T$^{-1}$. Two components are almost always unambiguously identified; based on the understanding obtained in the $bc$-plane rotation, we identify the higher and lower frequency components $F_1$ and $2F_2$, respectively, coming from $\gamma$ and $2\delta$ orbits.
At $\theta$ = 90$^{\circ}$ (i.e. in the longitudinal configuration $B \parallel I$), the SdH ocillations are blurred and identifying the frequency was not possible.
The strong dispersion shown in Fig. S9(e) indicates a significantly elongated torus. An anisotropy of 16 in the Fermi velocity along the $a$ and $b$ axes explains the observed dispersion very well.

\subsubsection{Extraction of the Fermi velocities} \label{sec:vels}

The relations between band parameters and the extremal orbits of a torus Fermi surface relevant to quantum oscillations have been disucssed by Yang {\it et al.} \cite{Yang2018}. Also, an experimental investigation of the torus Fermi surface in CaAgAs has been reported by Kwan {\it et al.} \cite{Kwan2020}. Following these previous works, we found that the most reasonable interpretation of our SdH-oscillation data is to identify the frequencies $F_1$, $F_2$, and $F_3$ to be due to $\gamma$, $\delta$, and $\beta$ orbits shown in the inset of Fig. 3(c) of the main text. At low values of $\theta$, the frequency $F_2$ in our case is so low that it is buried in the background noise; we indicated the expected positions of $F_2$ with red dots in Fig. 3(f) of the main text and in Fig. S9(d). At larger values of $\theta$ close to the critical angle (which in our case is 86$^{\circ}$), the two $\delta$ orbitals on either side of the torus get closer, leading to magnetic breakdown \cite{Alexandradinata2018}; this causes the quantum oscillations with the $2F_2$ frequency to become more prominent than the $F_2$ frequency. In fact, in our FT data shown in Fig. 3(f) of the main text, the $2F_2$ frequency is the most prominent just before the critical angle of 86$^{\circ}$ is reached. The reason for the dominance of the second harmonic ($2F_2$) over the first harmonic ($F_2$) even at low angles is not clear at the moment. Nevertheless, we clearly observed the $F_2$ peak at least for $\theta$ = 75$^{\circ}$ (see Fig. 3(e) in the main text), so the interpretation of the $2F_2$ peaks as the second harmonic of $F_2$ seems justified.

It turns out that we can obtain the relative ratios of the Fermi velocities $v_a$, $v_b$, and $v_c$ directly from fitting the SdH frequencies. For the rotation in the $ab$-plane we obtain the ratio of $v_a/v_b \approx$ 16 from the fact the frequency $F_2$ obeys $F_2(\phi)= \frac{2\hbar S(\phi)}{2\pi e }=\frac{\mu^2}{e \hbar v_c v(\phi)}$, where $v(\phi)=\sqrt{v_a^2 \cos^2 \phi+v_b^2 \sin^2 \phi}$ \cite{Yang2018}. A fit of these frequencies is shown in Fig. S9(e). The $F_1$ frequencies in the $ab$- and $bc$-rotation are more complex and require numerical calculation of the extremal areas. This means they must be fitted numerically, and the result is shown in Fig. 3(c) of the main text and in Fig. S9(e), with the key parameter being the ratio $v_c/v_b \approx 4$. We therefore find that the velocities satisfy $v_a:v_b:v_c = 16:1:4$.

To obtain the absolute magnitude of the Fermi velocities we need to use the cyclotron mass --- defined as $m_c=\frac{\hbar^2}{2\pi} \frac{\partial S}{\partial \mu}$ --- and values of the frequency. This is easiest done for $\vec B \| \hat{\vec c}$ where the frequency corresponding to the $\beta$ orbit ($F_3$) and the cyclotron mass are given by \cite{Yang2018}
\begin{equation} \label{eq:F3andmc}
F_3=\frac{(\Delta-\mu)^2}{2 e v_a v_b \hbar } \hspace{2em} \& \hspace{2em} m_c=\frac{\Delta-\mu}{v_a v_b}.
\end{equation}
Utilising our experimental values of $F_3=5.0$ T and $m_c=0.089 m_0$ and the relation $v_a v_b=\frac{2e \hbar F_3 }{m_c^2}$, we obtain $v_b=0.43\times10^5$ m/s and therefore that $v_a=6.9\times10^5$ m/s and $v_c=1.7\times10^5$ m/s. These are consistent with the values proposed in previous studies of ${\rm ZrTe_5}$ \cite{Li2016, Tang2019, Sun2020}, although these works did not consider its torus Fermi surface.

\subsubsection{Extraction of symmetry breaking, chemical potential, and estimate of carrier density} \label{sec:paras}

The remaining parameters of the torus Fermi surface are $\mu$ and $\Delta$. The chemical potential $\mu$ can be obtained directly from the frequency $2F_2=0.32$ T in the $b$-direction such that $\mu=\sqrt{e\hbar v_c v_a F_2}\approx4.9$ meV.
To obtain $\Delta$ we can again utilise the equations \eqref{eq:F3andmc} to obtain $\Delta - \mu = \frac{2 e \hbar F_3}{m_c}\approx14.2$ meV, which implies that $\Delta \approx 19.1$ meV. The critical angle of rotation in the $bc$-plane, $\theta_c$, must satisfy $\frac{v_b \mu}{v_c \Delta}= \tan \left(\frac{\pi}{2}-\theta_c \right)$, which for our parameters gives $\theta_c\approx 86^{\rm o}$, as found in the experiment.

Utilising all of these values, we find a carrier density $n= \frac{\Delta \mu^2}{4\pi \hbar^3 v_a v_b v_c}\approx 2.3 \times10^{16}\;{\rm cm}^{-3}$.

\subsubsection{Extraction of the scattering time to govern the quantum oscillations} \label{sec:DingleT}

\begin{figure}[h]
	\centering
	\includegraphics[width=0.55\textwidth]{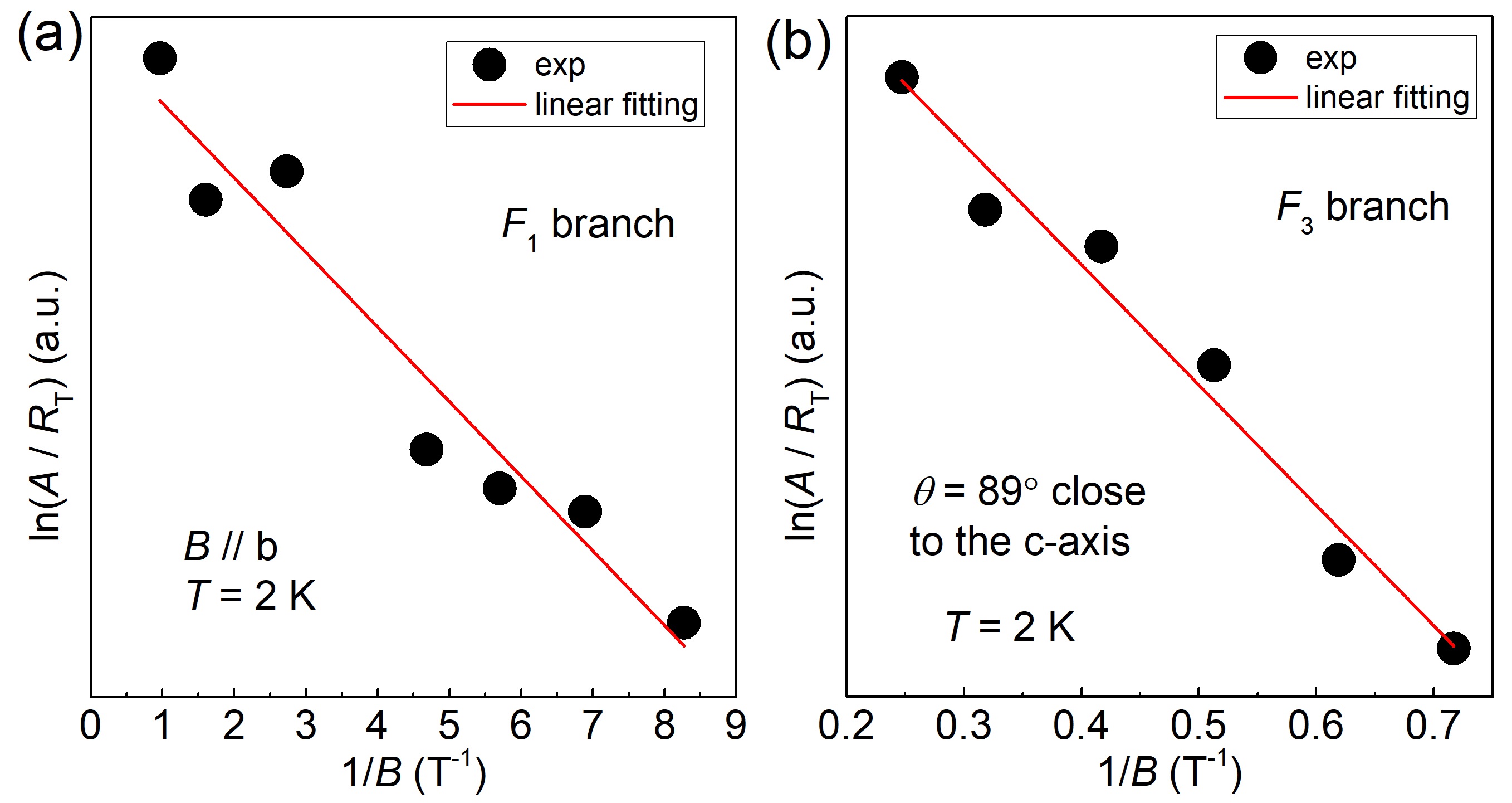}
	\caption{\textbf{Dingle plots of the SdH oscillations in sample B.} (a),(b) Dingle plots of the $F_1$ branch in $B \parallel b$ (panel (a)) and the $F_3$ branch for $\theta$ = 89$^{\circ}$ close to the $c$-axis (panel (b)), both measured at $T$ = 2 K.
	}
	\label{fig:S10}
\end{figure}

According to Eq. (S2), the extrema in the SdH oscillations occur at the $1/B$ values that satisfy $\cos \left[2\pi\left(\frac{F}{B}-\frac{1}{2}+\beta \pm \frac{1}{8}\right)\right]=\pm 1$, where the oscillation amplitude $A$ obeys $A \propto R_{T} R_{D}$ (note that $ R_{S}$ is independent of $B$ and $T$). Hence, using the relation $A / R_{T} \propto  R_{D} = \exp \left[-\left(\alpha  n_{c} T_{D} / B\right)\right]$ with $\alpha$ = 14.96 T/K and $n_{c} \equiv m_{c} / m_{0}$, one obtains the Dingle temperature $T_{\rm D}$ from the plot of $\ln(A / R_{T})$ vs $1/B$ (called Dingle plot), where the linear slope corresponds to $-\alpha n_{c} T_{D}$. Such plots for the $F_1$ branch (in $B \parallel b$) and for the $F_3$ branch (at $\theta$ = 89$^{\circ}$ which is close to the $c$-axis) are shown in Fig. S10. The amplitudes, $A$, shown in Figs. S10(a) and S10(b) are taken from the oscillation data shown in Figs. S5(c) and S8(a), respectively. For $B \parallel b$, we restricted the $B^{-1}$ range to $B^{-1} <$ 9 T$^{-1}$ to avoid the complications coming from other components. These analyses give $T_{\rm D}$ of 2.4 and 3.7 K for  $B \parallel b$ and $B \parallel c$, respectively. These $T_{\rm D}$'s correspond to the scattering times $\tau_{\rm D} [= \hbar /\left(2 \pi k_{B} T_{D}\right)]$ of 500 and 320 fs, respectively. This scattering time reflects the scattering events in all directions, since the cyclotron motion is circular.

\subsubsection{Critical angle beyond 90$^{\circ}$ in the $bc$-plane rotation} \label{sec:critical}

\begin{figure}[h]
	\centering
	\includegraphics[width=0.6\textwidth]{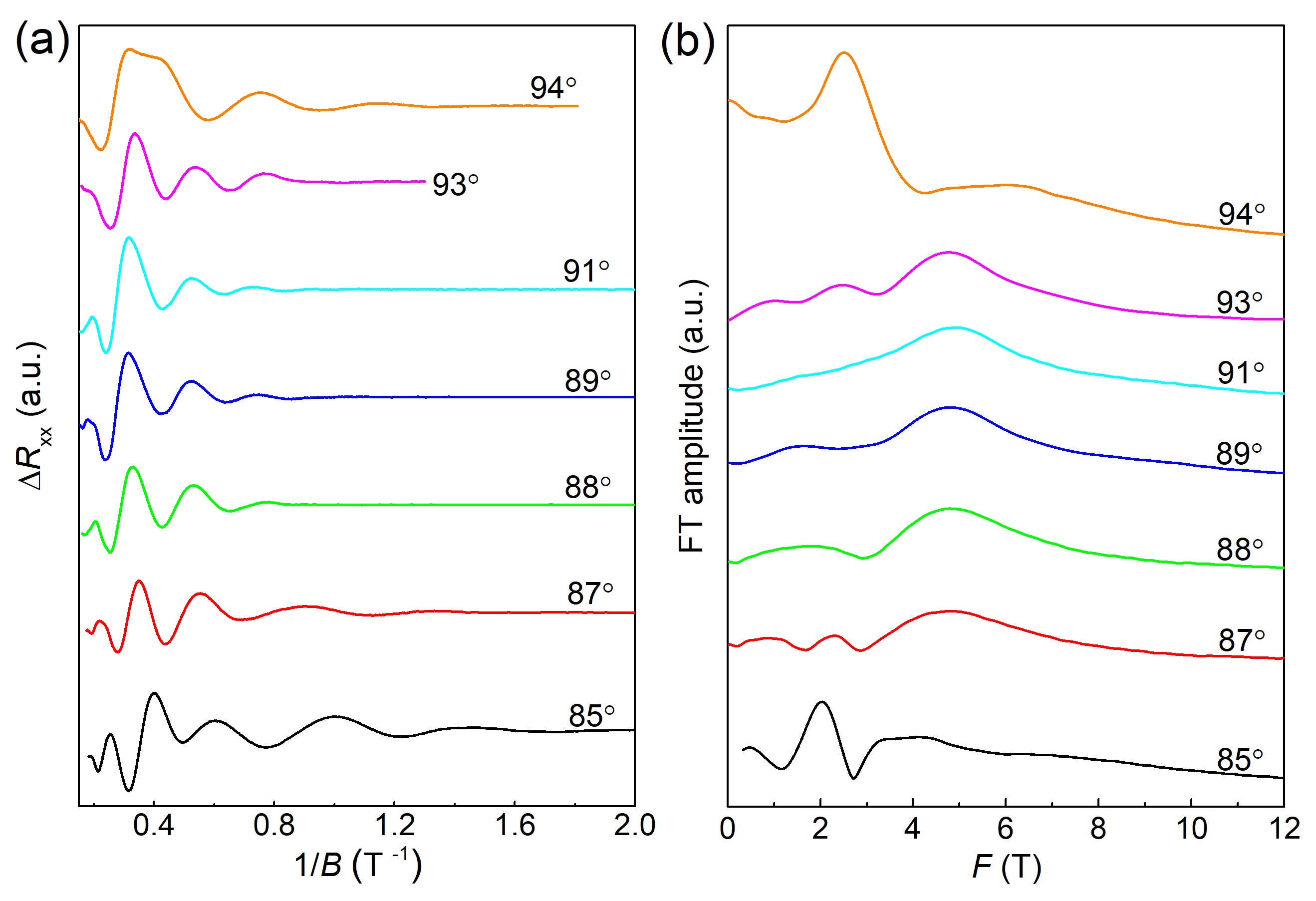}
	\caption{\textbf{SdH oscillations in sample B for the magnetic field orientation close to the $c$-axis.} (a) $\Delta R_{xx}$ vs $B^{-1}$ at 2 K for magnetic-field angles close to the $c$ axis rotated in the $bc$-plane. (b) Results of the FT analyses of the data shown in panel (a).
	}
	\label{fig:beyond90deg}
\end{figure}

Regarding the critical angle $\theta_c$ where the SdH-oscillation frequencies change discontinuously due to the switching of the orbits on the torus-shaped Fermi surface, we checked for its possible asymmetry across 90$^{\circ}$. If the torus is not lying exactly in the $ab$ plane, one would expect to see an asymmetry. 
Figure S11 shows the relevant SdH-oscillation data for $\theta$ between 85$^{\circ}$ and 94$^{\circ}$. 
At the angles of $\pm$3$^{\circ}$ from 90$^{\circ}$ (i.e. at 87$^{\circ}$ and 93$^{\circ}$), the SdH frequency is $F_3$ which comes from the $\beta$ orbit. Our data show that the SdH frequency changes suddenly to $2F_2$ (coming from the $\delta$ orbit) at $\pm$4-5$^{\circ}$ from 90$^{\circ}$ (i.e. at 85$^{\circ}$ and 94$^{\circ}$), indicating that $\theta_c$ is symmetric within an experimental error of 1$^{\circ}$.

\subsection{Anomalous Hall effect}

\begin{figure}[h]
	\centering
	\includegraphics[width=0.45\textwidth]{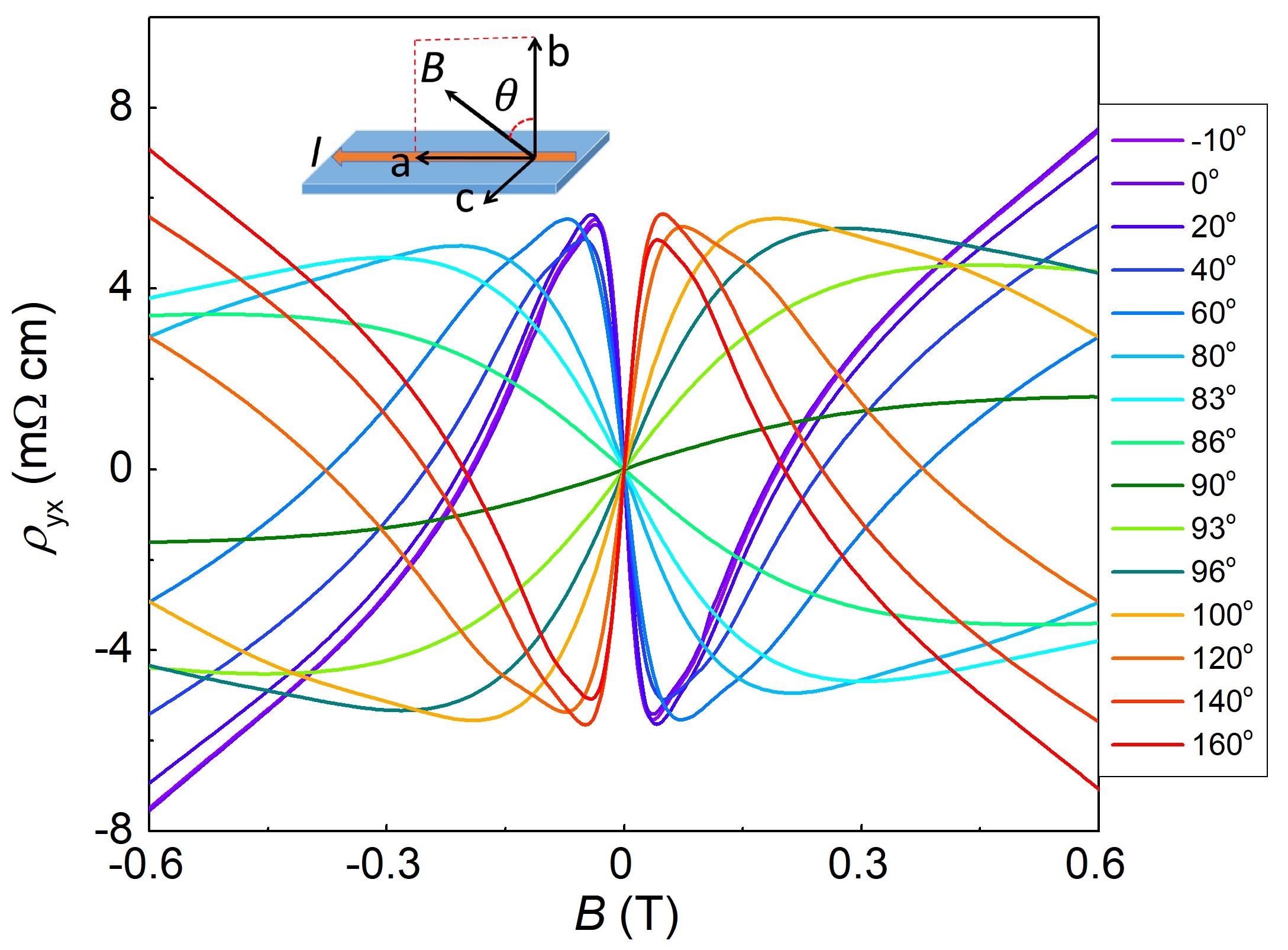}
	\caption{\textbf{Anomalous Hall effect observed in sample A.} Magnetic-field dependencies of $\rho_{yx}$ at 3 K for $B$ rotated in the $ab$-plane; the magnetic-field angle $\theta$ is defined in the inset. The small but finite $\rho_{yx}$ at $\theta$ = 90$^{\circ}$ is due to a misalignment of the sample.
	}
	\label{fig:AHE}
\end{figure}

For the semimetallic state of ZrTe$_5$ near the topological quantum phase transition, a magnetic-field-induced anomalous Hall effect (AHE) has been reported \cite{Sun2020, Liang2018}. This AHE is reproduced in our samples. For example, the plots of the Hall resistivity $\rho_{yx}$ vs $B$ for various angle of the magnetic field rotated in the $ab$-plane are shown in Fig. S12. This result is essentially consistent with what is reported in Refs. \cite{Sun2020, Liang2018} and points to the existence of magnetic-field-induced Berry curvature. We note that the existence of the AHE compont makes it difficult to identify the slope of $\rho_{yx}(B)$ governed by the ordinary Hall effect, which can in principle be used for the estimation of the carrier density based on the semi-classical theory.

\subsection{Nonreciprcal response in samples having a finite $T_{\rm p}$}\label{sec:finiteTp}

\begin{figure}[h]
	\centering
	\includegraphics[width=0.75\textwidth]{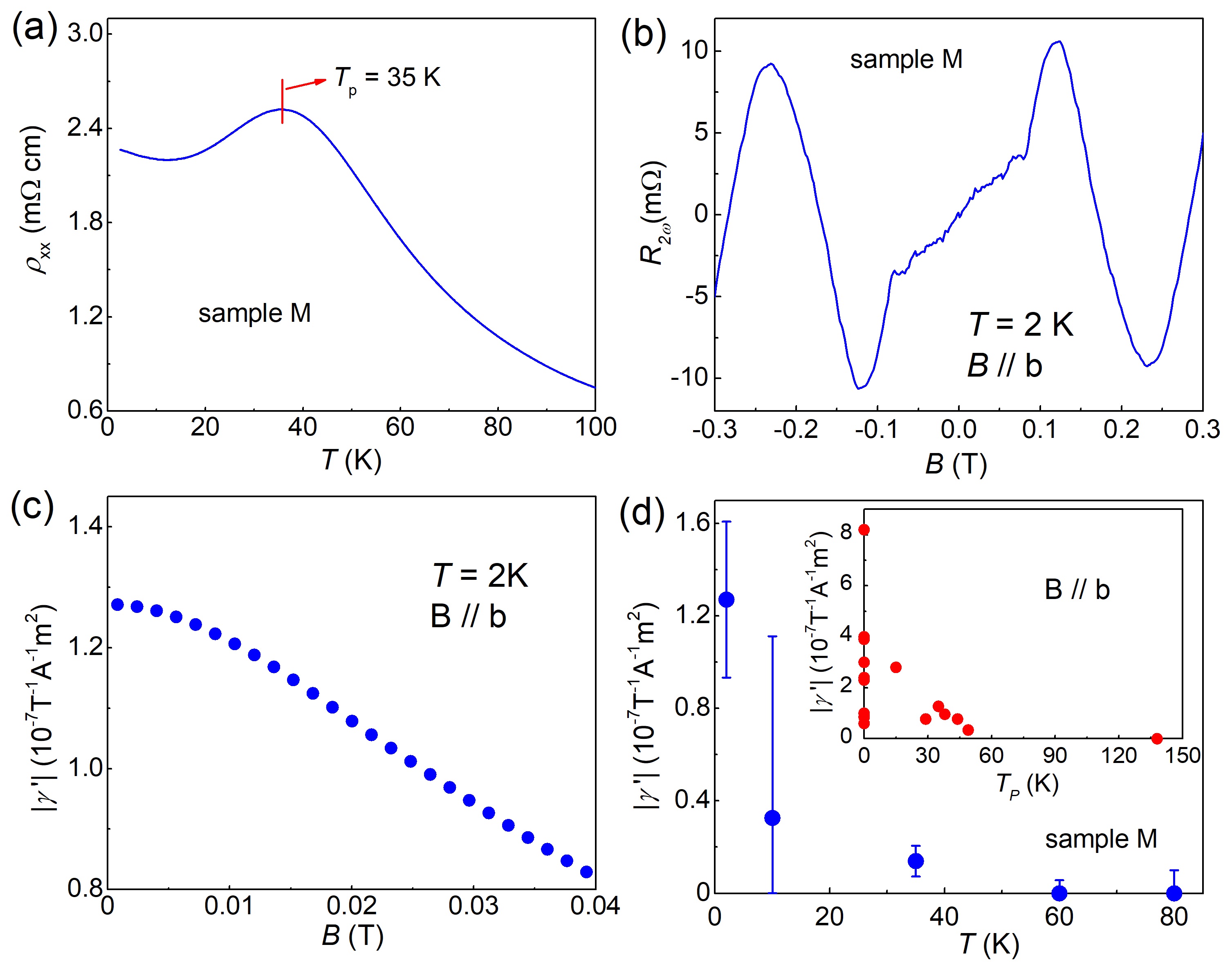}
	\caption{\textbf{Nonreciprcal response in a sample M with $T_{\rm p}$ = 29 K.} (a)--(d) The observed behaviours of $\rho_{xx}(T)$, $R_{2\omega}(B)$, $|\gamma'(B)|$ and $|\gamma'(T)|$ are plotted for sample M. The inset of panel (d) shows the $|\gamma'|$ value at base temperature for all the samples studied as a function of $T_{\rm p}$.
	}
	\label{fig:sampleM}
\end{figure}

As summarized in Table I, we have measured 8 additional samples having a finite $T_{\rm p}$. They were grown with different conditions described in Materials and Methods.
The samples with $T_{\rm p}$ in the range of 15 -- 50 K presented a finite nonreciprocal response with the $|\gamma'|$ values which are generally smaller than those in $T_{\rm p}$ = 0 K samples. As an example, the data set of $\rho_{xx}(T)$, $R_{2\omega}(B)$, $|\gamma'(B)|$ and $|\gamma'(T)|$ are shown for sample M in Fig. S13, where one can see that the nonreciprocal response is qualitatively the same as that in the $T_{\rm p}$ = 0 K samples. In particular, $|\gamma'|$ is maximum at the lowest temperature --- if the origin of the nonreciprocal response was tied to a gapless dispersion or a gap closing, one would expect the maximum in $|\gamma'|$ to occure at $T_{\rm p}$; however, it was not observed. 
Most likely, the suppression of $|\gamma'|$ arises from the larger carrier density in these samples, which also suppresses the formation of charge inhomogeneities, see Sec. \ref{sec:puddle}.
In fact, the $T_{\rm p}$ = 138 K samples, in which the nonreciprocal response was not observed, had an order of magnitude lower residual resistivity (see Table I), suggesting that the carrier density in these samples was much higher.
The inset of Fig. S13(d) shows that there is a trend that $|\gamma'|$ weakens with increasing $T_{\rm p}$.

\subsection{Single-crystal XRD analysis of the crystal structure of ZrTe$_5$}

\begin{table}[b]
	\caption{{\bf Summary of the XRD analysis for ZrTe$_5$.}
		Bragg reflection intensities were averaged according to Laue class $mmm$ and refinements were performed with the Prometheus program package. Anisotropic displacement parameters $U_{ij}$ are given in 10$^{-4}$\AA$^2$ and $R$ values in \%. There are two refinements with the $Cm2m$ model: one considering only 4 distortion parameters describing deviations breaking the $c$ glide-mirror plane, and the other considering only Te3 $z$ distortion (see text). The refinement with the $Cm$ model considered only 4 distortion parameters. Note that $y$ and $z$ in this table are different from the rest of this paper and correspond to the $b$- and $c$-axis coordinates, respectively.
	}
	\label{tab:character_table}
	{\scriptsize
		\begin{ruledtabular}
			\begin{tabular}{c|c |c |c }
				~~~ &  30\,K &  100\,K  &  290\,K \\
				\hline
				refl.            &  13886   &  17211  & 121760 \\
				ind. refl.        &  1042    &   1182  &   2317 \\
				a (\AA )     &  3.9794(5)    & 3.9745(2)   &     3.9916(2)       \\
				b (\AA )     &  14.4759(19)  & 14.4760(10) &     14.5332(6)      \\
				c (\AA )     &  13.6564(17)  & 13.6747(8)  &     13.7387(5)      \\ \hline
				space group  & $Cmcm$    & $Cmcm$ & $Cmcm$  \\
				$R_w$  $R_{all}$       &  7.326 11.471  &   7.314 13.133&  3.996  8.13\\
				Zr1 y        &.31581(19) &.31566(18) &.31540(5)   \\
				~ ~ U$_{11}$ $_{22}$ &  79(10)  182(13)  & 82(9) 191(13)   &  107(2) 171(3)   \\
				~ ~ U$_{33}$ &  68(11)   &  63(11)   &   103(3)   \\
				Te1 y        &.66300(12) &.66322(12) &.66348(3)   \\
				~ ~ U$_{11}$ $_{22}$ &  85(7)   165(9)  &  101(6)  163(9)  &  130(2) 164(2)   \\
				~ ~ U$_{33}$ &   67(8)   &   71(8)   &  153(2)    \\
				Te2 y        &.93219(9)  &.93187(9)  & .93152(3)  \\
				~ ~ z        &.14938(9)  &.14950(9)  & .14982(3)  \\
				~ ~ U$_{11}$ $_{22}$  & 88(5)    166(6)  &   113(5)  179(6) &  159(1)  216(2)  \\
				~ ~ U$_{33}$ $_{23}$ &    87(2) 11(5) &   89(6) 23(5)   &  193(2)  59(1)   \\
				Te3 y        & .20950(9) &.20966(9)  &.20979(3)   \\
				~ ~ z        & .43567(8) &.43566(8)  &.43556(2)   \\
				~ ~ U$_{11}$ $_{22}$& 87(5) 178(6)     &  115(5) 195(6)  & 170(1) 242(2)    \\
				~ ~ U$_{33}$ $_{23}$& 61(6)     -6(5)   &   59(5)  -8(5) &  119(1)  13(1)   \\
				\hline
				space group  & $Cm2m$    & $Cm2m$ & $Cm2m$  \\
				$R_w$  $R_{all}$   &  7.246 11.159  &  7.253 12.742&  3.901 7.707\\
				$\Delta$Zr  y&-.0002(18)   &  -.0007(15)&.00009(46) \\
				$\Delta$Te1 y&.0001(12)    & .0000(9)   &-.00010(34)\\
				$\Delta$Te3 y&.0009(9)     & .0014(7)   &.00060(34) \\
				$\Delta$Te3 z&.0018(4)     & .0017(4)   & .00141(10)\\
				\hline
				$R_w$  $R_{all}$     &7.252 11.173    &   7.274   12.90    &   3.904  7.746   \\
				$\Delta$Te3 z&   .00183(36)   & .00173(49) &   .00141(11)   \\
				\hline
				space group  & $Cm$        & $Cm$ & $Cm$  \\
				$R_w$ $R_{all}$        &  7.244  11.169    & 7.250  12.71     &  3.898  7.702 \\
				Te$1'_z$     & .74979(25)  & .75011(22) & .74981(7)   \\
				$\Delta$Te2 z& -.00042(76) & -.00085(62)&-.00027(22)  \\
				$\Delta$Te3 y&  .00075(86) & .00140(55) & .00058(30)  \\
				$\Delta$Te3 z&  .00181(37) & .00154(43) & .00139(11)  \\
			\end{tabular}
		\end{ruledtabular}
	}
\end{table}

The crystal structure of ZrTe$_5$ was originally described in space group $Cmcm$ by Furuseth {\it et al.} \cite{Furuseth1973}, but later experiments yielded conflicting results. For example, Skelton {\it et al.} reported strong temperature dependence of x-ray diffraction intensities at (0 0 $l$) reflections with odd $l$ that seem to follow the anomaly in the electrical resistance \cite{Skelton1982}, but these peaks were attributed to multiple diffraction in a later work \cite{Sambongi1986}. Powder diffraction experiments do not yield evidence for a phase transition between 10 and $\sim$830\,K \cite{Fjellvag1986}.

To search for possible symmetry breaking, we studied a ZrTe$_5$ single crystal with $T_{\rm p}$ = 0 K having a rectangular shape with dimensions of 174, 71 and 91\,$\mu$m along $a$, $b$ and $c$, respectively, by X-ray diffraction (XRD). Complete sets of Bragg reflection intensities were taken at room temperature and at 100 and 30\,K.  The structural parameters refined in space group $Cmcm$ at these three temperatures are shown in Table II. The quality of the refinements is high and error-bars of positional and atomic displacement parameters (ADP) are quite low. The three data sets are fully consistent with each other and reveal tiny shifts within the $Cmcm$ structure upon cooling that can be associated with the change in electronic structure \cite{YZhang2017}. With the room-temperature data set, the occupation of the Zr was refined to 0.994(5) without a significant improvement in the $R$ values and without changes in the parameter beyond their error bars. Hence, ZrTe$_5$ presenting a zero-temperature resistance peak is nearly stoichiometric. The ADP along the $b$ direction, $U_{22}$, appear rather large and do not considerably shrink upon cooling. Along the $b$ direction, there is only weak van-der-Waals bonding in ZrTe$_5$ so that low phonon frequencies for modes polarized along $b$ can partially explain this observation in addition to disorder.

The symmetry can be lowered by breaking one or more of the mirror planes in space group $Cmcm$. Following the report by Skelton \cite{Skelton1982}, we first studied reflections excluded by the $c$ glide-mirror plane perpendicular to orthorhombic $b$ [selection rule $l$ even for ($h$ 0 $l$)] at 30 K. By counting several minutes,
we find some weak intensity at (0 0 5) that remained almost unchanged upon heating to 100\,K. This symmetry reduction results in the non-centrosymmetric spacegroup $Cm2m$ and, indeed, the refinements within this model yielded an improvement of the $R$ values at all temperatures as shown in Table II. The completeness of the room-temperature data set allowed us to refine the structural model in space group $Cm2m$ without any constraints, but the same quality of the fit was also reached by using the $Cmcm$ average structure and refining only four distortion parameters: Without the glide-mirror planes, all atomic sites split into two, and we tried to refine the distortions by displacing the atom out of the $Cmcm$ positions with shifts violating the glide-mirror symmetry, atom $A$ at (0,$y{+}\Delta_y$,$z{+}\Delta_z$) and $A'$ at (0,${-}y{+}\Delta_y$,$\frac{1}{2}{+}z{-}\Delta_z$) with (0,$y$,$z$) the position of the atom obtained with the $Cmcm$ refinement. (Note that in this XRD section, we deviate from the definition of $(x,y,z)$ used in the rest of the work, so that $y$ and $z$ are $b$- and $c$-axis coordinates, respectively; furthermore, we use a non-conventional setting of space group $Cm2m$ in order to keep the origin unchanged.) We further tried to identify the main distortion by restricting the distortion to the Te3 $z$ parameter, and the obtained result was almost of the same fit quality. This Te3 $z$ distortion amounts to $\pm$0.00141(11)$\times c$ at 290\,K and is well beyond the error of the refinement. We therefore conclude that the precise crystal structure of ZrTe$_5$ breaks inversion symmetry already at room temperature. The pattern of the Te3 $z$ distortion identified in this analysis is shown in Fig. S14. In passing, we also refined the distortion excluding the ($h$ 0 $l$) reflections that are forbidden in $Cmcm$ and obtained the same values within the error, which  excludes the possibility that this distortion arises from multiple diffraction contamination of these reflections.

\begin{figure}[b]
	\centering
	\includegraphics[width=0.35\textwidth]{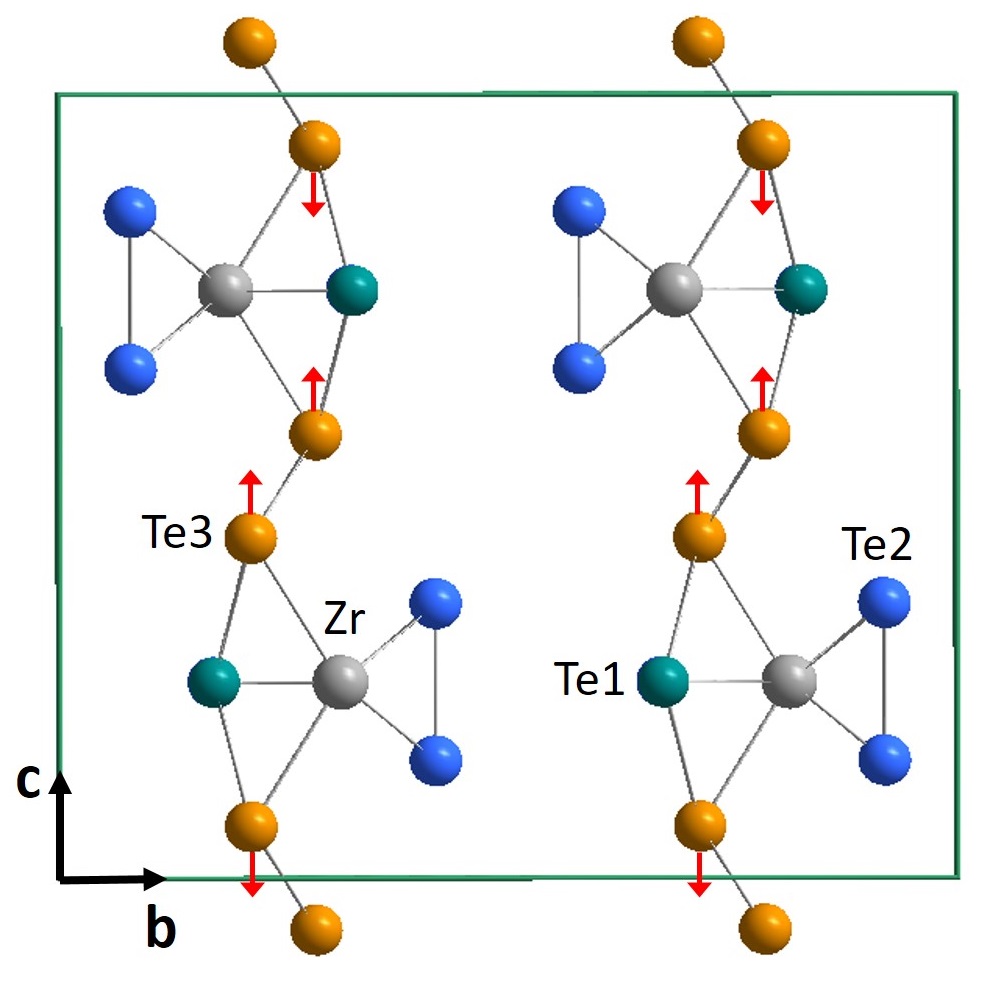}
	\caption{\textbf{Distortion pattern of the Te3 atoms breaking inversion symmetry.} The single-crystal XRD analysis identified this pattern of distortion beyond the error. The staggered displacement of the Te3 atoms is along the $c$ axis which results in weak XRD intensity at (0 0 $l$) with odd $l$, which is forbidden in the $Cmcm$ symmetry but was observed by Skelton {\it et al.} \cite{Skelton1982} as well as in our study.}
	\label{fig:distortion}
\end{figure}

While the above analysis identified the Te3 $z$ distortion beyond the error and this distortion already breaks inversion symmetry, the highest symmetry compatible with this distortion is $Cm2m$ which preserves $ab$ mirror symmetry. Given our transport results which clearly indicate that mirror symmetry is broken with respect to both $ab$ and $ac$ planes, we tried to refine a model with one-step lower symmetry, $Cm$. (Again we use non-conventional setting with $a$ being the monoclinic axis to keep the analogy with the other models, and we ignore the monoclinic angle whose deviation from 90$^{\circ}$ is impossible to observe in our experiment due to the expected twinning). In this structure model, there are twelve independent sites if unconstrained; however, considering just 4 distortion parameters shown in Table II, a significant reduction of the $R$ value is achieved compared to the $Cmcm$ model; 
there is also a small reduction of the $R$ value compared to the $Cm2m$ model. The inversion-symmetry-breaking distortion of the Te3 atoms detected here is staggered and is along the $c$ axis (Fig. S14). For a more comprehensive understanding, a higher-resolution XRD study using synchrotron facilities would be desirable. Nevertheless, one can say at this stage that the XRD result is at least consistent with the transport results for symmetry breaking.

We note that we have also performed similar measurements on crystals having $T_{\rm p}$ = 40 K and 140\,K and obtained similar results; namely, the inversion symmetry is broken with the peculiar Te3 $z$ distortion. Hence, the lower symmetry seems to be not restricted to samples with $T_{\rm p}$ = 0 K. We further note that, in view of the intrinsic inhomogeneity of ZrTe$_5$, the observation of the structural symmetry breaking can be limited by the size of the corresponding domains. If the size of the domains associated with the distortion is below a few hundred\,\AA, XRD will only sense the averaged structure and the implied diffuse scattering remains small, so it cannot be detected.


\section{Theoretical description}

\subsection{Experimental symmetries}
As discussed in the main text and above, ${\rm ZrTe}_{\rm 5}$ nominally belongs to the $Cmcm$ (${\rm D^{17}_{2h}}$) space group \cite{Weng2014}, with the basic lattice structure consisting of sheets in the $ac$-plane coupled along the $\hat{\vec b}$ direction by van-der-Waals interactions. Therefore, nominally, the point group of ${\rm ZrTe}_{\rm 5}$ (which determines its low-energy electronic structure) contains three mirror planes: $m_{ab}$, $m_{bc}$, and $m_{ac}$ and hence also inversion symmetry as well as
two-fold rotation symmetry around all three axes. Our experimental results indicate clearly that several of these purported symmetries are broken. Since the absence of these symmetries is crucial for understanding any potential mechanism for the large magnetochiral anisotropy (MCA)  found in ${\rm ZrTe}_{\rm 5}$, we begin this theoretical section by highlighting  which symmetries our experiments indicate to be absent. In later subsections we will analyse the consequences of these broken symmetries for the low energy physics of ${\rm ZrTe}_{\rm 5}$ and for transport.

We begin with the presence of the MCA itself: The experimental presence of an MCA in the $\hat{\vec a}$ direction which, as shown by the angular dependences in Fig.~2 of the main text, exists predominantly for a magnetic field in the $\hat{\vec b}$ direction, $j_a \propto E_a^2 B_b$, indicates that inversion symmetry must be broken. More precisely, the mirror symmetry $m_{ab}$ (which maps $B_b$ to $-B_b$ as $\vec B$ is a pseudo-vector) and the two-fold rotations around the $\hat{\vec a}$  and $\hat{\vec b}$ axis are broken. This is also consistent with the symmetries previously observed in the Hall-effect measurements of Ref.~\cite{Liang2018}.

Next we consider the XRD experiments: From XRD we also conclusively determine that inversion symmetry is absent. The data clearly show that the mirror-plane $m_{ac}$ is absent and are consistent with the breaking of $m_{ab}$ suggested by the MCA. The breaking of $m_{ac}$ also results in the absence of the two-fold rotation about $\hat{\vec c}$.

Finally, we consider the symmetries suggested by the angular dependence of magnetoresistance, as shown in Fig.~\ref{fig:S1}.  It should, however, be noted that the highly anisotropic nature of transport in ${\rm ZrTe}_{\rm 5}$ means that small misalignment effects can result in sizeable contributions to resistivity and so the broken symmetries suggested by magnetoresistance should only be relied upon in combination with the other experimental observations outlined above. Nonetheless, for example, we find that the values of $\rho_{xx}(\theta)$ (Fig.~\ref{fig:S1}(a)) at high fields differ significantly for $\theta=0^{\rm o}$ and $\theta=180^{\rm o}$; since magnetic field is a pseudo-vector, this suggests a breaking of the mirror-plane $m_{ab}$, as expected from the MCA. 
Additionally, $\rho_{xx}(\theta)$ in the $bc$-plane rotation (Fig.~\ref{fig:S1}(b)) and $\rho_{xx}(\varphi)$ in the $ac$-plane rotation (Fig.~S1(c)) both show deviations for fields parallel and anti-parallel to the $c$-axis -- i.e. there is a difference in the magnitude of magnetoresistance between $\theta$ or $\varphi$ of 90$^{\rm o}$ and 270$^{\rm o}$ for both cases -- this implies, as expected from XRD, that the mirror-plane $m_{ac}$ is absent.

To conclude, all our experimental probes suggest that inversion symmetry is broken. More precisely, all measurements either require or are consistent with the absence of both the mirror-planes $m_{ac}$ and $m_{ab}$. In the following subsection we will analyse the consequences these broken symmetries have on the low-energy Hamiltonian of the system.

\subsection{Hamiltonian and the torus Fermi surface}
To construct the low-energy Hamiltonian, we take a small momentum expansion about the minimal band gap, which occurs at the $\Gamma$ point \cite{Weng2014}. 
Following Ref.~\cite{Weng2014} (see also main text), the four low lying bands are written in the basis $\ket{\Psi_{\vec k}}=\{\ket{\psi^\uparrow_+},\ket{\psi^\uparrow_-},\ket{\psi^\downarrow_+},\ket{\psi^\downarrow_-}\}$. These states are Kramers pairs (labeled by $\uparrow/\downarrow$) of linear combinations of Te $p_y$ orbitals, chosen to have eigenvalues of the mirror symmetry $m_{ab}=\mp1$. The resulting low-energy Hamiltonian is therefore written in terms of $4\times4$ matrices of the form $\sigma_\alpha \otimes \tau_\beta$ where $\sigma_\alpha$ indexes the spin in the $b$-direction and $\tau_\beta$ the parity.

First implementing all nominal symmetries to obtain the resulting Dirac Hamiltonian, we find that the symmetry operations in this basis are as follows:
\begin{itemize}
\item Reflection $m_{ab}$ acts like $i \sigma_y\otimes \tau_z$
\item Reflection $m_{bc}$ acts like $i \sigma_x\otimes \mathbb{1}$
\item Reflection $m_{ac}$ acts like $i \sigma_z\otimes \mathbb{1}$
\item Inversion acts like $ \mathbb{1} \otimes\tau_z$
\item Time reversal acts like $(i\sigma_y \otimes\mathbb{1})K_c$ (with $K_c$ complex conjugation).
\end{itemize}

\begin{figure}[b]
	\centering
  \includegraphics[width=0.9\textwidth]{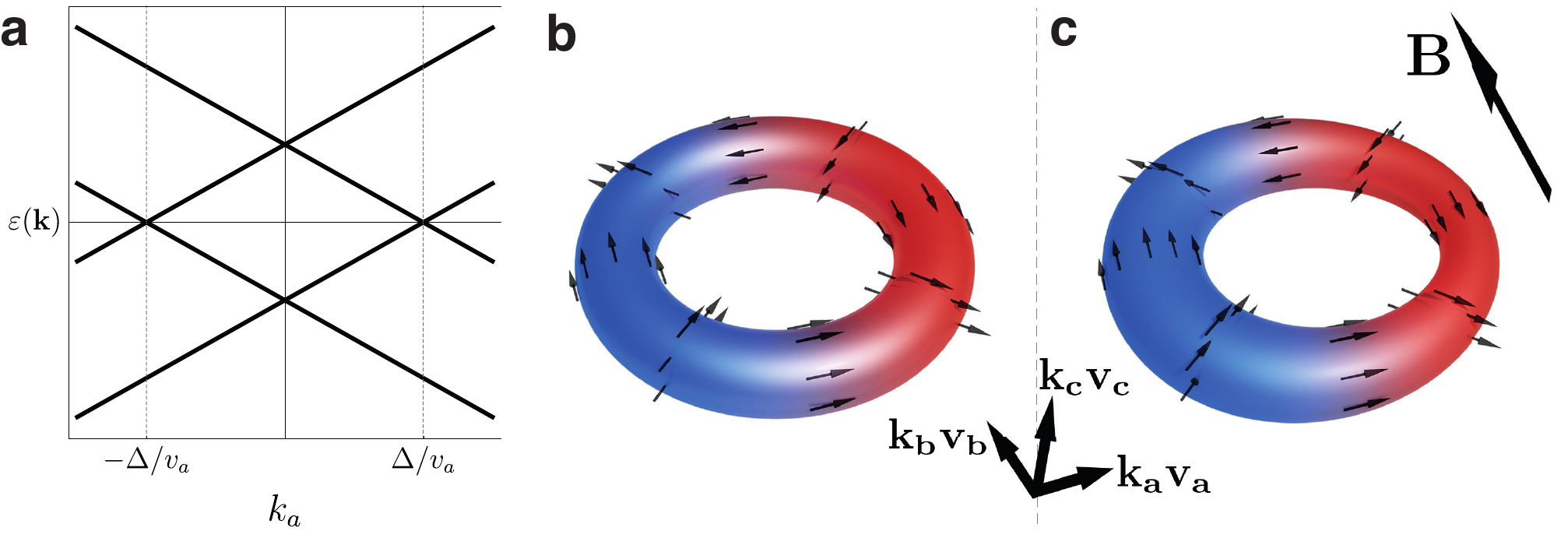}
	\caption{
	{\bf Dispersion, Fermi surface, and spin texture.} (a) The dispersion along the $k_a$ direction for $m=0$ and $\xi=0$. Two bands meet at $k_a=\pm \Delta/v_a$. The dispersion in the $k_b$ direction is similar and results in a nodal line in the $ab$-plane. (b) The Fermi surface (FS) for a finite chemical potential $\mu$ is a torus in the $ab$-plane. The torus is elongated and squashed depending upon the relative values of the Fermi velocities $v_a$, $v_b$, and $v_c$ (see text). On opposite sides of the FS in the $ab$-plane, spins are anti-parallel; red and blue colours signify the opposite spin polarization in the $\hat b$ direction. (c) In a finite magnetic field, the FS is enlarged (shrunk) for elements of the FS with spins parallel (anti-parallel) to the magnetic field due to the Zeeman effect.
	}
	\label{fig:FS}
\end{figure}

These symmetries place strong conditions on the Hamiltonian to linear order in $\vec k$, such that the only allowed terms are \cite{RYChen2015}
\begin{equation}
H_{\rm 0}=m \mathbb{1}\otimes \tau_z+\hbar (v_a k_a \sigma_z \otimes \tau_x +v_b k_b  \sigma_x \otimes \tau_x +v_c k_c \mathbb{1} \otimes \tau_y)+\mu  \mathbb{1},\label{eq:ham0}
\end{equation}
where $m$ is the band-gap at the $\Gamma$-point and $\mu$ the chemical potential. A sign change of $m$ drives the transition from a weak to a strong topological insulator.  Our ${\rm ZrTe}_{\rm 5}$ is located in close proximity to this quantum phase transition \cite{Xu2018} and therefore $m$ is expected to be very small. This nominal Hamiltonian describes a three-dimensional (3D) Dirac semimetal. 
We know, however, that the $ab$-mirror symmetry $m_{ab}$ and $ac$-mirror symmetry $m_{ac}$ are absent. Taking the remaining mirror symmetry and the time-reversal symmetry into account, two extra $k$-independent term are allowed: $\mathbb{1} \otimes \tau_x$ and $\sigma_x \otimes \tau_y$ due to the breaking of $m_{ab}$ and $m_{ac}$, respectively. These terms transform the Dirac point into a nodal-line (see below). 
Adding the terms, we obtain the Hamiltonian
\begin{equation}
H=m \mathbb{1}\otimes \tau_z+\hbar (v_a k_a \sigma_z \otimes \tau_x +v_b k_b  \sigma_x \otimes \tau_x +v_c k_c \mathbb{1} \otimes \tau_y)+\Delta \mathbb{1} \otimes \tau_x +\xi \sigma_x \otimes \tau_y. \label{eq:ham}
\end{equation}
The eigenenergies of this Hamiltonian are
\begin{equation}
\varepsilon(\vec k)= \pm \sqrt{m^2 + (\hbar K_2)^2 +\left(\hbar \sqrt{(v_a k_a)^2+(K_1)^2}\pm \sqrt{\Delta^2+\xi^2}\right)^2 },\label{eq:energy}
\end{equation}
with the two $\pm$ distinct from each other, resulting in four separate energy bands. Here we made the coordinate rotation in the (rescaled) $bc$-plane such that $K_1=v_b k_b \cos\phi+v_c k_c \sin\phi$ and $K_2=-v_b k_b \sin\phi+v_c k_c \cos\phi$, where the angle $\phi$ is defined by $\cos\phi=\Delta/\sqrt{\Delta^2+\xi^2}$. Two of these bands have a gap of $\pm \sqrt{m^2+\Delta^2+\xi^2}$, whereas the other two cross for $m=0$ when $\sqrt{(v_a k_a)^2+(K_1)^2}=\sqrt{\Delta^2+\xi^2}$ and $K_2=0$, forming a nodal line (see Fig.~\ref{fig:FS}(a)) and hence the Fermi surface is a torus. This means the nodal line lies in the plane defined by $K_2=0$, i.e. the plane rotated about the $a$-axis from the $ab$-plane by the angle $\theta_{\rm tilt}$, defined via $\cos\theta_{\rm tilt}=\frac{\Delta}{\sqrt{\Delta^2+v_b^2 \xi^2/v_c^2}}\approx1-\frac{v_b^2 \xi^2}{2\Delta^2 v_c^2}$. Since the Fermi-velocities satisfy $v_c \gg v_b$ the angle $\theta_{\rm tilt}$ can be expected to be very small --- this is confirmed by the fact the quantum oscillation experiments find the critical angles in the $bc$-plane rotation $|\theta^\pm_c-90^{\rm o}|$ differ by less than $1^{\rm o}$ (see Sec. \ref{sec:critical}) which also suggests that $\xi\ll\Delta$. As such the impact of the inversion symmetry breaking on the Fermi surface and hence transport is entirely dominated by the energy scale of the $m_{ab}$ breaking term, $\Delta$, which is consistent with the experimental observation that the MCA is predominantly due to magnetic fields in the $\hat{\vec b}$ direction (see previous subsection). In what follows we therefore consider only the experimentally relevant case of $\theta_{\rm tilt} \approx 0$, meaning that $\xi\approx0$, $K_1 \approx v_b k_b$, and $K_2 \approx v_c k_c$, which gives the Hamiltonian Eq. (1) of the main text. The spin texture of the torus Fermi surface is then such that the spin points perpendicular to the momentum projected onto the $ab$-plane (see Figs.~\ref{fig:FS}(b) \& \ref{fig:FS}(c)).

\subsection{First and second order current due to Zeeman effect}

The peculiar spin texture and Fermi surface of ${\rm ZrTe}_{\rm 5}$ are the most obvious candidates to explain the large MCA found in our experiment. Therefore, in the following sections we evaluate the Boltzmann equation to first and second order in electric field to ascertain if there is an easy theoretical explanation for the gigantic experimental MCA from either Zeeman or orbital effects of the magnetic field. We will find that only the Zeeman effect can result in any MCA and, whilst the theoretical effect is sizable in comparison to other materials, the predicted effect is significantly smaller than that found in our experiment.

In the following, we will mainly use the relaxation time approximation \cite{Ideue2017} to solve the Boltzmann equation
\begin{equation}
e (\vec E +{\rm \bf v}_{\vec k}\times \vec B)\cdot \frac{\partial f}{\partial \vec k}=\frac{f-f_0}{\tau},\label{relax}
\end{equation}
where we expand the distribution function $f=f_0+f_1+f_2+\dots$, in powers of the electric field,  where $f_n$ is the $n$th order response proportional to $E^n$ and the static distribution function $f_0$ the Fermi-Dirac distribution function with a chemical potential $\mu$. Effects not described by this simple relaxation time approximation are also discussed below; for instance, in Sec.~\ref{S:anisotropic} we consider anisotropic scattering and our results for orbital fields (Sec.~\ref{S:orbital}) require only that the collision rate respects time-reversal symmetry. Importantly, the magnetic field $\vec B$ enters in two different ways: the orbital contribution is incorporated directly into \eqref{relax}, while the Zeeman term changes directly the electronic dispersion $\epsilon_{\vec k}$ (and the scattering, see Sec.~\ref{S:anisotropic}). We will mainly discuss effects linear in $B$, where by symmetry only the component of $B$ parallel to the $\hat{\vec b}$ direction contributes to the MCA, which is consistent with the experiment.

To describe the torus it is easiest to switch to polar coordinates $\vec k=\{k_r,\phi\}$ such that in the $ab$-plane $k_a=k_r \cos\phi$ and $k_b=k_r \sin\phi$. In this section, in addition to the symmetry breaking term $\Delta(\mathbb{1} \otimes \tau_x)$, we include a Zeeman term for a magnetic field pointing into the $\hat b$ direction, $\frac{1}{2} g_b \mu_B B\sigma_z$, where $g_b$ is the $g$-factor for magnetic fields along the $b$-axis. Below we write $g$ instead of $\frac{1}{2} g_b \mu_B$, to simplify notations.
Including this term, the lowest energy band (for electron doping, $\mu>0$) for isotropic velocities is given by
\begin{equation}
\varepsilon_{\vec k} \approx \sqrt{m^2+(\hbar v k_c)^2+(\hbar v k_r-\Delta)^2}-g B\cos\phi,
\end{equation}
where, without loss of generality, we will always assume energy is positive and we have only taken the first order in Zeeman energy $g B$.

In all cases, we will first present the calculation for the isotropic and massless case, i.e. $v_a=v_b=v_c=v$ and $m=0$, since many of these results can be evaluated analytically. The anisotropic case can then  easily be calculated by using a rescaling of the $k$-space coordinates by $v_i k_i \rightarrow \tilde{k}_i$ in the integral to map to the isotropic case. We also find numerically that our results are largely independent of $m$ --- assuming $|m|\ll \mu$, where $\mu$ is the chemical potential. To begin with, we will discuss effects due to the Zeeman term before investigating orbital effects. It should be noted that there can be no extra contribution to MCA linear in $B$ from an anomalous velocity arising from Berry curvatures, $\vec \Omega \times \vec E$, since this is always transversal to the electric field $\vec E$.

\subsubsection{1st order current with Zeeman term}

Within the relaxation time approximation, \eqref{relax}, the 1st-order distribution function at zero temperature is given by (setting $\hbar=1$ until the end of the calculation)
\begin{equation}
f_1 =e \tau E \frac{\partial f_0}{\partial k_a} =e \tau E\left(-\frac{v k_r-\Delta}{\sqrt{v^2 k_c^2+(v k_r-\Delta)^2}} v \cos\phi\;+\frac{g B \sin^2 \phi}{k_r} \right)\delta(\mu -\epsilon_{\mathbf{k}}).
\end{equation}
The velocity in the $\hat{\vec a}$ direction is given by
\begin{align}
{\rm v}_a=\frac{\partial \varepsilon_{\vec k}}{\partial k_a}=\left(\frac{v k_r-\Delta}{\sqrt{v^2 k_c^2+(v k_r-\Delta)^2}}\cos\phi -\frac{g B \sin^2 \phi}{k_r}\right).
\end{align}
From this, we can obtain the first-order current,
\begin{align}
j^{(1)}_a=&-e \int \frac{d^3{\vec k}}{(2\pi)^3}\; {\rm v}_a f_1 =e^2 E \tau \int \frac{dk_r d\phi dk_c}{(2\pi)^3} k_r \frac{\partial \varepsilon_{\vec k}}{\partial k_a} \frac{\partial f_0}{\partial k_a},\\
&=\frac{e^2 E \tau \Delta}{v(2\pi)^3} \int_0^{2\pi} d\phi \;\pi  \left\{(\mu +g \cos \phi)  \Delta \cos ^2 \phi+  2 g^2 B^2 \sin ^4(\phi ) \sqrt{\frac{(g B \cos \phi +\mu)^2}{\Delta^2 -(g B \cos \phi +\mu)^2 }}\right\}\nonumber\\
&\approx \frac{e^2 E \tau \Delta}{v(2\pi \hbar)^3} \pi^2 \mu \left(1+ \frac{3g^2 B^2}{2\Delta \sqrt{\Delta^2-\mu^2}}\right), \nonumber
\end{align}
where the approximation takes up to 2nd order in $g B$. This shows that the first-order current with only a Zeeman contribution is almost entirely independent of $B$, since $\frac{g^2 B^2}{\Delta^2  \sqrt{\Delta^2-\mu^2}} \ll 1$ in the experimental region of interest $|B|\lesssim0.1$ T.

For anisotropic velocities in the $ab$-plane from the rescaling of the integral measure (see above), we find the zero-field conductivity is
\begin{equation}
\sigma^{(1)}=\frac{v_a e^2 \tau  \mu \Delta}{v_b v_c 8\pi \hbar^3} \label{eq:cond}.
\end{equation}

Comparing this to the experimental reciprocal resistivity in zero field, $R_0$ = 8.9 m$\Omega$cm of sample B, and using the values of $\Delta$ and $\mu$ obtained from SdH oscillations in the same sample (see above), this suggests a (transport) scattering time of $\tau^{\rm tr} \approx 57$ fs or, equivalently, a mean free path of $\ell=v_a \tau^{\rm tr} \approx 40$ nm.

\subsubsection{2nd-order current with Zeeman term}

Continuing with the relaxation time approximation to the next order in electric field, $E^2$, the distribution function at this order is given by 
\begin{align}\label{2ndorder}
f_2 =e \tau E \frac{\partial f_1}{\partial k_a}&=e^2 \tau^2 E^2 \frac{\partial^2 f_0}{\partial k_a^2}
= \frac{e^2}{\hbar^2} E^2  \tau^2 \left( \frac{\partial^2 \epsilon_{\mathbf{k}}}{\partial k_a^2} \frac{\partial f_0}{\partial \epsilon_{\mathbf{k}}} + \left( \frac{\partial \epsilon_{\mathbf{k}}}{\partial k_a}\right)^2 \frac{\partial^2 f_0}{\partial \epsilon_{\mathbf{k}}^2} \right).
\end{align} 
The corresponding second-order current density is for $\mu \lesssim \Delta$
\begin{equation}
j^{(2)}_{\rm Zee}=E^2 \sigma^{(2)}=-e \int \frac{d^3{\vec k}}{(2\pi)^3}\;{\rm v}_a f_2 =e^3 E^2 \tau^2 \int \dfrac{d\phi dk_c d k_r}{(2\pi)^3} k_r  \dfrac{\partial \epsilon_{\mathbf{k}}}{\partial k_a}  \frac{\partial^2 f_0}{\partial k_a^2}\approx \frac{3 e^3 E^2 B \tau^2}{128 \pi \hbar^3} g\mu_B.
\end{equation}
For anisotropic velocities, the extra $k_a$-derivative leads to the result multiplied by a factor $v_a$ in the numerator compared to the first-order current, such that the second-order conductivity $\sigma^{(2)}=j^{(2)}_{\rm Zee}/E^2$ is 
\begin{equation}
|\sigma^{(2)}|\approx \frac{3 e^3 g B \tau^2 v_a^2}{128 \pi \hbar^3 v_b v_c}.\label{eq:sigma2}
\end{equation}
Remarkably, the size of  $\sigma^{(2)}$ is essentially independent of the size of $\Delta$ and the value of the chemical potential for $\mu \ll \Delta$. Both parameters can, however, influence the value of $\tau$. Note that the sign $\sigma^{(2)}$ of the conductivity does, however, depend on the sign of $\Delta$, which determines the direction of the spin texture around the Fermi surface. Furthermore, the sign also changes when $\mu$ changes sign which leads to a hole torus instead of a electron torus.

As in Ref.~\cite{Ideue2017}, the first- and second-order contributions result in a total current density $j=\sigma^{(1)} E+\sigma^{(2)} E^2$ and can be related to the magnitude of $\gamma'$ in the experiment via 
\begin{equation}
|\gamma'|=\frac{|\sigma_{\rm Zee}^{(2)}|}{|B| (\sigma^{(1)})^2}\approx \frac{3 \hbar^3 \pi g v_b  v_c}{2 \Delta^2 e \mu^2}=\frac{3 g}{8 e \Delta v_a n}\sim1\times10^{-11}\;{\rm m^2 A^{-1}T^{-1}}.
\end{equation}
Note that $\gamma'$ is independent of the scattering rate, and we used the relevant paramters estimated from SdH oscillations (see Secs.~\ref{sec:vels} and \ref{sec:paras}) along with the $g$-factor of $\sim$20 estimated in Ref.~\citenum{RYChen2015}. We see that $\gamma'$ diverges rapidly for $\mu \to 0$ and is thus very large for our weakly doped system. Indeed, our numerical estimate is large compared to all other known bulk materials, but it is four orders of magnitude {\em smaller} than the $\gamma'$ experimentally found in ${\rm ZrTe}_{\rm 5}$.

\subsubsection{Anisotropic scattering}\label{S:anisotropic}

Another possible source of the large MCA could be anisotropic scattering around the Fermi surface occurring due to the novel spin texture. Such an effect goes beyond the simple relaxation time approximation of \eqref{relax} and allows for an angular dependence of the scattering time $\tau(\phi)$ induced by the external magnetic field. As a concrete example, we will consider the anisotropic scattering due to the Zeeman effect. The reason for the scattering anisotropy in this case will be the change in density of states seen by different portions of the Fermi surface and the fact that the matrix elements of the Fermi surface favour small angle scattering; both elements result from the novel spin texture. We will, however, more generally show that unless such an anisotropic scattering is extremely large, it cannot explain the gigantic MCA of our experiment. 

We start with Fermi's golden rule which tells us that the scattering rate on the Fermi surface is given by
\begin{equation}
\frac{1}{\tau_{\vec k}(\mu)}=\int \frac{d^3 \vec k'}{(2\pi)^3} |\bra{\vec k'}U\ket{\vec k}|^2 \delta(\varepsilon_{\vec k}-\varepsilon_{\vec k'})\delta(\varepsilon_{\vec k'}-\mu).\label{golden}
\end{equation}
Assuming diagonal, $u_0\mathbb{1}\otimes \mathbb{1}$, impurities --- any anisotropy is actually largely unaffected if impurities also involve a $\tau_i$ component --- and including a Zeeman term in \eqref{eq:ham} due to a field in the $\hat{\vec b}$-direction, this evaluates to 
\begin{equation}
\frac{1}{\tau_{\vec k}(\mu)}\approx \frac{n_{\rm imp}u_0^2}{4}\nu_0(\mu)\left(1+\frac{g B}{2\mu}\cos \phi\right),
\end{equation}
with $\phi$ the angle to the $\hat{\vec a}$-direction, $\nu_0(\mu)=\frac{\Delta\;\mu}{2\pi  \hbar^3 v^3}$ the zero-field density of states, and the factor $1/4$ arises form matrix-element effects around the torus. This means that the scattering time is dependent on angle such that
\begin{equation}
\tau_{\vec k}\approx \tau(\phi)\approx \tau_0\left(1-\frac{g B}{2\mu} \cos\phi\right),
\end{equation}
where the approximation is true to linear order in magnetic field and $\tau_0$ is the zero-field scattering time. Even for the largest fields of interest $\sim 0.1$ T, the anisotropy of scattering is small, $\sim$1\%. For the case of anisotropic velocities, the density of states becomes $\nu_0(\mu)=\frac{\Delta\;\mu}{2\pi  \hbar^3 v_a v_b v_c}$, but the relative anisotropy in the scattering time is unaffected.

The calculation of the 2nd-order conductivity proceeds as above and care only needs to be taken with the derivatives of $\tau(\phi)$. The second-order distribution function, \eqref{2ndorder}, becomes
\begin{align}
f_2 &=e \tau(\phi) E \frac{\partial f_1}{\partial k_a}=e^2 \tau(\phi) E^2 \frac{\partial}{\partial k_a} \left(\tau(\phi)  \frac{\partial}{\partial k_a}  f_0\right)
\\
&= \frac{e^2}{\hbar^2} E^2  \tau(\phi) \left( \tau(\phi) \left(\frac{\partial^2 \epsilon_{\mathbf{k}}}{\partial k_a^2} \frac{\partial f_0}{\partial \epsilon_{\mathbf{k}}} + \left( \frac{\partial \epsilon_{\mathbf{k}}}{\partial k_a}\right)^2 \frac{\partial^2 f_0}{\partial \epsilon_{\mathbf{k}}^2}\right)+ \frac{\partial \tau(\phi)}{\partial k_a} \frac{\partial \epsilon_{\mathbf{k}}}{\partial k_a} \frac{\partial f_0}{\partial \epsilon_{\mathbf{k}}} \right).\nonumber
\end{align} 
Focussing only on the contributions due to this anisotropic scattering, we find that the second-order current $\sigma^{(2)}$ is three times as large as that given by the Fermi-surface deformation, such that
\begin{equation}
|\sigma^{(2)}_\tau|\approx \frac{9 e^3 B \tau_0^2 g v_b v_c}{128 \pi \hbar^3}.
\end{equation}
The prefactor will be modified by factors of order $1$ when instead of the single-particle relaxation rate of \eqref{golden} a transport relaxation rate (or a full solution of the Boltzmann equation for local impurities)  is considered, but this will not change the conclusion that anisotropic scattering rates arising from matrix-element effects cannot explain the gigantic second-harmonic signal observed experimentally.

\subsection{Orbital contributions}\label{S:orbital}
In the previous section, we considered only the Zeeman term, which results in a large MCA but much smaller than that found in the experiment. We now perform a similar investigation of the impact of the orbital contribution, ${\rm \bf v_k} \times \vec B$. Orbital corrections to transport are organised in powers of $\omega_c \tau$, where $\omega_c\propto B$ is the cyclotron frequency. For clean systems with large $\tau$, they therefore are typically much larger than corrections from Zeeman terms.
We will confirm this well-known result for the linear resistivity, but we will show that the leading correction to the second-harmonic transport expected from this argument, $\sigma_2 \propto \tau^3 B$, vanishes. It is therefore unlikely that orbital effects can explain the large MCA found in our experiments.

Starting from \eqref{relax}, we expand again the distribution function $f=f_0+f_1+f_2+...$, where the order $f_i$ now refers to the sum of the powers of $E$ and $B$ fields, with
\begin{equation}
e \vec{E}\  \dfrac{\partial f_{i-1}}{\partial \mathbf{k}}  + e \left({\rm \bf v}_{\vec k}  \times \vec{B}\right) \dfrac{\partial f_{i-1}}{\partial \mathbf{k}} = \dfrac{f_i}{\tau}.
\end{equation}
The first non-zero contributions to the linear magneto-conductivity due to the orbital effects come from the third order distribution function $f_3$. At this order, we are interested in the terms proportional to $E B^2$ and $E^2 B$, which result in a first- and second-order contribution to the current, respectively.

\subsubsection{First-order current}

The contribution of the orbital term to the first-order current is given by the term $\sim E B^2$ in the third-order distribution function, which is proportional to $\tau^3$ and takes the form
\begin{align}
f^{(1)}_3=&e^3 E B^2 \tau^3 \left( \left( \dfrac{\partial \epsilon_{\mathbf{k}}}{\partial k_c} \right)^2 \dfrac{\partial^3 \epsilon_{\mathbf{k}}}{\partial k_a^3} - \dfrac{\partial \epsilon_{\mathbf{k}}}{\partial k_a} \dfrac{\partial \epsilon_{\mathbf{k}}}{\partial k_c} \dfrac{\partial^3 \epsilon_{\mathbf{k}}}{\partial k_a^2 \partial k_c} - \dfrac{\partial \epsilon_{\mathbf{k}}}{\partial k_a} \dfrac{\partial^2 \epsilon_{\mathbf{k}}}{\partial k_c^2} \dfrac{\partial^2 \epsilon_{\mathbf{k}}}{\partial k_a^2} -\dfrac{\partial \epsilon_{\mathbf{k}}}{\partial k_a} \dfrac{\partial \epsilon_{\mathbf{k}}}{\partial k_c} \dfrac{\partial^3 \epsilon_{\mathbf{k}}}{\partial k_a^2 \partial k_c} \right.\\
 & \left. \qquad \qquad \qquad \quad + \dfrac{\partial \epsilon_{\mathbf{k}}}{\partial k_a} \left( \dfrac{\partial^2 \epsilon_{\mathbf{k}}}{\partial k_a \partial k_c} \right)^2 + \left( \dfrac{\partial \epsilon_{\mathbf{k}}}{\partial k_a} \right)^2 \dfrac{\partial^3 \epsilon_{\mathbf{k}}}{\partial k_a \partial k_c^2} \right) \dfrac{\partial f_0}{\partial \epsilon_{\mathbf{k}}}. \nonumber
\end{align}

From this distribution function, we calculate the orbital contribution to the current at first order in electric field by taking the integral
\begin{align}
	j_{a,\text{orb}}^{(1)} = e \int \frac{\mathrm{d}^3 \vec k}{(2\pi)^3} \ \dfrac{\partial \epsilon_{\mathbf{k}}}{\partial k_a} f^{(1)}_3 = \dfrac{3 e^4 E B^2 \tau^3 v^3}{2^4 \pi \mu^3 } \Delta \left(\Delta \sqrt{\left( \Delta^2 - \mu^2 \right)} -\Delta^2 \right).
\end{align}
Neglecting the much smaller Zeeman contribution and reinserting $\hbar$, the resulting magnetoconductivity is
\begin{equation}
\sigma_{aa} = \dfrac{j_a}{E_a} \approx  \dfrac{e^2 \tau \Delta}{(2\pi \hbar)^3 v} \pi^2 \mu -\frac{3 e^4 \tau ^3 v^3 \Delta}{ 2^5 \pi  \hbar^3 \mu } B^2,
\end{equation}
where the approximation assumes $\mu \ll \Delta$.

To calculate the longitudinal magnetoresistivity, we need the full conductivity tensor 
\begin{equation}
\sigma = \left(
\begin{array}{cc}
\sigma_{aa} & \sigma_{ac} \\
\sigma_{ca} & \sigma_{cc} \\
\end{array}
\right)=
\left(
\begin{array}{cc}
\dfrac{\Delta  e^2 \mu  \tau }{8 \pi  \hbar^3 v}-\dfrac{3  \Delta  e^4 \tau ^3 v^3}{2^5 \pi  \hbar^3 \mu } B^2& - \dfrac{\Delta  e^3 \tau ^2 v}{8 \pi  \hbar^3}B \\
\dfrac{\Delta  e^3 \tau ^2 v}{8 \pi  \hbar^3}B & \dfrac{\Delta  e^2 \mu  \tau }{4 \pi  \hbar^3 v}-\dfrac{ \Delta  e^4 \tau ^3 v^3}{8 \pi  \hbar^3 \mu }B^2 \\
\end{array}
\right).
\end{equation}
Adding the effects of anisotropic velocities using again a scaling analysis, we obtain
\begin{equation}
\rho_{aa} = \left(\sigma^{-1} \right)_{aa} \approx \frac{8 \pi  \hbar^3 v_b v_c }{v_a \Delta  e^2 \mu  \tau }+\frac{2 \pi \hbar^3 \tau  v_a v_b v_c^3}{\Delta  \mu ^3} B^2 = \frac{8 \pi  \hbar^3 v_b v_c }{v_a \Delta  e^2 \mu  \tau }\left( 1+ \left(\frac{e B v_a v_c  \tau}{2 \mu}\right)^2 \right).\label{eq:Resistivity}
\end{equation}

\subsubsection{Second-order current}

The term proportional to $E^2 B$ in the third-order distribution function is the candidate to explain our MCA. This term has the form
\begin{align}
f^{(2)}_3 &= \dfrac{e^3}{\hbar^3} E^2 B \tau^3 \left[ \dfrac{\partial}{\partial k_a} \left(  \left(- \dfrac{\partial \epsilon_{\mathbf{k}}}{\partial k_c} \dfrac{\partial^2 \epsilon_{\mathbf{k}}}{\partial k_a^2} + \dfrac{\partial \epsilon_{\mathbf{k}}}{\partial k_a} \dfrac{\partial^2 \epsilon_{\mathbf{k}}}{\partial k_a \partial k_c}     \right) \dfrac{\partial f_0}{\partial \epsilon_{\mathbf{k}}} \right) \right. \\ 
&\quad + \left. \left( - \rm{v}_c \dfrac{\partial}{\partial k_a} + \rm{v}_a \dfrac{\partial}{\partial k_c} \right) \left( \dfrac{\partial^2 \epsilon_{\mathbf{k}}}{\partial k_a^2} \dfrac{\partial f_0}{\partial \epsilon_{\mathbf{k}}} + \left( \dfrac{\partial \epsilon_{\mathbf{k}}}{\partial k_a}\right)^2 \dfrac{\partial^2 f_0}{\partial \epsilon_{\mathbf{k}}^2} \right) \right].  \nonumber
\end{align}

From this we can calculate the current $\sim E^2 B$, i.e. second order in electric field resulting from the orbital effects
\begin{align}
j_{a,\text{orb}}^{(2)} &= e^4 E^2 B \tau^3 \int \dfrac{\rm{d}\phi \rm{d} k_c \rm{d}k_r}{(2\pi \hbar)^3} k_r 
\left[ \dfrac{\partial \epsilon_{\mathbf{k}}}{\partial k_a} \left(\dfrac{\partial \epsilon_{\mathbf{k}}}{\partial k_c} \dfrac{\partial^3 \epsilon_{\mathbf{k}}}{\partial k_a^3} -\dfrac{\partial \epsilon_{\mathbf{k}}}{\partial k_a} \dfrac{\partial^3 \epsilon_{\mathbf{k}}}{\partial k_a^2 \partial k_c} -3 \dfrac{\partial^2 \epsilon_{\mathbf{k}}}{\partial k_a^2} \dfrac{\partial^2 \epsilon_{\mathbf{k}}}{\partial k_a \partial k_c} \right) \right. \\
& \quad + \left. 3 \dfrac{\partial \epsilon_{\mathbf{k}}}{\partial k_c} \left( \dfrac{\partial^2 \epsilon_{\mathbf{k}}}{\partial k_a^2}\right)^2 \right] \delta(\mu -\epsilon_{\mathbf{k}}) = e^4 E^2 B \tau^3 I^{(2)}. \nonumber \label{eq:IntegralOrbital}
\end{align}

Surprisingly, the integral $I^{(2)}$ gives exactly zero due to time-reversal symmetry of the underlying bandstructure (evaluated for $B=0$ as we consider only effects linear in $B$ here). One can show this by using the transformation $\vec{k} \rightarrow -\vec{k}$ of the integral $I^{(2)}$ and the relation $\epsilon_{\mathbf{-k}} = \epsilon_{\mathbf{k}}$ for a time-reversal symmetric energy. Since each term in the integral $I^{(2)}$ contains an odd number of momentum derivatives, this forces $I^{(2)}=j_{a,\text{orb}}^{(2)} = 0$. Importantly, this result does not depend on the relaxation-time approximation and can be generalized to any time-reversal symmetric collision  rate, $\left.\frac{\partial f_{\vec k}}{\partial t}\right|_{\text{coll}} =\sum_{\vec k'} M_{\vec k \vec k'} \delta f_{\vec k'}$ such that the matrix elements satisfy $M_{\vec k,\vec k'}=M_{-\vec k',-\vec k}$. We therefore find that at least within the Boltzmann equation, the MCA from orbital effects vanishes exactly and can thus not explain the experimental result.

\subsection{Further effects beyond the Boltzmann equation and relaxation-time approximation}

In this section, we discuss briefly several effects not included in the calculations presented above and for which a theory has not yet been developed to our knowledge.
First, external electric and magnetic fields in general do not only affect the left-hand side but also the right-hand side of the Boltzmann equation which can give rise to a large number of effects contributing to the MCA. One such effect arising from matrix-element effects has been calculated in Sec.~\ref{S:anisotropic}. Another possible mechanism is that  the external $B$ field can trigger skew scattering processes \cite{Smit1958} even in non-magnetic materials which should also provide a contribution to the MCA. Similarly, an external electric field may deform the screening cloud around a charged impurity. Interaction effects, for example the proximity to a ferroelectric transitions, can enhance such effects. Similarly, all effects arising from Zeeman fields can be enhanced by the proximity to a ferromagnetic instability.
Furthermore, effects may become important which cannot be described by the Boltzmann equation. For example, the vanishing of the orbital contribution of the magnetic field to the MCA is an artifact of the Boltzmann equation. The Boltzmann equation is, however, exact in leading order in $\tau$ and therefore our calculation shows that there is no contribution $\sigma^{(2)}$ proportional to $\tau^3$ and hence no correction to $\gamma'$ which is proportional to $\tau$.

For all those mechanisms, it is not easy to see how an enhancement of the MCA by four orders of magnitude can be achieved.
As discussed in the next paragraph, we argue that a giant enhancement may arise from large-scale inhomogeneities which naturally occur in the presence of charged impurities in systems with low density and weak screening.

\subsection{Enhancement of nonlinear transport by large-scale inhomogeneities}
\label{sec:puddle}

Due to the tiny electron density of $n\approx  2.3 \times10^{16}\;{\rm cm}^{-3}$, which corresponds to just $4.7\times 10^{-6}$ electrons per formula unit, one can expect that the unavoidable presence of charged impurities leads to the formation of large-scale inhomogeneities, so-called puddles \cite{Skinner2012,Borgwardt2016,Breunig2017}, which we can describe by smooth variations in the chemical potential $\mu(\vec r)$. Puddle formation and a percolation transition of puddles driven by a magnetic field has, for example, been observed in weakly-doped topological insulators \cite{Breunig2017} even for electron densities exceeding ours by at least one order of magnitude.

If the apparent resistivity in such an inhomogeneous system is governed by regions with a high resistivity (and possibly, intrinsic $pn$-junctions), this can give rise to a gigantic enhancement of nonlinear transport as the inhomogeneous local electric field can become very large in high-resistivity areas (see below). 

A comparison of resistivity and quantum-oscillation data provides strong experimental evidence for the presence of inhomogeneous transport. The observation of quantum oscillations demonstrates that our system is rather clean. From the Dingle analysis of our quantum-oscillation data, presented above in Sec.~\ref{sec:DingleT}, we extract the relevant scattering times, $\tau_{\rm D}$, of $320$\,fs  for $\vec B \| \hat c$  (F$_3$ branch) and $\sim$500\,fs  for $\vec B \| \hat b$ (F$_1$ branch). In contrast, the transport scattering time, obtained by fitting \eqref{eq:cond} to the measured conductivity, is much smaller, $\tau_{\rm tr} \simeq 57\,$fs. The observation that $ \tau_{\rm tr}\ll \tau_{\rm D} $ is highly anomalous. In homogeneous systems, transport scattering times are usually \emph{larger} than the scattering time controlling quantum oscillations (i.e.  $\tau_{\rm D}$), as small-angle forward scattering does not contribute to transport but does lead to a decay of quantum oscillations. The fact that the observed resistivity is almost an order of magnitude larger than expected from our fits to the quantum oscillation data strongly suggests that the resistivity (and therefore also nonlinear transport) is dominated by local regions of low conductivity in a system with large-scale inhomogeneities. 

The short transport scattering time also implies that regions of low conductivity are not simply short-circuited by regions of high conductivity.
Here, an important factor is that the transport in ZrTe$_5$ is mainly along ZrTe$_3$ chains oriented along the $a$ direction \cite{Lv2017} which is also reflected in the 
highly anisotropic Fermi velocities, $v_a:v_b:v_c = 16:1:4$, as discussed in Sec.~\ref{sec:vels}. These anisotropies suppress the flow of electrons around obstacles and electrons cannot easily avoid regions of low conductivity.

In passing, we note that our observation of the resistivity to be higher than that expected from quantum oscillations speaks against the contribution of a parallel conduction channel, which would work to reduce the resistivity. In this regard, even though topological surface states are expected to exist in topological semimetals, it is unlikely that they are providing a measurable contribution to the transport properties in ZrTe$_5$. The absence of any additional SdH-oscillation frequency to point to the existence of a 2D Fermi surface also supports the lack of contribution from the surface states in our experiment.

The quasi one-dimensional nature of transport motivates us to investigate a highly simplified setup 
 where the local chemical potential varies only parallel to the direction of current flow, $\mu=\mu(x)$, where we denote the coordinate in the $\hat a$ direction by $x$. If we assume that the variations of $\mu$ occur on a length scale much larger than the Fermi wavelength and the mean-free path, we can calculate the local electric fields simply from
\begin{eqnarray}
j=\sigma^{(1)}(x) E(x)+ \sigma^{(2)}(x) E(x)^2,
\end{eqnarray}
where the current density $j$ is constant in space within our simplified setup. For small $j$ this gives rise to a voltage drop
\begin{eqnarray}
\Delta V=  \int E(x) dx \approx j \int \frac{1}{\sigma^{(1)}(x)} dx E(x)- j^2 \int \frac{\sigma^{(2)}(x)}{\sigma^{(1)}(x)^3}dx,
\end{eqnarray}
from which we obtain for the amplitude of the MCA
\begin{eqnarray}
\gamma'=  - \frac{\int \frac{\sigma^{(2)}(x)}{B (\sigma^{(1)}(x))^3}d x}{\int \frac{1}{\sigma^{(1)}(x)} dx }.
\end{eqnarray}
According to our Boltzmann results, \eqref{eq:cond} and \eqref{eq:sigma2},  $\sigma^{(1)}$ is linear in $|\mu|$ while $\sigma^{(2)}$ remains independent of $\mu$ for $1/\tau \ll \mu \ll \Delta$ but changes sign for negative $\mu$. Denoting the average chemical potential by $\bar \mu$ and the value of $\gamma'$ for the homogeneous system by $\bar \gamma'$ and assuming for simplicity a space-independent scattering time $\tau$, we find that $\gamma'$ is enhance by a factor $A$
\begin{eqnarray}
\gamma' \approx A \bar \gamma'  \qquad \text{with } A= \frac{\int \left(\frac{\bar \mu}{\mu(x)}\right)^3 d x}{\int  \frac{\bar \mu}{|\mu(x)|} dx }.\label{eq:enhancement}
\end{eqnarray}
Due to the $1/\mu(x)^3$ term the integral is strongly divergent when $\mu$ approaches $0$ and thus the nodal line. 
In a real system, this divergence will be cut off by a number of effects (e.g., scattering rates or geometry effects arising from the three-dimensional current flow) and thus \eqref{eq:enhancement} does not provide a quantitative prediction of the enhancement effect. It demonstrates, however, that a strong enhancement of nonlinear transport can be expected in the presence of large-scale inhomogeneities if transport is dominated by areas with a low conductivity. The local formation of intrinsic $pn$-junctions may enhance the effect, again due to an increase of the local electric fields.

As our analysis of scattering times supports this scenario, as discussed above, we think that the enhancement of nonlinear transport by large-scale inhomogeneities is the most likely mechanism to explain our data.

\subsection{Discussion of the Berry curvature and the anomalous Hall effect for the torus Fermi surface} \label{S:berry}

The Berry curvature in the direction perpendicular to the $i,j$ plane due to a band $n$ is most easily calculated numerically from Eq.~(1) in the main text by using the formula
\begin{equation}
\Omega_{{ij},n}(\vec k)=i\sum_{n'\neq n}{\langle n|(\partial H/\partial k_i) |n'\rangle\langle n'|(\partial H/\partial k_j) | n\rangle-\langle n|(\partial H/\partial k_j) |n'\rangle\langle n'|(\partial H/\partial k_i) | n\rangle\over(\varepsilon_n-\varepsilon_{n'})^2},
\end{equation}
where the sum over $n'$ is of all bands, not including the band $n$. 

We find numerically that at zero magnetic field the Berry curvature remains exactly zero and that the Zeeman terms of the form $\sigma_i \otimes \mathbb{1}$ only results in a finite Berry curvature for a magnetic field in the $\hat{\vec c}$ direction. Nonetheless, other possible Zeeman-{\it like} terms such as $\sigma_z \otimes \tau_z$ are possible; such terms induce Berry curvatures also for fields pointing in the $\hat{\vec b}$ direction and require further investigation.


\begin{thebibliography}{39}%
\makeatletter
\providecommand \@ifxundefined [1]{%
 \@ifx{#1\undefined}
}%
\providecommand \@ifnum [1]{%
 \ifnum #1\expandafter \@firstoftwo
 \else \expandafter \@secondoftwo
 \fi
}%
\providecommand \@ifx [1]{%
 \ifx #1\expandafter \@firstoftwo
 \else \expandafter \@secondoftwo
 \fi
}%
\providecommand \natexlab [1]{#1}%
\providecommand \enquote  [1]{``#1''}%
\providecommand \bibnamefont  [1]{#1}%
\providecommand \bibfnamefont [1]{#1}%
\providecommand \citenamefont [1]{#1}%
\providecommand \href@noop [0]{\@secondoftwo}%
\providecommand \href [0]{\begingroup \@sanitize@url \@href}%
\providecommand \@href[1]{\@@startlink{#1}\@@href}%
\providecommand \@@href[1]{\endgroup#1\@@endlink}%
\providecommand \@sanitize@url [0]{\catcode `\\12\catcode `\$12\catcode
  `\&12\catcode `\#12\catcode `\^12\catcode `\_12\catcode `\%12\relax}%
\providecommand \@@startlink[1]{}%
\providecommand \@@endlink[0]{}%
\providecommand \url  [0]{\begingroup\@sanitize@url \@url }%
\providecommand \@url [1]{\endgroup\@href {#1}{\urlprefix }}%
\providecommand \urlprefix  [0]{URL }%
\providecommand \Eprint [0]{\href }%
\providecommand \doibase [0]{https://doi.org/}%
\providecommand \selectlanguage [0]{\@gobble}%
\providecommand \bibinfo  [0]{\@secondoftwo}%
\providecommand \bibfield  [0]{\@secondoftwo}%
\providecommand \translation [1]{[#1]}%
\providecommand \BibitemOpen [0]{}%
\providecommand \bibitemStop [0]{}%
\providecommand \bibitemNoStop [0]{.\EOS\space}%
\providecommand \EOS [0]{\spacefactor3000\relax}%
\providecommand \BibitemShut  [1]{\csname bibitem#1\endcsname}%
\let\auto@bib@innerbib\@empty
\bibitem [{\citenamefont {Tokura}\ and\ \citenamefont
  {Nagaosa}(2018)}]{Tokura2018}%
  \BibitemOpen
  \bibfield  {author} {\bibinfo {author} {\bibfnamefont {Y.}~\bibnamefont
  {Tokura}}\ and\ \bibinfo {author} {\bibfnamefont {N.}~\bibnamefont
  {Nagaosa}},\ }\href {https://doi.org/10.1038/s41467-018-05759-4} {\bibfield
  {journal} {\bibinfo  {journal} {Nat. Commun.}\ }\textbf {\bibinfo {volume}
  {9}},\ \bibinfo {pages} {3740} (\bibinfo {year} {2018})}\BibitemShut
  {NoStop}%
\bibitem [{\citenamefont {Rikken}\ \emph {et~al.}(2001)\citenamefont {Rikken},
  \citenamefont {Folling},\ and\ \citenamefont {Wyder}}]{Rikken2001}%
  \BibitemOpen
  \bibfield  {author} {\bibinfo {author} {\bibfnamefont {G.~L. J.~A.}\
  \bibnamefont {Rikken}}, \bibinfo {author} {\bibfnamefont {J.}~\bibnamefont
  {Folling}},\ and\ \bibinfo {author} {\bibfnamefont {P.}~\bibnamefont
  {Wyder}},\ }\href {https://doi.org/10.1103/PhysRevLett.87.236602} {\bibfield
  {journal} {\bibinfo  {journal} {Phys. Rev. Lett.}\ }\textbf {\bibinfo
  {volume} {87}},\ \bibinfo {pages} {236602} (\bibinfo {year}
  {2001})}\BibitemShut {NoStop}%
\bibitem [{\citenamefont {Rikken}\ and\ \citenamefont
  {Wyder}(2005)}]{Rikken2005}%
  \BibitemOpen
  \bibfield  {author} {\bibinfo {author} {\bibfnamefont {G.~L. J.~A.}\
  \bibnamefont {Rikken}}\ and\ \bibinfo {author} {\bibfnamefont
  {P.}~\bibnamefont {Wyder}},\ }\href
  {https://doi.org/10.1103/PhysRevLett.94.016601} {\bibfield  {journal}
  {\bibinfo  {journal} {Phys. Rev. Lett.}\ }\textbf {\bibinfo {volume} {94}},\
  \bibinfo {pages} {016601} (\bibinfo {year} {2005})}\BibitemShut {NoStop}%
\bibitem [{\citenamefont {Ideue}\ \emph {et~al.}(2017)\citenamefont {Ideue},
  \citenamefont {Hamamoto}, \citenamefont {Koshikawa}, \citenamefont {Ezawa},
  \citenamefont {Shimizu}, \citenamefont {Kaneko}, \citenamefont {Tokura},
  \citenamefont {Nagaosa},\ and\ \citenamefont {Iwasa}}]{Ideue2017}%
  \BibitemOpen
  \bibfield  {author} {\bibinfo {author} {\bibfnamefont {T.}~\bibnamefont
  {Ideue}}, \bibinfo {author} {\bibfnamefont {K.}~\bibnamefont {Hamamoto}},
  \bibinfo {author} {\bibfnamefont {S.}~\bibnamefont {Koshikawa}}, \bibinfo
  {author} {\bibfnamefont {M.}~\bibnamefont {Ezawa}}, \bibinfo {author}
  {\bibfnamefont {S.}~\bibnamefont {Shimizu}}, \bibinfo {author} {\bibfnamefont
  {Y.}~\bibnamefont {Kaneko}}, \bibinfo {author} {\bibfnamefont
  {Y.}~\bibnamefont {Tokura}}, \bibinfo {author} {\bibfnamefont
  {N.}~\bibnamefont {Nagaosa}},\ and\ \bibinfo {author} {\bibfnamefont
  {Y.}~\bibnamefont {Iwasa}},\ }\href {https://doi.org/10.1038/nphys4056}
  {\bibfield  {journal} {\bibinfo  {journal} {Nat. Phys.}\ }\textbf {\bibinfo
  {volume} {13}},\ \bibinfo {pages} {578} (\bibinfo {year} {2017})}\BibitemShut
  {NoStop}%
\bibitem [{\citenamefont {Rikken}\ and\ \citenamefont
  {Avarvari}(2019)}]{Rikken2019}%
  \BibitemOpen
  \bibfield  {author} {\bibinfo {author} {\bibfnamefont {G.~L. J.~A.}\
  \bibnamefont {Rikken}}\ and\ \bibinfo {author} {\bibfnamefont
  {N.}~\bibnamefont {Avarvari}},\ }\href
  {https://doi.org/10.1103/PhysRevB.99.245153} {\bibfield  {journal} {\bibinfo
  {journal} {Phys. Rev. B}\ }\textbf {\bibinfo {volume} {99}},\ \bibinfo
  {pages} {245153} (\bibinfo {year} {2019})}\BibitemShut {NoStop}%
\bibitem [{\citenamefont {Morimoto}\ and\ \citenamefont
  {Nagaosa}(2016)}]{Morimoto2016}%
  \BibitemOpen
  \bibfield  {author} {\bibinfo {author} {\bibfnamefont {T.}~\bibnamefont
  {Morimoto}}\ and\ \bibinfo {author} {\bibfnamefont {N.}~\bibnamefont
  {Nagaosa}},\ }\href {https://doi.org/10.1103/PhysRevLett.117.146603}
  {\bibfield  {journal} {\bibinfo  {journal} {Phys. Rev. Lett.}\ }\textbf
  {\bibinfo {volume} {117}},\ \bibinfo {pages} {146603} (\bibinfo {year}
  {2016})}\BibitemShut {NoStop}%
\bibitem [{\citenamefont {Weng}\ \emph {et~al.}(2014)\citenamefont {Weng},
  \citenamefont {Dai},\ and\ \citenamefont {Fang}}]{Weng2014}%
  \BibitemOpen
  \bibfield  {author} {\bibinfo {author} {\bibfnamefont {H.}~\bibnamefont
  {Weng}}, \bibinfo {author} {\bibfnamefont {X.}~\bibnamefont {Dai}},\ and\
  \bibinfo {author} {\bibfnamefont {Z.}~\bibnamefont {Fang}},\ }\href@noop {}
  {\bibfield  {journal} {\bibinfo  {journal} {Phys. Rev. X}\ }\textbf {\bibinfo
  {volume} {4}},\ \bibinfo {pages} {011002} (\bibinfo {year}
  {2014})}\BibitemShut {NoStop}%
\bibitem [{\citenamefont {Xu}\ \emph {et~al.}(2018)\citenamefont {Xu},
  \citenamefont {Zhao}, \citenamefont {Marsik}, \citenamefont {Sheveleva},
  \citenamefont {Lyzwa}, \citenamefont {Dai}, \citenamefont {Chen},
  \citenamefont {Qiu},\ and\ \citenamefont {Bernhard}}]{Xu2018}%
  \BibitemOpen
  \bibfield  {author} {\bibinfo {author} {\bibfnamefont {B.}~\bibnamefont
  {Xu}}, \bibinfo {author} {\bibfnamefont {L.~X.}\ \bibnamefont {Zhao}},
  \bibinfo {author} {\bibfnamefont {P.}~\bibnamefont {Marsik}}, \bibinfo
  {author} {\bibfnamefont {E.}~\bibnamefont {Sheveleva}}, \bibinfo {author}
  {\bibfnamefont {F.}~\bibnamefont {Lyzwa}}, \bibinfo {author} {\bibfnamefont
  {Y.~M.}\ \bibnamefont {Dai}}, \bibinfo {author} {\bibfnamefont {G.~F.}\
  \bibnamefont {Chen}}, \bibinfo {author} {\bibfnamefont {X.~G.}\ \bibnamefont
  {Qiu}},\ and\ \bibinfo {author} {\bibfnamefont {C.}~\bibnamefont
  {Bernhard}},\ }\href {https://doi.org/10.1103/PhysRevLett.121.187401}
  {\bibfield  {journal} {\bibinfo  {journal} {Phys. Rev. Lett.}\ }\textbf
  {\bibinfo {volume} {121}},\ \bibinfo {pages} {187401} (\bibinfo {year}
  {2018})}\BibitemShut {NoStop}%
\bibitem [{\citenamefont {Chen}\ \emph {et~al.}(2015)\citenamefont {Chen},
  \citenamefont {Chen}, \citenamefont {Song}, \citenamefont {Schneeloch},
  \citenamefont {Gu}, \citenamefont {Wang},\ and\ \citenamefont
  {Wang}}]{RYChen2015}%
  \BibitemOpen
  \bibfield  {author} {\bibinfo {author} {\bibfnamefont {R.~Y.}\ \bibnamefont
  {Chen}}, \bibinfo {author} {\bibfnamefont {Z.~G.}\ \bibnamefont {Chen}},
  \bibinfo {author} {\bibfnamefont {X.~Y.}\ \bibnamefont {Song}}, \bibinfo
  {author} {\bibfnamefont {J.~A.}\ \bibnamefont {Schneeloch}}, \bibinfo
  {author} {\bibfnamefont {G.~D.}\ \bibnamefont {Gu}}, \bibinfo {author}
  {\bibfnamefont {F.}~\bibnamefont {Wang}},\ and\ \bibinfo {author}
  {\bibfnamefont {N.~L.}\ \bibnamefont {Wang}},\ }\href
  {https://doi.org/10.1103/PhysRevLett.115.176404} {\bibfield  {journal}
  {\bibinfo  {journal} {Phys. Rev. Lett.}\ }\textbf {\bibinfo {volume} {115}},\
  \bibinfo {pages} {176404} (\bibinfo {year} {2015})}\BibitemShut {NoStop}%
\bibitem [{\citenamefont {Li}\ \emph {et~al.}(2016)\citenamefont {Li},
  \citenamefont {Kharzeev}, \citenamefont {Zhang}, \citenamefont {Huang},
  \citenamefont {Pletikosic}, \citenamefont {Fedorov}, \citenamefont {Zhong},
  \citenamefont {Schneeloch}, \citenamefont {Gu},\ and\ \citenamefont
  {Valla}}]{Li2016}%
  \BibitemOpen
  \bibfield  {author} {\bibinfo {author} {\bibfnamefont {Q.}~\bibnamefont
  {Li}}, \bibinfo {author} {\bibfnamefont {D.~E.}\ \bibnamefont {Kharzeev}},
  \bibinfo {author} {\bibfnamefont {C.}~\bibnamefont {Zhang}}, \bibinfo
  {author} {\bibfnamefont {Y.}~\bibnamefont {Huang}}, \bibinfo {author}
  {\bibfnamefont {I.}~\bibnamefont {Pletikosic}}, \bibinfo {author}
  {\bibfnamefont {A.~V.}\ \bibnamefont {Fedorov}}, \bibinfo {author}
  {\bibfnamefont {R.~D.}\ \bibnamefont {Zhong}}, \bibinfo {author}
  {\bibfnamefont {J.~A.}\ \bibnamefont {Schneeloch}}, \bibinfo {author}
  {\bibfnamefont {G.~D.}\ \bibnamefont {Gu}},\ and\ \bibinfo {author}
  {\bibfnamefont {T.}~\bibnamefont {Valla}},\ }\href
  {https://doi.org/10.1038/nphys3648
  http://www.nature.com/nphys/journal/v12/n6/abs/nphys3648.html#supplementary-information}
  {\bibfield  {journal} {\bibinfo  {journal} {Nat. Phys.}\ }\textbf {\bibinfo
  {volume} {12}},\ \bibinfo {pages} {550} (\bibinfo {year} {2016})}\BibitemShut
  {NoStop}%
\bibitem [{\citenamefont {Zhang}\ \emph {et~al.}(2017)\citenamefont {Zhang},
  \citenamefont {Wang}, \citenamefont {Yu}, \citenamefont {Liu}, \citenamefont
  {Liang}, \citenamefont {Huang}, \citenamefont {Nie}, \citenamefont {Sun},
  \citenamefont {Zhang}, \citenamefont {Shen}, \citenamefont {Liu},
  \citenamefont {Weng}, \citenamefont {Zhao}, \citenamefont {Chen},
  \citenamefont {Jia}, \citenamefont {Hu}, \citenamefont {Ding}, \citenamefont
  {Zhao}, \citenamefont {Gao}, \citenamefont {Li}, \citenamefont {He},
  \citenamefont {Zhao}, \citenamefont {Zhang}, \citenamefont {Zhang},
  \citenamefont {Yang}, \citenamefont {Wang}, \citenamefont {Peng},
  \citenamefont {Dai}, \citenamefont {Fang}, \citenamefont {Xu}, \citenamefont
  {Chen},\ and\ \citenamefont {Zhou}}]{YZhang2017}%
  \BibitemOpen
  \bibfield  {author} {\bibinfo {author} {\bibfnamefont {Y.}~\bibnamefont
  {Zhang}}, \bibinfo {author} {\bibfnamefont {C.}~\bibnamefont {Wang}},
  \bibinfo {author} {\bibfnamefont {L.}~\bibnamefont {Yu}}, \bibinfo {author}
  {\bibfnamefont {G.}~\bibnamefont {Liu}}, \bibinfo {author} {\bibfnamefont
  {A.}~\bibnamefont {Liang}}, \bibinfo {author} {\bibfnamefont
  {J.}~\bibnamefont {Huang}}, \bibinfo {author} {\bibfnamefont
  {S.}~\bibnamefont {Nie}}, \bibinfo {author} {\bibfnamefont {X.}~\bibnamefont
  {Sun}}, \bibinfo {author} {\bibfnamefont {Y.}~\bibnamefont {Zhang}}, \bibinfo
  {author} {\bibfnamefont {B.}~\bibnamefont {Shen}}, \bibinfo {author}
  {\bibfnamefont {J.}~\bibnamefont {Liu}}, \bibinfo {author} {\bibfnamefont
  {H.}~\bibnamefont {Weng}}, \bibinfo {author} {\bibfnamefont {L.}~\bibnamefont
  {Zhao}}, \bibinfo {author} {\bibfnamefont {G.}~\bibnamefont {Chen}}, \bibinfo
  {author} {\bibfnamefont {X.}~\bibnamefont {Jia}}, \bibinfo {author}
  {\bibfnamefont {C.}~\bibnamefont {Hu}}, \bibinfo {author} {\bibfnamefont
  {Y.}~\bibnamefont {Ding}}, \bibinfo {author} {\bibfnamefont {W.}~\bibnamefont
  {Zhao}}, \bibinfo {author} {\bibfnamefont {Q.}~\bibnamefont {Gao}}, \bibinfo
  {author} {\bibfnamefont {C.}~\bibnamefont {Li}}, \bibinfo {author}
  {\bibfnamefont {S.}~\bibnamefont {He}}, \bibinfo {author} {\bibfnamefont
  {L.}~\bibnamefont {Zhao}}, \bibinfo {author} {\bibfnamefont {F.}~\bibnamefont
  {Zhang}}, \bibinfo {author} {\bibfnamefont {S.}~\bibnamefont {Zhang}},
  \bibinfo {author} {\bibfnamefont {F.}~\bibnamefont {Yang}}, \bibinfo {author}
  {\bibfnamefont {Z.}~\bibnamefont {Wang}}, \bibinfo {author} {\bibfnamefont
  {Q.}~\bibnamefont {Peng}}, \bibinfo {author} {\bibfnamefont {X.}~\bibnamefont
  {Dai}}, \bibinfo {author} {\bibfnamefont {Z.}~\bibnamefont {Fang}}, \bibinfo
  {author} {\bibfnamefont {Z.}~\bibnamefont {Xu}}, \bibinfo {author}
  {\bibfnamefont {C.}~\bibnamefont {Chen}},\ and\ \bibinfo {author}
  {\bibfnamefont {X.~J.}\ \bibnamefont {Zhou}},\ }\href
  {https://doi.org/10.1038/ncomms15512
  https://www.nature.com/articles/ncomms15512#supplementary-information}
  {\bibfield  {journal} {\bibinfo  {journal} {Nat. Commun.}\ }\textbf {\bibinfo
  {volume} {8}},\ \bibinfo {pages} {15512} (\bibinfo {year}
  {2017})}\BibitemShut {NoStop}%
\bibitem [{\citenamefont {Liang}\ \emph {et~al.}(2018)\citenamefont {Liang},
  \citenamefont {Lin}, \citenamefont {Gibson}, \citenamefont {Kushwaha},
  \citenamefont {Liu}, \citenamefont {Wang}, \citenamefont {Xiong},
  \citenamefont {Sobota}, \citenamefont {Hashimoto}, \citenamefont {Kirchmann},
  \citenamefont {Shen}, \citenamefont {Cava},\ and\ \citenamefont
  {Ong}}]{Liang2018}%
  \BibitemOpen
  \bibfield  {author} {\bibinfo {author} {\bibfnamefont {T.}~\bibnamefont
  {Liang}}, \bibinfo {author} {\bibfnamefont {J.}~\bibnamefont {Lin}}, \bibinfo
  {author} {\bibfnamefont {Q.}~\bibnamefont {Gibson}}, \bibinfo {author}
  {\bibfnamefont {S.}~\bibnamefont {Kushwaha}}, \bibinfo {author}
  {\bibfnamefont {M.}~\bibnamefont {Liu}}, \bibinfo {author} {\bibfnamefont
  {W.}~\bibnamefont {Wang}}, \bibinfo {author} {\bibfnamefont {H.}~\bibnamefont
  {Xiong}}, \bibinfo {author} {\bibfnamefont {J.~A.}\ \bibnamefont {Sobota}},
  \bibinfo {author} {\bibfnamefont {M.}~\bibnamefont {Hashimoto}}, \bibinfo
  {author} {\bibfnamefont {P.~S.}\ \bibnamefont {Kirchmann}}, \bibinfo {author}
  {\bibfnamefont {Z.-X.}\ \bibnamefont {Shen}}, \bibinfo {author}
  {\bibfnamefont {R.~J.}\ \bibnamefont {Cava}},\ and\ \bibinfo {author}
  {\bibfnamefont {N.~P.}\ \bibnamefont {Ong}},\ }\href
  {https://doi.org/10.1038/s41567-018-0078-z} {\bibfield  {journal} {\bibinfo
  {journal} {Nat. Phys.}\ }\textbf {\bibinfo {volume} {14}},\ \bibinfo {pages}
  {451} (\bibinfo {year} {2018})}\BibitemShut {NoStop}%
\bibitem [{\citenamefont {Wang}\ \emph {et~al.}(2018)\citenamefont {Wang},
  \citenamefont {Liu}, \citenamefont {Li}, \citenamefont {Liu}, \citenamefont
  {Wang}, \citenamefont {Liu}, \citenamefont {Dai}, \citenamefont {Wang},
  \citenamefont {Li}, \citenamefont {Yan}, \citenamefont {Mandrus},
  \citenamefont {Xie},\ and\ \citenamefont {Wang}}]{HWang2018}%
  \BibitemOpen
  \bibfield  {author} {\bibinfo {author} {\bibfnamefont {H.}~\bibnamefont
  {Wang}}, \bibinfo {author} {\bibfnamefont {H.}~\bibnamefont {Liu}}, \bibinfo
  {author} {\bibfnamefont {Y.}~\bibnamefont {Li}}, \bibinfo {author}
  {\bibfnamefont {Y.}~\bibnamefont {Liu}}, \bibinfo {author} {\bibfnamefont
  {J.}~\bibnamefont {Wang}}, \bibinfo {author} {\bibfnamefont {J.}~\bibnamefont
  {Liu}}, \bibinfo {author} {\bibfnamefont {J.-Y.}\ \bibnamefont {Dai}},
  \bibinfo {author} {\bibfnamefont {Y.}~\bibnamefont {Wang}}, \bibinfo {author}
  {\bibfnamefont {L.}~\bibnamefont {Li}}, \bibinfo {author} {\bibfnamefont
  {J.}~\bibnamefont {Yan}}, \bibinfo {author} {\bibfnamefont {D.}~\bibnamefont
  {Mandrus}}, \bibinfo {author} {\bibfnamefont {X.~C.}\ \bibnamefont {Xie}},\
  and\ \bibinfo {author} {\bibfnamefont {J.}~\bibnamefont {Wang}},\ }\href
  {https://doi.org/10.1126/sciadv.aau5096} {\bibfield  {journal} {\bibinfo
  {journal} {Sci. Adv.}\ }\textbf {\bibinfo {volume} {4}},\ \bibinfo {pages}
  {eaau5096} (\bibinfo {year} {2018})}\BibitemShut {NoStop}%
\bibitem [{\citenamefont {Shahi}\ \emph {et~al.}(2018)\citenamefont {Shahi},
  \citenamefont {Singh}, \citenamefont {Sun}, \citenamefont {Zhao},
  \citenamefont {Chen}, \citenamefont {Lv}, \citenamefont {Li}, \citenamefont
  {Yan}, \citenamefont {Mandrus},\ and\ \citenamefont {Cheng}}]{Shahi2018}%
  \BibitemOpen
  \bibfield  {author} {\bibinfo {author} {\bibfnamefont {P.}~\bibnamefont
  {Shahi}}, \bibinfo {author} {\bibfnamefont {D.~J.}\ \bibnamefont {Singh}},
  \bibinfo {author} {\bibfnamefont {J.~P.}\ \bibnamefont {Sun}}, \bibinfo
  {author} {\bibfnamefont {L.~X.}\ \bibnamefont {Zhao}}, \bibinfo {author}
  {\bibfnamefont {G.~F.}\ \bibnamefont {Chen}}, \bibinfo {author}
  {\bibfnamefont {Y.~Y.}\ \bibnamefont {Lv}}, \bibinfo {author} {\bibfnamefont
  {J.}~\bibnamefont {Li}}, \bibinfo {author} {\bibfnamefont {J.~Q.}\
  \bibnamefont {Yan}}, \bibinfo {author} {\bibfnamefont {D.}~\bibnamefont
  {Mandrus}},\ and\ \bibinfo {author} {\bibfnamefont {J.~G.}\ \bibnamefont
  {Cheng}},\ }\href {https://doi.org/10.1103/PhysRevX.8.021055} {\bibfield
  {journal} {\bibinfo  {journal} {Phy. Rev. X}\ }\textbf {\bibinfo {volume}
  {8}},\ \bibinfo {pages} {021055} (\bibinfo {year} {2018})}\BibitemShut
  {NoStop}%
\bibitem [{\citenamefont {Tang}\ \emph {et~al.}(2019)\citenamefont {Tang},
  \citenamefont {Ren}, \citenamefont {Wang}, \citenamefont {Zhong},
  \citenamefont {Schneeloch}, \citenamefont {Yang}, \citenamefont {Yang},
  \citenamefont {Lee}, \citenamefont {Gu}, \citenamefont {Qiao},\ and\
  \citenamefont {Zhang}}]{Tang2019}%
  \BibitemOpen
  \bibfield  {author} {\bibinfo {author} {\bibfnamefont {F.}~\bibnamefont
  {Tang}}, \bibinfo {author} {\bibfnamefont {Y.}~\bibnamefont {Ren}}, \bibinfo
  {author} {\bibfnamefont {P.}~\bibnamefont {Wang}}, \bibinfo {author}
  {\bibfnamefont {R.}~\bibnamefont {Zhong}}, \bibinfo {author} {\bibfnamefont
  {J.}~\bibnamefont {Schneeloch}}, \bibinfo {author} {\bibfnamefont {S.~A.}\
  \bibnamefont {Yang}}, \bibinfo {author} {\bibfnamefont {K.}~\bibnamefont
  {Yang}}, \bibinfo {author} {\bibfnamefont {P.~A.}\ \bibnamefont {Lee}},
  \bibinfo {author} {\bibfnamefont {G.}~\bibnamefont {Gu}}, \bibinfo {author}
  {\bibfnamefont {Z.}~\bibnamefont {Qiao}},\ and\ \bibinfo {author}
  {\bibfnamefont {L.}~\bibnamefont {Zhang}},\ }\href
  {https://doi.org/10.1038/s41586-019-1180-9} {\bibfield  {journal} {\bibinfo
  {journal} {Nature}\ }\textbf {\bibinfo {volume} {569}},\ \bibinfo {pages}
  {537} (\bibinfo {year} {2019})}\BibitemShut {NoStop}%
\bibitem [{\citenamefont {Sun}\ \emph {et~al.}(2020)\citenamefont {Sun},
  \citenamefont {Cao}, \citenamefont {Cui}, \citenamefont {Zhu}, \citenamefont
  {Ma}, \citenamefont {Wang}, \citenamefont {Zhuo}, \citenamefont {Cheng},
  \citenamefont {Wang}, \citenamefont {Wan},\ and\ \citenamefont
  {Chen}}]{Sun2020}%
  \BibitemOpen
  \bibfield  {author} {\bibinfo {author} {\bibfnamefont {Z.}~\bibnamefont
  {Sun}}, \bibinfo {author} {\bibfnamefont {Z.}~\bibnamefont {Cao}}, \bibinfo
  {author} {\bibfnamefont {J.}~\bibnamefont {Cui}}, \bibinfo {author}
  {\bibfnamefont {C.}~\bibnamefont {Zhu}}, \bibinfo {author} {\bibfnamefont
  {D.}~\bibnamefont {Ma}}, \bibinfo {author} {\bibfnamefont {H.}~\bibnamefont
  {Wang}}, \bibinfo {author} {\bibfnamefont {W.}~\bibnamefont {Zhuo}}, \bibinfo
  {author} {\bibfnamefont {Z.}~\bibnamefont {Cheng}}, \bibinfo {author}
  {\bibfnamefont {Z.}~\bibnamefont {Wang}}, \bibinfo {author} {\bibfnamefont
  {X.}~\bibnamefont {Wan}},\ and\ \bibinfo {author} {\bibfnamefont
  {X.}~\bibnamefont {Chen}},\ }\href
  {https://doi.org/10.1038/s41535-020-0239-z} {\bibfield  {journal} {\bibinfo
  {journal} {npj Quantum Materials}\ }\textbf {\bibinfo {volume} {5}},\
  \bibinfo {pages} {36} (\bibinfo {year} {2020})}\BibitemShut {NoStop}%
\bibitem [{\citenamefont {Fu}\ \emph {et~al.}(2020)\citenamefont {Fu},
  \citenamefont {Wang},\ and\ \citenamefont {Shen}}]{Fu2020}%
  \BibitemOpen
  \bibfield  {author} {\bibinfo {author} {\bibfnamefont {B.}~\bibnamefont
  {Fu}}, \bibinfo {author} {\bibfnamefont {H.-W.}\ \bibnamefont {Wang}},\ and\
  \bibinfo {author} {\bibfnamefont {S.-Q.}\ \bibnamefont {Shen}},\ }\href
  {https://doi.org/10.1103/PhysRevLett.125.256601} {\bibfield  {journal}
  {\bibinfo  {journal} {Phys. Rev. Lett.}\ }\textbf {\bibinfo {volume} {125}},\
  \bibinfo {pages} {256601} (\bibinfo {year} {2020})}\BibitemShut {NoStop}%
\bibitem [{\citenamefont {Wang}(2021)}]{Wang2021}%
  \BibitemOpen
  \bibfield  {author} {\bibinfo {author} {\bibfnamefont {C.}~\bibnamefont
  {Wang}},\ }\href {https://doi.org/10.1103/PhysRevLett.126.126601} {\bibfield
  {journal} {\bibinfo  {journal} {Phys. Rev. Lett.}\ }\textbf {\bibinfo
  {volume} {126}},\ \bibinfo {pages} {126601} (\bibinfo {year}
  {2021})}\BibitemShut {NoStop}%
\bibitem [{SM()}]{SM}%
  \BibitemOpen
  \href@noop {} {\bibinfo  {journal} {See Supplemental Material for
  additional data and discussion, which includes Refs. \cite{Armitage2018,
  Ando2013, Shoenberg2009, Yang2018, Alexandradinata2018, Furuseth1973,
  Skelton1982, Sambongi1986, Fjellvag1986, Smit1958, Breunig2017, Lv2017}}\
  }\BibitemShut {NoStop}%
\bibitem [{\citenamefont {Hafez}\ \emph {et~al.}(2018)\citenamefont {Hafez},
  \citenamefont {Kovalev}, \citenamefont {Deinert}, \citenamefont {Mics},
  \citenamefont {Green}, \citenamefont {Awari}, \citenamefont {Chen},
  \citenamefont {Germanskiy}, \citenamefont {Lehnert}, \citenamefont
  {Teichert}, \citenamefont {Wang}, \citenamefont {Tielrooij}, \citenamefont
  {Liu}, \citenamefont {Chen}, \citenamefont {Narita}, \citenamefont {Müllen},
  \citenamefont {Bonn}, \citenamefont {Gensch},\ and\ \citenamefont
  {Turchinovich}}]{Hafez2018}%
  \BibitemOpen
\bibfield  {journal} {  }\bibfield  {author} {\bibinfo {author} {\bibfnamefont
  {H.~A.}\ \bibnamefont {Hafez}}, \bibinfo {author} {\bibfnamefont
  {S.}~\bibnamefont {Kovalev}}, \bibinfo {author} {\bibfnamefont {J.-C.}\
  \bibnamefont {Deinert}}, \bibinfo {author} {\bibfnamefont {Z.}~\bibnamefont
  {Mics}}, \bibinfo {author} {\bibfnamefont {B.}~\bibnamefont {Green}},
  \bibinfo {author} {\bibfnamefont {N.}~\bibnamefont {Awari}}, \bibinfo
  {author} {\bibfnamefont {M.}~\bibnamefont {Chen}}, \bibinfo {author}
  {\bibfnamefont {S.}~\bibnamefont {Germanskiy}}, \bibinfo {author}
  {\bibfnamefont {U.}~\bibnamefont {Lehnert}}, \bibinfo {author} {\bibfnamefont
  {J.}~\bibnamefont {Teichert}}, \bibinfo {author} {\bibfnamefont
  {Z.}~\bibnamefont {Wang}}, \bibinfo {author} {\bibfnamefont {K.-J.}\
  \bibnamefont {Tielrooij}}, \bibinfo {author} {\bibfnamefont {Z.}~\bibnamefont
  {Liu}}, \bibinfo {author} {\bibfnamefont {Z.}~\bibnamefont {Chen}}, \bibinfo
  {author} {\bibfnamefont {A.}~\bibnamefont {Narita}}, \bibinfo {author}
  {\bibfnamefont {K.}~\bibnamefont {Müllen}}, \bibinfo {author} {\bibfnamefont
  {M.}~\bibnamefont {Bonn}}, \bibinfo {author} {\bibfnamefont {M.}~\bibnamefont
  {Gensch}},\ and\ \bibinfo {author} {\bibfnamefont {D.}~\bibnamefont
  {Turchinovich}},\ }\href {https://doi.org/10.1038/s41586-018-0508-1}
  {\bibfield  {journal} {\bibinfo  {journal} {Nature}\ }\textbf {\bibinfo
  {volume} {561}},\ \bibinfo {pages} {507} (\bibinfo {year}
  {2018})}\BibitemShut {NoStop}%
\bibitem [{\citenamefont {Sun}\ \emph {et~al.}(2019)\citenamefont {Sun},
  \citenamefont {Yi}, \citenamefont {Song}, \citenamefont {Clark},
  \citenamefont {Huang}, \citenamefont {Shan}, \citenamefont {Wu},
  \citenamefont {Huang}, \citenamefont {Gao}, \citenamefont {Chen},
  \citenamefont {McGuire}, \citenamefont {Cao}, \citenamefont {Xiao},
  \citenamefont {Liu}, \citenamefont {Yao}, \citenamefont {Xu},\ and\
  \citenamefont {Wu}}]{Sun2019}%
  \BibitemOpen
  \bibfield  {author} {\bibinfo {author} {\bibfnamefont {Z.}~\bibnamefont
  {Sun}}, \bibinfo {author} {\bibfnamefont {Y.}~\bibnamefont {Yi}}, \bibinfo
  {author} {\bibfnamefont {T.}~\bibnamefont {Song}}, \bibinfo {author}
  {\bibfnamefont {G.}~\bibnamefont {Clark}}, \bibinfo {author} {\bibfnamefont
  {B.}~\bibnamefont {Huang}}, \bibinfo {author} {\bibfnamefont
  {Y.}~\bibnamefont {Shan}}, \bibinfo {author} {\bibfnamefont {S.}~\bibnamefont
  {Wu}}, \bibinfo {author} {\bibfnamefont {D.}~\bibnamefont {Huang}}, \bibinfo
  {author} {\bibfnamefont {C.}~\bibnamefont {Gao}}, \bibinfo {author}
  {\bibfnamefont {Z.}~\bibnamefont {Chen}}, \bibinfo {author} {\bibfnamefont
  {M.}~\bibnamefont {McGuire}}, \bibinfo {author} {\bibfnamefont
  {T.}~\bibnamefont {Cao}}, \bibinfo {author} {\bibfnamefont {D.}~\bibnamefont
  {Xiao}}, \bibinfo {author} {\bibfnamefont {W.-T.}\ \bibnamefont {Liu}},
  \bibinfo {author} {\bibfnamefont {W.}~\bibnamefont {Yao}}, \bibinfo {author}
  {\bibfnamefont {X.}~\bibnamefont {Xu}},\ and\ \bibinfo {author}
  {\bibfnamefont {S.}~\bibnamefont {Wu}},\ }\href
  {https://doi.org/10.1038/s41586-019-1445-3} {\bibfield  {journal} {\bibinfo
  {journal} {Nature}\ }\textbf {\bibinfo {volume} {572}},\ \bibinfo {pages}
  {497} (\bibinfo {year} {2019})}\BibitemShut {NoStop}%
\bibitem [{\citenamefont {Kovalev}\ \emph {et~al.}(2020)\citenamefont
  {Kovalev}, \citenamefont {Dantas}, \citenamefont {Germanskiy}, \citenamefont
  {Deinert}, \citenamefont {Green}, \citenamefont {Ilyakov}, \citenamefont
  {Awari}, \citenamefont {Chen}, \citenamefont {Bawatna}, \citenamefont {Ling},
  \citenamefont {Xiu}, \citenamefont {van Loosdrecht}, \citenamefont
  {Surówka}, \citenamefont {Oka},\ and\ \citenamefont {Wang}}]{Kovalev2020}%
  \BibitemOpen
  \bibfield  {author} {\bibinfo {author} {\bibfnamefont {S.}~\bibnamefont
  {Kovalev}}, \bibinfo {author} {\bibfnamefont {R.~M.~A.}\ \bibnamefont
  {Dantas}}, \bibinfo {author} {\bibfnamefont {S.}~\bibnamefont {Germanskiy}},
  \bibinfo {author} {\bibfnamefont {J.-C.}\ \bibnamefont {Deinert}}, \bibinfo
  {author} {\bibfnamefont {B.}~\bibnamefont {Green}}, \bibinfo {author}
  {\bibfnamefont {I.}~\bibnamefont {Ilyakov}}, \bibinfo {author} {\bibfnamefont
  {N.}~\bibnamefont {Awari}}, \bibinfo {author} {\bibfnamefont
  {M.}~\bibnamefont {Chen}}, \bibinfo {author} {\bibfnamefont {M.}~\bibnamefont
  {Bawatna}}, \bibinfo {author} {\bibfnamefont {J.}~\bibnamefont {Ling}},
  \bibinfo {author} {\bibfnamefont {F.}~\bibnamefont {Xiu}}, \bibinfo {author}
  {\bibfnamefont {P.~H.~M.}\ \bibnamefont {van Loosdrecht}}, \bibinfo {author}
  {\bibfnamefont {P.}~\bibnamefont {Surówka}}, \bibinfo {author}
  {\bibfnamefont {T.}~\bibnamefont {Oka}},\ and\ \bibinfo {author}
  {\bibfnamefont {Z.}~\bibnamefont {Wang}},\ }\href
  {https://doi.org/10.1038/s41467-020-16133-8} {\bibfield  {journal} {\bibinfo
  {journal} {Nature Communications}\ }\textbf {\bibinfo {volume} {11}},\
  \bibinfo {pages} {2451} (\bibinfo {year} {2020})}\BibitemShut {NoStop}%
\bibitem [{\citenamefont {Cheng}\ \emph {et~al.}(2020)\citenamefont {Cheng},
  \citenamefont {Kanda}, \citenamefont {Ikeda}, \citenamefont {Matsuda},
  \citenamefont {Xia}, \citenamefont {Schumann}, \citenamefont {Stemmer},
  \citenamefont {Itatani}, \citenamefont {Armitage},\ and\ \citenamefont
  {Matsunaga}}]{Cheng2020}%
  \BibitemOpen
  \bibfield  {author} {\bibinfo {author} {\bibfnamefont {B.}~\bibnamefont
  {Cheng}}, \bibinfo {author} {\bibfnamefont {N.}~\bibnamefont {Kanda}},
  \bibinfo {author} {\bibfnamefont {T.~N.}\ \bibnamefont {Ikeda}}, \bibinfo
  {author} {\bibfnamefont {T.}~\bibnamefont {Matsuda}}, \bibinfo {author}
  {\bibfnamefont {P.}~\bibnamefont {Xia}}, \bibinfo {author} {\bibfnamefont
  {T.}~\bibnamefont {Schumann}}, \bibinfo {author} {\bibfnamefont
  {S.}~\bibnamefont {Stemmer}}, \bibinfo {author} {\bibfnamefont
  {J.}~\bibnamefont {Itatani}}, \bibinfo {author} {\bibfnamefont {N.~P.}\
  \bibnamefont {Armitage}},\ and\ \bibinfo {author} {\bibfnamefont
  {R.}~\bibnamefont {Matsunaga}},\ }\href
  {https://doi.org/10.1103/PhysRevLett.124.117402} {\bibfield  {journal}
  {\bibinfo  {journal} {Physical Review Letters}\ }\textbf {\bibinfo {volume}
  {124}},\ \bibinfo {pages} {117402} (\bibinfo {year} {2020})}\BibitemShut
  {NoStop}%
\bibitem [{\citenamefont {Liu}\ \emph {et~al.}(2016)\citenamefont {Liu},
  \citenamefont {Yuan}, \citenamefont {Zhang}, \citenamefont {Jin},
  \citenamefont {Narayan}, \citenamefont {Luo}, \citenamefont {Chen},
  \citenamefont {Yang}, \citenamefont {Zou}, \citenamefont {Wu}, \citenamefont
  {Sanvito}, \citenamefont {Xia}, \citenamefont {Li}, \citenamefont {Wang},\
  and\ \citenamefont {Xiu}}]{Liu2016}%
  \BibitemOpen
  \bibfield  {author} {\bibinfo {author} {\bibfnamefont {Y.}~\bibnamefont
  {Liu}}, \bibinfo {author} {\bibfnamefont {X.}~\bibnamefont {Yuan}}, \bibinfo
  {author} {\bibfnamefont {C.}~\bibnamefont {Zhang}}, \bibinfo {author}
  {\bibfnamefont {Z.}~\bibnamefont {Jin}}, \bibinfo {author} {\bibfnamefont
  {A.}~\bibnamefont {Narayan}}, \bibinfo {author} {\bibfnamefont
  {C.}~\bibnamefont {Luo}}, \bibinfo {author} {\bibfnamefont {Z.}~\bibnamefont
  {Chen}}, \bibinfo {author} {\bibfnamefont {L.}~\bibnamefont {Yang}}, \bibinfo
  {author} {\bibfnamefont {J.}~\bibnamefont {Zou}}, \bibinfo {author}
  {\bibfnamefont {X.}~\bibnamefont {Wu}}, \bibinfo {author} {\bibfnamefont
  {S.}~\bibnamefont {Sanvito}}, \bibinfo {author} {\bibfnamefont
  {Z.}~\bibnamefont {Xia}}, \bibinfo {author} {\bibfnamefont {L.}~\bibnamefont
  {Li}}, \bibinfo {author} {\bibfnamefont {Z.}~\bibnamefont {Wang}},\ and\
  \bibinfo {author} {\bibfnamefont {F.}~\bibnamefont {Xiu}},\ }\href
  {https://doi.org/10.1038/ncomms12516} {\bibfield  {journal} {\bibinfo
  {journal} {Nat. Commun.}\ }\textbf {\bibinfo {volume} {7}},\ \bibinfo {pages}
  {12516} (\bibinfo {year} {2016})}\BibitemShut {NoStop}%
\bibitem [{\citenamefont {Kwan}\ \emph {et~al.}(2020)\citenamefont {Kwan},
  \citenamefont {Reiss}, \citenamefont {Han}, \citenamefont {Bristow},
  \citenamefont {Prabhakaran}, \citenamefont {Graf}, \citenamefont {McCollam},
  \citenamefont {Parameswaran},\ and\ \citenamefont {Coldea}}]{Kwan2020}%
  \BibitemOpen
  \bibfield  {author} {\bibinfo {author} {\bibfnamefont {Y.~H.}\ \bibnamefont
  {Kwan}}, \bibinfo {author} {\bibfnamefont {P.}~\bibnamefont {Reiss}},
  \bibinfo {author} {\bibfnamefont {Y.}~\bibnamefont {Han}}, \bibinfo {author}
  {\bibfnamefont {M.}~\bibnamefont {Bristow}}, \bibinfo {author} {\bibfnamefont
  {D.}~\bibnamefont {Prabhakaran}}, \bibinfo {author} {\bibfnamefont
  {D.}~\bibnamefont {Graf}}, \bibinfo {author} {\bibfnamefont {A.}~\bibnamefont
  {McCollam}}, \bibinfo {author} {\bibfnamefont {S.~A.}\ \bibnamefont
  {Parameswaran}},\ and\ \bibinfo {author} {\bibfnamefont {A.~I.}\ \bibnamefont
  {Coldea}},\ }\href {https://doi.org/10.1103/PhysRevResearch.2.012055}
  {\bibfield  {journal} {\bibinfo  {journal} {Phys. Rev. Research}\ }\textbf
  {\bibinfo {volume} {2}},\ \bibinfo {pages} {012055(R)} (\bibinfo {year}
  {2020})}\BibitemShut {NoStop}%
\bibitem [{\citenamefont {Skinner}\ \emph {et~al.}(2012)\citenamefont
  {Skinner}, \citenamefont {Chen},\ and\ \citenamefont
  {Shklovskii}}]{Skinner2012}%
  \BibitemOpen
  \bibfield  {author} {\bibinfo {author} {\bibfnamefont {B.}~\bibnamefont
  {Skinner}}, \bibinfo {author} {\bibfnamefont {T.}~\bibnamefont {Chen}},\ and\
  \bibinfo {author} {\bibfnamefont {B.~I.}\ \bibnamefont {Shklovskii}},\
  }\href@noop {} {\bibfield  {journal} {\bibinfo  {journal} {Phys. Rev. Lett.}\
  }\textbf {\bibinfo {volume} {109}},\ \bibinfo {pages} {176801} (\bibinfo
  {year} {2012})}\BibitemShut {NoStop}%
\bibitem [{\citenamefont {Borgwardt}\ \emph {et~al.}(2016)\citenamefont
  {Borgwardt}, \citenamefont {Lux}, \citenamefont {Vergara}, \citenamefont
  {Wang}, \citenamefont {Taskin}, \citenamefont {Segawa}, \citenamefont {van
  Loosdrecht}, \citenamefont {Ando}, \citenamefont {Rosch},\ and\ \citenamefont
  {Gr\"uninger}}]{Borgwardt2016}%
  \BibitemOpen
  \bibfield  {author} {\bibinfo {author} {\bibfnamefont {N.}~\bibnamefont
  {Borgwardt}}, \bibinfo {author} {\bibfnamefont {J.}~\bibnamefont {Lux}},
  \bibinfo {author} {\bibfnamefont {I.}~\bibnamefont {Vergara}}, \bibinfo
  {author} {\bibfnamefont {Z.}~\bibnamefont {Wang}}, \bibinfo {author}
  {\bibfnamefont {A.~A.}\ \bibnamefont {Taskin}}, \bibinfo {author}
  {\bibfnamefont {K.}~\bibnamefont {Segawa}}, \bibinfo {author} {\bibfnamefont
  {P.~H.~M.}\ \bibnamefont {van Loosdrecht}}, \bibinfo {author} {\bibfnamefont
  {Y.}~\bibnamefont {Ando}}, \bibinfo {author} {\bibfnamefont {A.}~\bibnamefont
  {Rosch}},\ and\ \bibinfo {author} {\bibfnamefont {M.}~\bibnamefont
  {Gr\"uninger}},\ }\href@noop {} {\bibfield  {journal} {\bibinfo  {journal}
  {Phys. Rev. B}\ }\textbf {\bibinfo {volume} {93}},\ \bibinfo {pages} {245149}
  (\bibinfo {year} {2016})}\BibitemShut {NoStop}%
\bibitem [{\citenamefont {Armitage}\ \emph {et~al.}(2018)\citenamefont
  {Armitage}, \citenamefont {Mele},\ and\ \citenamefont
  {Vishwanath}}]{Armitage2018}%
  \BibitemOpen
  \bibfield  {author} {\bibinfo {author} {\bibfnamefont {N.}~\bibnamefont
  {Armitage}}, \bibinfo {author} {\bibfnamefont {E.~J.}\ \bibnamefont {Mele}},\
  and\ \bibinfo {author} {\bibfnamefont {A.}~\bibnamefont {Vishwanath}},\
  }\href@noop {} {\bibfield  {journal} {\bibinfo  {journal} {Rev. Mod. Phys.}\
  }\textbf {\bibinfo {volume} {90}},\ \bibinfo {pages} {015001} (\bibinfo
  {year} {2018})}\BibitemShut {NoStop}%
\bibitem [{\citenamefont {Ando}(2013)}]{Ando2013}%
  \BibitemOpen
  \bibfield  {author} {\bibinfo {author} {\bibfnamefont {Y.}~\bibnamefont
  {Ando}},\ }\href@noop {} {\bibfield  {journal} {\bibinfo  {journal} {J. Phys.
  Soc. Jpn.}\ }\textbf {\bibinfo {volume} {82}},\ \bibinfo {pages} {102001}
  (\bibinfo {year} {2013})}\BibitemShut {NoStop}%
\bibitem [{\citenamefont {Shoenberg}(2009)}]{Shoenberg2009}%
  \BibitemOpen
  \bibfield  {author} {\bibinfo {author} {\bibfnamefont {D.}~\bibnamefont
  {Shoenberg}},\ }\href@noop {} {\emph {\bibinfo {title} {Magnetic oscillations
  in metals}}}\ (\bibinfo  {publisher} {Cambridge university press},\ \bibinfo
  {year} {2009})\BibitemShut {NoStop}%
\bibitem [{\citenamefont {Yang}\ \emph {et~al.}(2018)\citenamefont {Yang},
  \citenamefont {Moessner},\ and\ \citenamefont {Lim}}]{Yang2018}%
  \BibitemOpen
  \bibfield  {author} {\bibinfo {author} {\bibfnamefont {H.}~\bibnamefont
  {Yang}}, \bibinfo {author} {\bibfnamefont {R.}~\bibnamefont {Moessner}},\
  and\ \bibinfo {author} {\bibfnamefont {L.-K.}\ \bibnamefont {Lim}},\ }\href
  {https://doi.org/10.1103/PhysRevB.97.165118} {\bibfield  {journal} {\bibinfo
  {journal} {Phys, Rev, B}\ }\textbf {\bibinfo {volume} {97}},\ \bibinfo
  {pages} {165118} (\bibinfo {year} {2018})}\BibitemShut {NoStop}%
\bibitem [{\citenamefont {Alexandradinata}\ and\ \citenamefont
  {Glazman}(2018)}]{Alexandradinata2018}%
  \BibitemOpen
  \bibfield  {author} {\bibinfo {author} {\bibfnamefont {A.}~\bibnamefont
  {Alexandradinata}}\ and\ \bibinfo {author} {\bibfnamefont {L.}~\bibnamefont
  {Glazman}},\ }\href {https://doi.org/10.1103/PhysRevB.97.144422} {\bibfield
  {journal} {\bibinfo  {journal} {Phys. Rev. B}\ }\textbf {\bibinfo {volume}
  {97}},\ \bibinfo {pages} {144422} (\bibinfo {year} {2018})}\BibitemShut
  {NoStop}%
\bibitem [{\citenamefont {Furuseth}\ \emph {et~al.}(1973)\citenamefont
  {Furuseth}, \citenamefont {Brattas},\ and\ \citenamefont
  {Kjekshus}}]{Furuseth1973}%
  \BibitemOpen
  \bibfield  {author} {\bibinfo {author} {\bibfnamefont {S.}~\bibnamefont
  {Furuseth}}, \bibinfo {author} {\bibfnamefont {L.}~\bibnamefont {Brattas}},\
  and\ \bibinfo {author} {\bibfnamefont {A.}~\bibnamefont {Kjekshus}},\ }\href
  {https://doi.org/10.3891/acta.chem.scand.27-2367} {\bibfield  {journal}
  {\bibinfo  {journal} {Acta Chem. Scand.}\ }\textbf {\bibinfo {volume} {27}},\
  \bibinfo {pages} {2367} (\bibinfo {year} {1973})}\BibitemShut {NoStop}%
\bibitem [{\citenamefont {Skelton}\ \emph {et~al.}(1982)\citenamefont
  {Skelton}, \citenamefont {Wieting}, \citenamefont {Wolf}, \citenamefont
  {Fuller}, \citenamefont {Gubser}, \citenamefont {Francavilla},\ and\
  \citenamefont {Levy}}]{Skelton1982}%
  \BibitemOpen
  \bibfield  {author} {\bibinfo {author} {\bibfnamefont {E.~F.}\ \bibnamefont
  {Skelton}}, \bibinfo {author} {\bibfnamefont {T.~J.}\ \bibnamefont
  {Wieting}}, \bibinfo {author} {\bibfnamefont {S.~A.}\ \bibnamefont {Wolf}},
  \bibinfo {author} {\bibfnamefont {W.~W.}\ \bibnamefont {Fuller}}, \bibinfo
  {author} {\bibfnamefont {D.~U.}\ \bibnamefont {Gubser}}, \bibinfo {author}
  {\bibfnamefont {T.~L.}\ \bibnamefont {Francavilla}},\ and\ \bibinfo {author}
  {\bibfnamefont {F.}~\bibnamefont {Levy}},\ }\href
  {https://doi.org/https://doi.org/10.1016/0038-1098(82)91016-X} {\bibfield
  {journal} {\bibinfo  {journal} {Solid State Communications}\ }\textbf
  {\bibinfo {volume} {42}},\ \bibinfo {pages} {1} (\bibinfo {year}
  {1982})}\BibitemShut {NoStop}%
\bibitem [{\citenamefont {Sambongi}(1986)}]{Sambongi1986}%
  \BibitemOpen
  \bibfield  {author} {\bibinfo {author} {\bibfnamefont {T.}~\bibnamefont
  {Sambongi}},\ }\bibinfo {title} {Pentachalcogenides of transition metals},\
  in\ \href {https://doi.org/10.1007/978-94-009-4528-9_7} {\emph {\bibinfo
  {booktitle} {Crystal Chemistry and Properties of Materials with
  Quasi-One-Dimensional Structures: A Chemical and Physical Synthetic
  Approach}}},\ \bibinfo {editor} {edited by\ \bibinfo {editor} {\bibfnamefont
  {J.}~\bibnamefont {Rouxel}}}\ (\bibinfo  {publisher} {Springer Netherlands},\
  \bibinfo {address} {Dordrecht},\ \bibinfo {year} {1986})\ pp.\ \bibinfo
  {pages} {281--313}\BibitemShut {NoStop}%
\bibitem [{\citenamefont {Fjellvag}\ and\ \citenamefont
  {Kjekshus}(1986)}]{Fjellvag1986}%
  \BibitemOpen
  \bibfield  {author} {\bibinfo {author} {\bibfnamefont {H.}~\bibnamefont
  {Fjellvag}}\ and\ \bibinfo {author} {\bibfnamefont {A.}~\bibnamefont
  {Kjekshus}},\ }\href
  {https://doi.org/https://doi.org/10.1016/0038-1098(86)90536-3} {\bibfield
  {journal} {\bibinfo  {journal} {Solid State Commun.}\ }\textbf {\bibinfo
  {volume} {60}},\ \bibinfo {pages} {91} (\bibinfo {year} {1986})}\BibitemShut
  {NoStop}%
\bibitem [{\citenamefont {Smit}(1958)}]{Smit1958}%
  \BibitemOpen
  \bibfield  {author} {\bibinfo {author} {\bibfnamefont {J.}~\bibnamefont
  {Smit}},\ }\href
  {https://doi.org/https://doi.org/10.1016/S0031-8914(58)93541-9} {\bibfield
  {journal} {\bibinfo  {journal} {Physica}\ }\textbf {\bibinfo {volume} {24}},\
  \bibinfo {pages} {39} (\bibinfo {year} {1958})}\BibitemShut {NoStop}%
\bibitem [{\citenamefont {Breunig}\ \emph {et~al.}(2017)\citenamefont
  {Breunig}, \citenamefont {Wang}, \citenamefont {Taskin}, \citenamefont {Lux},
  \citenamefont {Rosch},\ and\ \citenamefont {Ando}}]{Breunig2017}%
  \BibitemOpen
  \bibfield  {author} {\bibinfo {author} {\bibfnamefont {O.}~\bibnamefont
  {Breunig}}, \bibinfo {author} {\bibfnamefont {Z.}~\bibnamefont {Wang}},
  \bibinfo {author} {\bibfnamefont {A.~A.}\ \bibnamefont {Taskin}}, \bibinfo
  {author} {\bibfnamefont {J.}~\bibnamefont {Lux}}, \bibinfo {author}
  {\bibfnamefont {A.}~\bibnamefont {Rosch}},\ and\ \bibinfo {author}
  {\bibfnamefont {Y.}~\bibnamefont {Ando}},\ }\href
  {https://doi.org/10.1038/ncomms15545
  https://www.nature.com/articles/ncomms15545#supplementary-information}
  {\bibfield  {journal} {\bibinfo  {journal} {Nat. Commun.}\ }\textbf {\bibinfo
  {volume} {8}},\ \bibinfo {pages} {15545} (\bibinfo {year}
  {2017})}\BibitemShut {NoStop}%
\bibitem [{\citenamefont {Lv}\ \emph {et~al.}(2017)\citenamefont {Lv},
  \citenamefont {Zhang}, \citenamefont {Zhang}, \citenamefont {Pang},
  \citenamefont {Yao}, \citenamefont {Chen}, \citenamefont {Ye}, \citenamefont
  {Zhou}, \citenamefont {Zhang},\ and\ \citenamefont {Chen}}]{Lv2017}%
  \BibitemOpen
  \bibfield  {author} {\bibinfo {author} {\bibfnamefont {Y.-Y.}\ \bibnamefont
  {Lv}}, \bibinfo {author} {\bibfnamefont {F.}~\bibnamefont {Zhang}}, \bibinfo
  {author} {\bibfnamefont {B.-B.}\ \bibnamefont {Zhang}}, \bibinfo {author}
  {\bibfnamefont {B.}~\bibnamefont {Pang}}, \bibinfo {author} {\bibfnamefont
  {S.-H.}\ \bibnamefont {Yao}}, \bibinfo {author} {\bibfnamefont {Y.~B.}\
  \bibnamefont {Chen}}, \bibinfo {author} {\bibfnamefont {L.}~\bibnamefont
  {Ye}}, \bibinfo {author} {\bibfnamefont {J.}~\bibnamefont {Zhou}}, \bibinfo
  {author} {\bibfnamefont {S.-T.}\ \bibnamefont {Zhang}},\ and\ \bibinfo
  {author} {\bibfnamefont {Y.-F.}\ \bibnamefont {Chen}},\ }\href
  {https://doi.org/10.1016/j.jcrysgro.2016.04.042} {\bibfield  {journal}
  {\bibinfo  {journal} {J. Crystal Growth}\ }\textbf {\bibinfo {volume}
  {457}},\ \bibinfo {pages} {250} (\bibinfo {year} {2017})}\BibitemShut
  {NoStop}%
\end{thebibliography}
\end{document}